\documentclass{ws-ijmpa}

\newcommand{\be}{\begin{equation}}
\newcommand{\ee}{\end{equation}}
\newcommand{\beq}{\begin{eqnarray}}
\newcommand{\eeq}{\end{eqnarray}}
\def\CA{{\cal A}}

\def\CD{{\cal D}}

\def\CF{{\cal F}}
\def\CG{{\cal G}}

\def\CL{{\cal L}}
\def\CM{{\cal M}}
\def\CN{{\cal N}}
\def\CO{{\cal O}}

\def\CS{{\cal S}}

\def\draftnote{\relax}

\newcommand{\RR}{\mathbb{R}}
\newcommand{\ZZ}{\mathbb{Z}}
\newcommand{\SL}{\mathrm{SL}}
\newcommand{\U}{\mathrm{U}}
\newcommand{\SO}{\mathrm{SO}}

\def\Label#1{\label{#1}%
  \smash{\hbox to0pt{\raise1ex\hbox{\tiny[#1]}\hss}}}
\def\noLabels{\let\Label=\label}
\def\nobbibitem{\let\bbibitem=\bibitem}

\begin{document}

\markboth{Joan Sim\'on}
{Extremal black holes, Holography \& Coarse graining}

%
\catchline{}{}{}{}{}
%

\title{EXTREMAL BLACK HOLES, HOLOGRAPHY \& COARSE GRAINING}

\author{JOAN SIMON}

\address{School of Mathematics and Maxwell Institute for Mathematical Sciences\\
King's Buildings, University of Edinburgh\\
Edinburgh, EH9 3JZ, United Kingdom\\
j.simon@ed.ac.uk}

\maketitle


\begin{abstract}
I review some of the concepts at the crossroads of gravitational thermodynamics, holography and quantum mechanics. First, the origin of gravitational thermodynamics due to coarse graining of quantum information is exemplified using the half-BPS sector of ${\cal N}=4$ SYM and its LLM description in type IIB supergravity. The notion of black holes as effective geometries, its relation to the fuzzball programme and some of the puzzles raising for large black holes are discussed.  
Second, I review recent progress for extremal black holes, both microscopically, discussing a constituent model for stationary extremal non-bps black holes, and semiclassically, discussing the extremal black hole/CFT conjecture. The latter is examined from the AdS${}_3$/CFT${}_2$ perspective. Third, I review the importance of the holographic principle to encode non-local gravity features allowing us to relate the gravitational physics of local observers with thermodynamics and the role causality plays in these arguments by identifying horizons (screens) as diathermic walls. I speculate with the emergence of an approximate CFT in the deep IR close to any horizon and its relation with an effective dynamical description of the degrees of freedom living on these holographic screens.


\keywords{black holes; holography; conformal field theory.}
\end{abstract}

\ccode{PACS numbers: 11.25.Hf, 123.1K}

\section{Introduction}	

The equivalence principle linked gravitational physics with the geometry of spacetime \cite{Einstein:1916vd}. The extensive research on solutions to Einstein's equations, or generalisations thereof, and the study of their properties gave rise to many interesting facts and puzzles, especially when interpreted in the light of other branches of physics
\begin{itemlist}
\item the connection between {\it black hole} physics and {\it thermodynamics}\cite{Bardeen:1973gs,Hawking:1974sw,Bekenstein:1973ur}
\item the existence of curvature singularities \cite{Penrose:1964wq,Hawking:1969sw,Hawking:1973uf} and observer dependent horizons
\item the quantum nature of spacetime and its emergence in the classical limit.
\end{itemlist}

General Relativity is viewed as an effective field theory. This follows, for example, from its lack of renormalizability or the existence of singularities. It suggests that a proper understanding of gravity requires the identification of the relevant degrees of freedom in the ultraviolet (UV). The same conclusion may be reached using its connection to thermodynamics, through black hole physics. Thermodynamics is a {\it universal} branch of physics relatively independent of the microscopic details of the system under consideration. The birth of statistical mechanics, initiated with 
Boltzmann's work explaining the properties of macroscopic systems in thermal equilibrium in terms of the statistical averages of their microscopic degrees of freedom \cite{boltzmann}, further motivates the search for a quantum theory of gravity. The assignment of entropy to a classical spacetime raises the question as for what the microscopic degrees of freedom responsible for it are, {\it i.e.} what the analogue of the molecules in a gas is for spacetime.

The universality of gravity, in the sense that any energy source gravitates, may suggest that its proper formulation should follow from a first principle capturing the intricate structure one sees in its classical limit. This is what the {\it holographic principle} attempts \cite{'tHooft:1993gx,Susskind:1994vu}. It states that the number of degrees of freedom $N$, understood as independent quantum states, describing a region $B$ of spacetime, understood as an emergent structure from a more fundamental theory of matter and Lorentzian geometries, is bounded by the area $A(B)$ of its boundary $\partial B$
\begin{equation}
  N \leq \frac{A(B)}{4G_N}\,,
\end{equation}
where $G_N$ stands for Newton's constant.  For a review on the holographic principle, where a more mathematically accurate statement is given in terms of the {\it covariant entropy bound} \cite{Bousso:1999xy} can be found in \cite{Bousso:2002ju}.

\begin{figure}[pb]
\centerline{\psfig{file=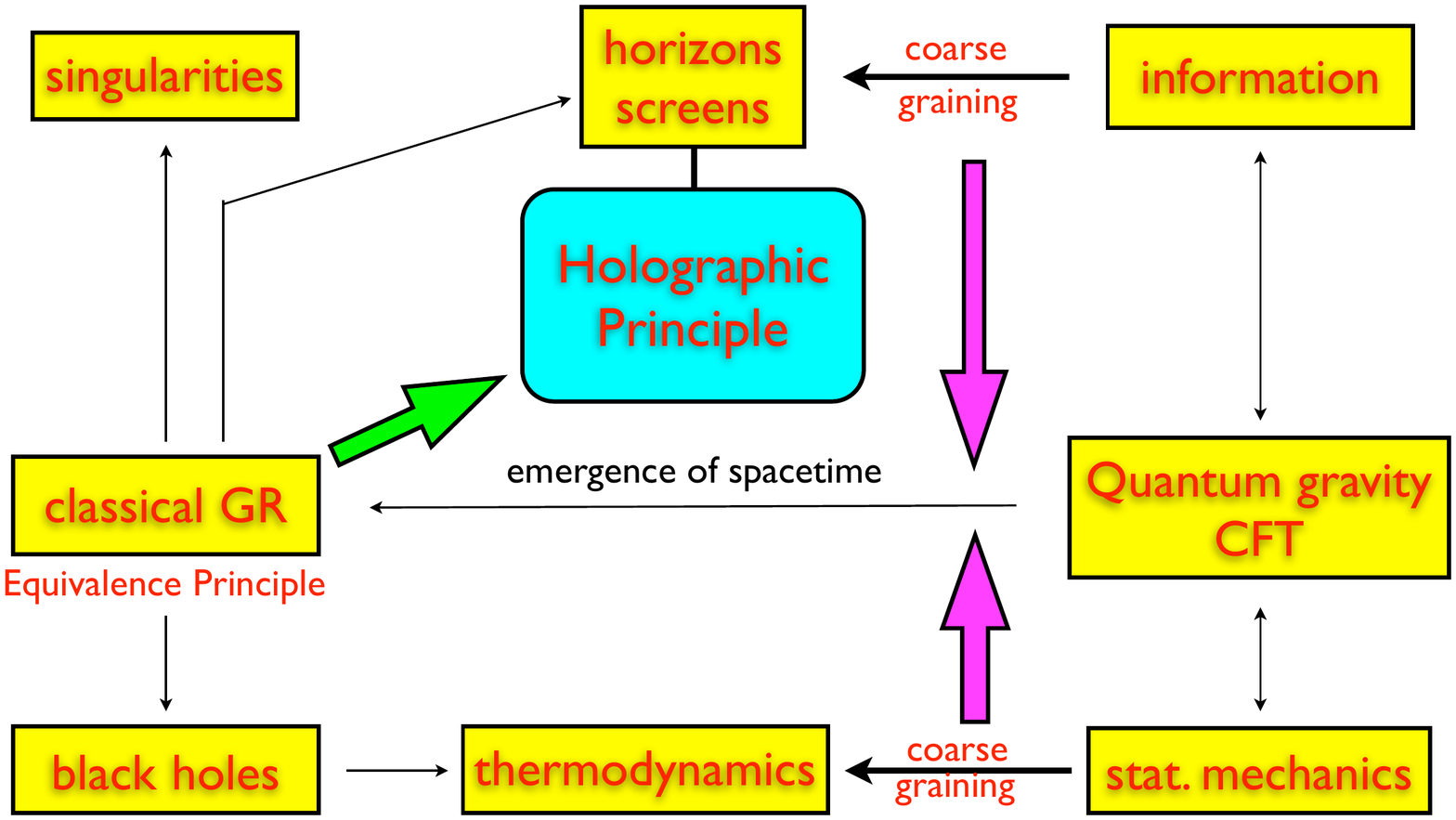,width=15cm}}
\vspace*{8pt}
\caption{Set of concepts and relations discussed in these notes. \label{fig1}}
\end{figure}

The holographic principle challenges the standard quantum field theory description of matter, stresses the non-local nature that gravity manifests in black hole physics, extends it to a general principle, going beyond the notion of event horizons, and emphasises that we can associate entropy and consequently, information, to {\it any} region of spacetime. Importantly, it does not provide an answer for what the degrees of freedom responsible for this entropy are.

String theory provides a mathematically well defined framework where to test some of these ideas.
The importance of duality symmetries \cite{Hull:1994ys,Witten:1995ex}, the discovery of D-branes as capturing non-perturbative aspects of string theory \cite{Polchinski:1995mt} and the formulation of the anti de Sitter (AdS)- conformal field theory (CFT) correspondence conjecture \cite{Maldacena:1997re,Witten:1998qj,Gubser:1998bc} are among the most important developments that have allowed to make both technical and conceptual progress in some of these issues. For example, string theory does provide with additional UV degrees of freedom, it allows to view certain black holes as bound states of D-branes \cite{Strominger:1996sh} and the AdS/CFT correspondence provides an explicit realisation of the holographic principle itself.

In this review, I will mainly be concerned with
\begin{itemlist}
\item the origin of gravitational thermodynamics in black hole physics through the coarse graining of quantum information and the use of the holographic principle to argue that such information loss is not necessarily confined to the black hole singularity, allowing us to view a black hole as a coarse grained object matching its standard thermal state interpretation. These ideas will be exemplified using the half-BPS sector of ${\cal N}=4$ SYM and its LLM description in type IIB supergravity to describe the emergence of classical spacetime, singularities and entropy through coarse graining defined as a renormalization group (RG) transformation in a phase space description of quantum mechanics. The exposition will briefly cover the relation to the fuzzball programme, some speculative technical remarks on the information paradox \cite{Hawking:1976ra} and will conclude with a discussion on some of the difficulties and puzzles appearing when trying to extend these ideas to large black holes.
\item the description of the microscopics and semiclassical methods that have recently been developed for {\it extremal} black holes in an attempt to understand more realistic black holes, explaining their macroscopic entropy, given by the universal Bekenstein-Hawking formula
\begin{equation}
  S_{\text{BH}} = \frac{A}{4G_N}\,,
\label{eq:beh-haw}
\end{equation}
where $A$ stands for the area of the black hole event horizon, in terms of the degrees of freedom living on a 2-dimensional (2d) CFT related to the black hole horizon, whose number of independent quantum states is universally controlled (at large temperatures) by Cardy's formula \cite{Cardy:1986ie}
\begin{equation}
  S_{\text{CFT}} = 2\pi\sqrt{\frac{c}{6}\left(L_0-\frac{c}{24}\right)} +
 2\pi\sqrt{\frac{{\bar c}}{6}\left(\bar{L}_0-\frac{{\bar c}}{24}\right)}\,,
\label{eq:cardy}
\end{equation}
where $c,\,{\bar c}$ stand for the 2d non-chiral CFT central charges and $L_0\pm {\bar L}_0$ are related to the conformal dimension and spin of the quantum states. 
\item the importance of the holographic principle as a unifying principle allowing us to extend the connection between classical gravity and thermodynamics to more general situations than black hole physics. This will relate Einstein's equations to local equilibrium thermodynamics, as perceived by a local observer, to identify stretched horizons (screens) as information storage devices and use semiclassical methods to argue the emergence of approximate chiral CFTs describing the effective dynamics very close to these screens, suggesting a universal link between Bekenstein-Hawking formula \eqref{eq:beh-haw} and Cardy's formula \eqref{eq:cardy}. 
\end{itemlist}

\section{Black Holes \& Thermodynamics}
\label{sec:bhthermo}

The main goal of this section is to explore the {\it origin of gravitational thermodynamics} in the context of black hole physics, focusing in the relation between entropy and the emergence of spacetime and classical singularities through {\it coarse graining} of quantum information at microscopic scales.

The connection between black holes physics and thermodynamics has long been known \cite{Bardeen:1973gs,Hawking:1974sw,Bekenstein:1973ur}. The latter is a branch of physics dealing with systems of an effective infinite number of degrees freedom whose individual interactions are not measurable by a macroscopic observer. They are instead replaced by a coarse grained description involving an effective infinite reduction in the number of degrees of freedom at the price of introducing {\it entropy}, a magnitude measuring the amount of {\it information lost} in the reduction. 

This last remark assumes the existence of a different physical description of the system at smaller scales not available to the macroscopic observer. One way to motivate quantum gravity is certainly to appeal to the universal link between statistical mechanics and thermodynamics when studying the black hole-thermodynamics relation. In black hole physics, it has long believed that the information loss about the true microscopic state of the system, responsible for the existence of entropy, is fully localised at the curvature singularity lying in the deep interior of the black hole. But this expectation is challenged by the holographic principle. Indeed, information takes space, and for a black hole, it involves a classical scale, the horizon scale. This would suggest that information about the state of the black hole, even if typically encoded in Planck scale $(\ell_p)$ physics, may be spread over macroscopic scales, such as the horizon scale, and not being merely localised to the singularity \cite{Mathur:2005ai}.

It is interesting to explore this observation a bit further. Whenever a quantum mechanical formulation is available, black holes are described by a  {\it density matrix} $\rho$. The latter carries an intrinsic entropy
\begin{equation}
  S_\rho = -{\rm Tr}\left(\rho\log \rho\right)\,.
\end{equation}
This is a thermal description of the system, which differs from a microcanonical one, in which one would account for the black hole entropy by counting  {\it pure states} $\{|\Psi_A\rangle\}$ having the same charges as the black hole. Quantum mechanically, density matrices and pure states are distinct. In principle, one can tell their difference apart by computing expectation values of operators $\CO$
\begin{equation}
  {\rm Tr}\left(\rho\,\CO\right) \quad \quad \text{vs} \quad \quad \langle\Psi_A|\CO|\Psi_A\rangle\,.
\end{equation}
But, how large are these differences ? More importantly for the current discussion, do they manifest in classical gravitational physics ? The answer to this question {\it must} necessarily be related to whether the information on the state of the black hole is fully encoded in the singularity or whether it spreads all the way to the horizon, as suggested by the holographic principle, though typically in $\ell_p$ cells. 

The above digression suggests a very conservative approach to the origin of gravitational thermodynamics : the existence of some scale $L$ such that after coarse graining all the microscopic information at smaller scales, all individual microstates would typically look alike. 
If true, a black hole should be understood as a {\it coarse grained} object, matching the standard density matrix description in quantum mechanics. Once more, one may be tempted to associate the scale $L$ with $\ell_p$, but this may well depend on the degeneracy of states encoded in the holographic relation dictated by the Bekenstein-Hawking formula \eqref{eq:beh-haw}. In string theory, if we generically denote the number of constituents of a given system by $N$, the scale that controls the classical gravitational curvature is proportional to $\ell_p\,N^{|\alpha|}$. This was checked by explicit calculations in some particularly symmetric configurations (see \cite{Mathur:2005ai} and references therein).

\begin{figure}[pb]
\centerline{\psfig{file=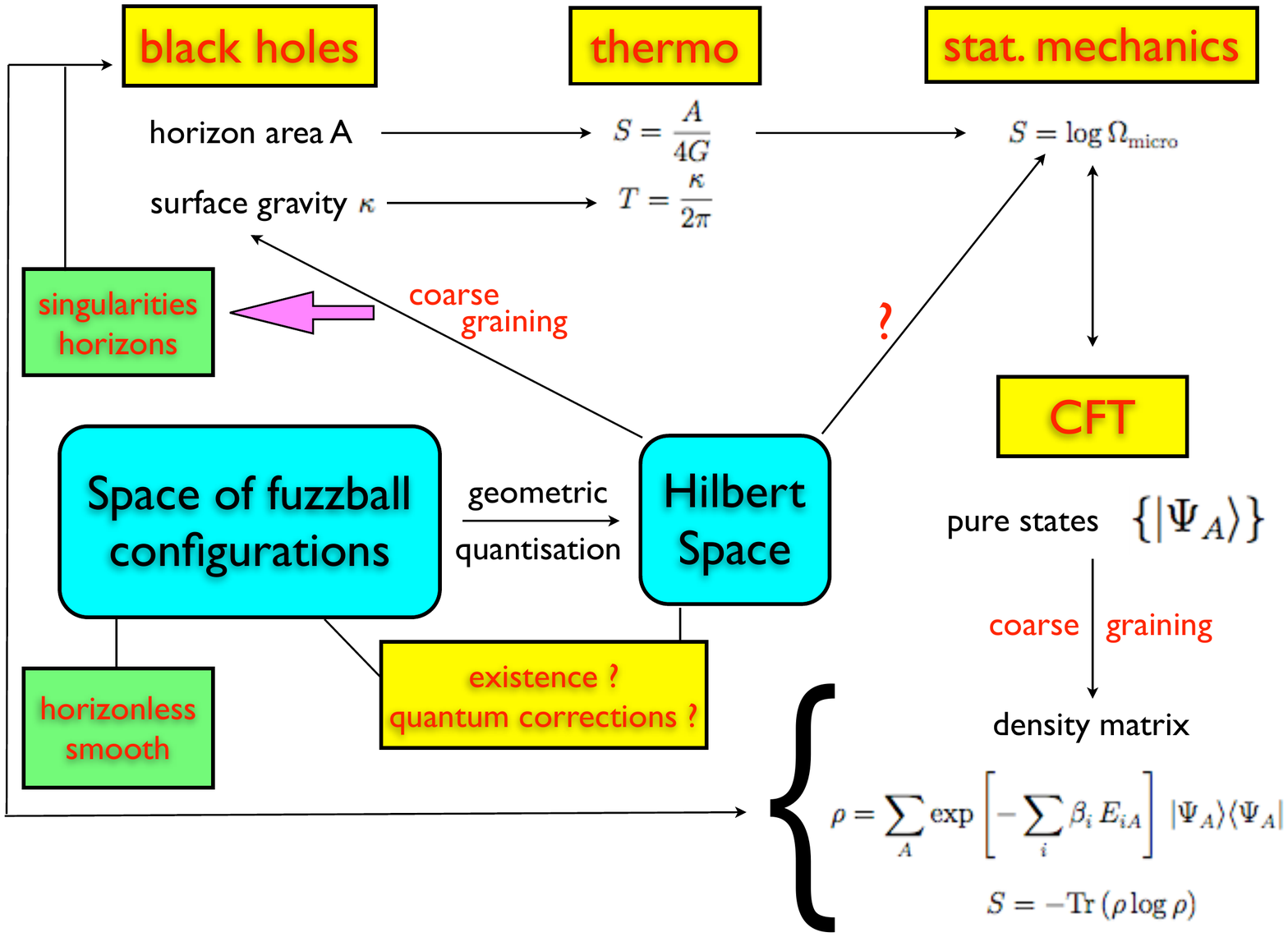,width=15cm}}
\vspace*{8pt}
\caption{From black holes to statistical mechanics \& CFT and back using fuzzball ideas. \label{fig2}}
\end{figure}

Given the relation between black holes and thermal states, it is natural to wonder whether pure states allow any kind of reliable classical description in gravity. Generically, we would not expect this, but in the presence of enough supersymmetry, dynamics may be constrained enough so that as classical gravity becomes reliable, the information on some of these states may remain \footnote{The language used in this argument may induce some readers to think of a transition between an open string (gauge theory) description to a closed string (gravitational) description. This may indeed be helpful, but the argument is more generic. If one assumes the existence of a fully quantum mechanical description of gravity, there is no guarantee that a typical pure state in such Hilbert space allows a reliable description in terms of a classical geometry when taking the classical limit.} . If these geometries exist, one would expect them to have the remarkable global property of being {\it horizonless}, to carry {\it no} entropy, matching their pure state nature. 

This research programme was initiated in \cite{first,oleg-juan}. The starting mathematical problem consists in determining the classical moduli space of configurations having the same asymptotics and charges as a black hole, but having no horizon, being smooth and free of causal closed curves. Any configuration satisfying these requirements will be referred to as a {\it fuzzball} configuration. There exists an extensive literature on the subject. I refer the readers to some excellent reviews (and references therein) : for the importance of fuzzball ideas to the resolution of the information paradox, see \cite{fuzzball,samir-rev}; for the construction and interpretation of supergravity multi-center configurations, see \cite{bena-warner-rev,kostas-marika}; for the use of phase space quantisation, the importance of typicality and the nature of black holes as effective geometries, see \cite{vijay-jan,kostas-marika}. The intuitive idea consists in searching for configurations with the same asymptotics as a given black hole, but whose interior differs by removing the horizon and replacing the singularity by a smooth capped space. In some vague sense, the potentially complicated structure emerging in this inner region is reminiscent of Wheeler's ideas \cite{wheeler,vijay-vishnu}.

Even if such classical moduli space exists, the connection to black hole entropy is still not apparent. This requires to quantise this moduli space, constructing a Hilbert space enabling us to count states with a given set of charges. This step can in principle be achieved through {\it geometric quantisation}, following Crnkovic and Witten \cite{Crnkovic:1986ex} by quantising the phase space of such configurations. This is the approach reviewed in \cite{vijay-jan}. A priori, there is no guarantee this programme may work in a general non-extremal situation. On the contrary, the addition of generic non-extremal excitations may suggest the appearance of singularities in classical gravity. But, for highly supersymmetric configurations, where these solutions have been constructed explicitly, it has provided important insights into the nature of gravitational thermodynamics and the resolution of the information paradox.

In the following, we first discuss a universal statistical feature emerging when describing the differences between pure states quantum mechanically, correlating the difficulty in telling the individual states apart with the entropy of the system. Then, we will review the half-BPS sector of ${\cal N}=4$ SYM and type IIB supergravity in AdS${}_5\times$S${}^5$ where the amount of supersymmetry will allow us to explore the emergence of classical spacetime and singularities as quantum information about the precise state of the system is lost. This discussion will provide explicit evidence that in the case $L\to 0$ there exists an intricate "spacetime" fuzzball, which generically does {\it not} allow for a reliable classical description, but which can effectively be replaced by a singular configuration whose quantum mechanical description agrees with a coarse grained description of the exact quantum mechanical description of the system and which can not be told apart, by a semiclassical observer, from a typical pure state characterised by statistical mechanics considerations. We will conclude with a discussion regarding the important difficulties emerging when trying to extend these ideas to large black holes, i.e. $L\gg \ell_p$. 

\subsection{Distinction of states, typicality and fuzzballs}
\label{sec:typstates}

By first principles, quantum mechanical density matrices and pure states are different. Their differences can be encoded in expectation values (observables). Assume a large black hole allows a quantum mechanical description. There will exist a large degeneracy of pure microstates encoded in its entropy $S$. Consequently, there must exist an statistical description of these states. The latter should allow us to mathematically determine the notion of {\it typical} state, {\it i.e.} how most of these degenerate pure microstates look like. Even more, in principle, it should be possible to determine the typical differences in observables among these typical pure states and the thermal state (density matrix), in an statistical sense. 

These are quantum mechanical questions. Here one will also be interested in encoding the amount of information that survives the classical limit giving rise to a spacetime description that one calls black hole. This seems particularly relevant given the semiclassical nature of the calculations usually used to probe black hole physics. The main idea behind these statistical considerations is to provide some mathematically sounded ground where to compute the deviations from the averaged thermal answers and interpret them gravitationally : the black hole, viewed as a coarse grained effective geometry, will capture the averaged observable answer, but will differ from the exact quantum mechanical one in a given pure microstate. One would like to know by how much and what the most efficient observables are to highlight these differences.

It should not come as a surprise, that such deviations are highly suppressed. It was shown in \cite{mukundus} that the variances in local observables over the relevant Hilbert space of pure states are suppressed by a power of $e^{-S}$. 
To see this, consider a basis of the quantum mechanical Hilbert space of states with energy eigenvalues between $E$ and $E+\Delta E$
\begin{equation}
\CM_{bas} = \left\{ \  |s\rangle  \ : \ H |s \rangle = e_s |s \rangle ~~~;~~~ E \leq e_s \leq E + \Delta E \
\right\} \,.
\label{Eeigenstates}
\end{equation}
The full set of states in this sector of the theory is
\begin{equation}
\mathcal{M}_{sup} =
\left\{|\Psi\rangle = \sum_s c^\psi_s  |s\rangle\,\,,\quad \quad \sum_s |c_s|^2 =1 \right\} \, .
\label{allstates}
\end{equation}
If the entropy of the system is $S(E)$, then the basis in (\ref{Eeigenstates}) has
dimension $e^{S(E)}$:
\begin{equation}
1+ \dim{\CM_{sup}} = | \CM_{bas} | = e^{S(E)}\, .
\label{numstates}
\end{equation}

Take any local operator $\CO$ and compute local observables of the form
\begin{equation}
c_\psi^k(x^1,\ldots , x^k) = \langle \Psi| \CO(x^1) \cdots \CO(x^k) | \Psi \rangle \, .
\end{equation}
To measure how these vary over the ensemble $\mathcal{M}_{sup}$, define their averages and variances as
\begin{eqnarray}
\langle c(x^1,\ldots , x^k) \rangle_{\CM_{sup}} &=& \int D\Psi  \, c_\psi(x^1,\ldots , x^k)
\label{moment2} \\
{\rm var}[c(x^1,\ldots , x^k)] _{\CM_{sup}}  &=&
 \int D\Psi \,\left[(c_\psi(x^1,\ldots , x^k))^2 -
 \langle c(x^1,\ldots , x^k) \rangle_{\CM_{sup}} ^2\right]
 \label{var2}
\end{eqnarray}
The differences between states in $\mathcal{M}_{sup}$ in their responses to $\CO$ will be captured by the standard-deviation to mean ratios
\begin{equation}
\frac{\sigma[c(x^1,\ldots , x^k)]_{\CM_{sup}}}{\langle c(x^1,\ldots , x^k) \rangle_{\CM_{sup}}}
=
\frac{\sqrt{{\rm var}[c(x^1,\ldots , x^k)]_{\CM_{sup}}}}{\langle c(x^1,\ldots , x^k) \rangle_{\CM_{sup}}}
\end{equation}
It was shown in \cite{mukundus} that
\begin{equation}
  {\rm var}[c(x^1,\ldots , x^k)] _{\CM_{sup}} < \frac{1}{e^S + 1}\, {\rm var}[c(x^1,\ldots , x^k)] _{\CM_{bas}}\,.
\end{equation}

This is a general result arising merely from statistical considerations. In particular, it is independent on the dynamical details of the theory. There are two ways to overcome this statistical suppression : given a fixed operator $\CO$, one can wait for long time scales or given a fixed time scale, one can probe the state with operators having large statistical responses.
Concerning the first option, it was pointed out that in lorentzian signature there was generically no time scale smaller than the Poincar\'e recurrence time to overcome these statistical factors \cite{mukundus}, in agreement with previous claims in the literature \cite{poincare-others}. On the other hand, extending previous work 
\cite{kos}, it was shown that the analytic structure of these correlation functions, when working with euclidean signature in the complex plane, allowed the reduction of this time scale. It is unfortunately not clear whether these euclidean correlations can actually be measured by a single observer. The second option can provide a slightly different perspective on the comparison between {\it exact} quantum correlations and results derived from {\it semiclassical} considerations. In particular, the average correlation (\ref{moment2}) will equal the thermal answer. The latter can be computed in a semiclassical approximation by doing a quantum field theory calculation in a black hole background. Our considerations above suggest
\begin{equation}
  \langle \Psi| \Phi(x^1) \cdots \Phi(x^N) | \Psi \rangle_{\text{exact}} = \langle \Psi| \Phi(x^1) \cdots \Phi(x^N) | \Psi \rangle_{\text{eft}} +   
\CO\left(e^{-\left(S-N\right)}\right)
\end{equation}
This observation was also made in \cite{ArkaniHamed:2007ky}, building on \cite{Page:1993wv}. It must be interpreted with care. It means that for a {\it fixed} time scale below the Poincar\'e recurrence, one should expect to find deviations from the thermal answer when the size of the probe operator is comparable to the entropy of the system. In the case of a CFT, size could stand for the conformal dimension of the probe operator, and operators generating black hole microstates themselves would be intuitive examples for this mechanism to be realised.

The importance of this statistical $e^{-S}$ suppression in the context of the information paradox has been recently emphasised in \cite{Balasubramanian:2011dm}. Readers interested in a review on the information paradox itself and the perspective offered by the fuzzball ideas are referred to \cite{Mathur:2008wi,Mathur:2009hf}. For recent discussions regarding the physics of infalling observers in this context, see \cite{Mathur:2010kx,Mathur:2011wg}. 

These field theoretic considerations suggest that black hole backgrounds provide very accurate descriptions of the physics accessible to a classical observer \cite{cg-oleg,masaki-vijay,vijay-lcurve,vijay-jan}. It is natural to ask, especially in an AdS/CFT context, whether the deviations in correlation functions mathematically described in terms of variances translate into some non-trivial spacetime scale and whether the latter survives the semiclassical limit. As mentioned before, a priori, this sounds improbable, since one would expect all these effects to be confined to the Planck scale. It is also technically hard to find reliable and precise results in the field theory side. Even if we restrict to highly supersymmetric sectors, the machinery for computing correlation functions in heavy states is still not fully developed, but there has been important progress towards achieving this goal in \cite{CJR,Koch:2008cm,deMelloKoch:2009jc}. In the context of large black holes in the AdS${}_5$/CFT${}_4$ correspondence, these heavy states will have conformal dimension $\Delta\sim N^2$. Thus, even if working in a large N limit, the degeneracy of states is so large that standard large N perturbative diagrammatic counting arguments are not guaranteed to apply.

\subsection{Half-BPS states in ${\cal N}=4$ SYM vs LLM geometries}

As an explicit example of the ideas outlined above, I will review the gauge and gravity descriptions for half-BPS states in ${\cal N}=4$ Super Yang-Mills (SYM) and type IIB supergravity, respectively. These states are characterised by their R-charge $J$, since supersymmetry forces them to saturate the bound $\Delta=J$, where $\Delta$ is their conformal dimension. This set-up has two important advantages : its large amount of symmetry and its microscopic interpretation in terms of spherical rotating D-branes. The first guarantees that perturbative gauge theory states can be compared with their strongly coupled descriptions in supergravity. The second will help us to establish a dictionary between these two different descriptions. Unfortunately, the degeneracy of these states is not large enough to generate a macroscopic horizon. Hence, this sector of  ${\cal N}=4$ SYM will not be good enough to test our ideas for large black holes.

The particular half-BPS black holes we will be interested in were first found in \cite{original}. These are type IIB $\left(\U(1)\right)^2\times \SO(4)\times \SO(4)$ invariant supergravity configurations with constant dilaton and non-trivial metric and Ramond-Ramond (RR) 4-form potential $C_4$
\begin{eqnarray}
  ds^2 &=& -\frac{\sqrt{\gamma}}{H}\,f\,dt^2 +
  \frac{\sqrt{\gamma}}{f}\, dr^2 + \sqrt{\gamma}\,r^2\,ds^2_{S^3} +
  \sqrt{\gamma}\,L^2\,d\theta^2 + \frac{L^2}{\sqrt{\gamma}}\,
  \sin^2\theta\,ds^2_{\tilde{S}^3} \nonumber \\
  & & + \frac{1}{\sqrt{\gamma}}\cos^2\theta\,[
  L\,d\phi + (H^{-1}-1)\,dt]^2~,
 \label{eq:ssmetric} \\
 C_4 &=& -\frac{r^4}{L}\,\gamma\,dt\wedge d^3\Omega -L\,q\,\cos^2\theta\left(Ld\phi-dt\right)~,
\label{eq:ssmetric1} 
\end{eqnarray}
where $H=1+q/r^2$, $f=1+r^2\,H/L^2$, $\gamma=1+q\sin^2\theta/r^2$ and $L^4=4\pi\,g_s N\,l_s^4$ is the radius of AdS${}_5$, with $g_s$ the string coupling and $l_s$ the string scale. These are asymptotically global AdS${}_5\times$S${}^5$ singular configurations with vanishing horizon size carrying charge 
\begin{equation}
  \Delta = J = \omega\,\frac{N^2}{2}~, \quad \omega = \frac{q}{L^2}~,
 \label{eq:superstardata}
\end{equation}
These were coined {\it superstars} in \cite{myers}, where they were interpreted as a distribution of giant gravitons \cite{susskind-john}. A single giant graviton corresponds to a D3-brane wrapping $\tilde{S}^3$ while rotating at the speed of light in the $\phi$ direction. They preserve the same half of the supersymmetries as a point particle graviton, but they carry an R-charge of order N, i.e. $J\propto N$. Such N scaling is easy to understand : the dimensionless mass carried by the giant $\Delta_{\text{giant}}$ must be proportional to the D3-brane tension $T_{\text{D3}}=1/(8\pi^3g_s\,l_s^4)$ and its worldvolume
\begin{equation}
  \Delta_{\text{giant}} \propto T_{\text{D3}}\,L^4 \propto N\,.
\end{equation}
Physically, a pointlike graviton carrying R-charge of order N expands to an spherical D3-brane, through Myers' effect \cite{Myers:1999ps}. The solution \eqref{eq:ssmetric}-\eqref{eq:ssmetric1} sources a certain number $N_C$ of such giants that can be determined through the flux quantisation condition
\begin{equation}
   N_C = \frac{1}{16\pi G_{10}T_{\text{D3}}}\int_{S^5} F_5\,d^5\Omega = \omega\,N~,
 \label{eq:superstardata1}
\end{equation}
where $F_5=dC_4$ is the RR 5-form field strength.

In the forthcoming sections, our main goal is to provide evidence that the superstar (\ref{eq:ssmetric}) corresponds to a coarse grained configuration emerging from integrating the quantum data of the exact quantum mechanical wave function.

\subsubsection{Gauge theory description \& Typicality}

Due to the state--operator correspondence, highest weight half-BPS states in ${\cal N}=4$ SYM correspond to multi-trace operators, ${\cal O} = \prod_{n,m}\left(\text{Tr}\left(X^m\right)\right)^n$, built of a single real scalar field $X$ transforming in the adjoint representation. In \cite{CJR,Berenstein}, it was shown that the degrees of freedom describing these states are equivalent to $N$ fermions $\{q_1,\,\dots ,q_N\}$ in a one dimensional harmonic potential.
The number of these operators with conformal dimension $\Delta\sim N^2$ at very small chemical potential $\beta$ \cite{david,Kinney:2005ej}
\begin{equation}
  S_{\textrm{1/2-BPS}} \propto N \log N\,,
\end{equation}
captures the large temperature behaviour of N harmonic oscillators plus an $1/N!$ statistical factor, in agreement with its fermionic interpretation. Given the exact nature of the half-BPS partition function, one can extrapolate this answer to strong coupling and estimate the size of an stretched horizon $\rho_h$ by comparing the field theory entropy with the Bekenstein-Hawking entropy \eqref{eq:beh-haw}
\begin{equation}
  S_\text{grav} =  S_{\text{1/2-BPS}} \sim N^2\,\left(\frac{\rho_h}{L}\right)^3 \quad \Rightarrow \quad \frac{\rho_h}{L}\ll 1\,.
 \label{eq:shorizon}
\end{equation}
One concludes the degeneracy of states is not large enough to generate a large horizon.

The fermionic description corresponds to an integrable system with ground state, the filled Fermi sea, consisting of fermions with energies $E^g_i = (i-1)\hbar + \hbar/2$ for $i=1,\ldots, N$. Every excitation corresponds to a half-BPS state where the energy of the individual fermions is $E_i = f_i  \hbar + \hbar/2$. Thus, there exists a correspondence between states and a set of unique non-negative ordered integers $f_i$. Exchanging these with a new set of integers $r_i=f_i-i+1$, describing the excitations above the ground state, one establishes a correspondence between states and Young diagrams with $N$ rows having as many boxes $r_i$ as the excitation spectrum. 

These diagrams are equally determined using the set of variables 
\begin{equation}
c_N = r_1 ~~~~;~~~~ c_{N-i} = r_{i+1} - r_{i} ~~~~;~~~~ i=1,2,\ldots, (N-1) \, .
\label{columnvars}
\end{equation}
These $c_j$ count the number of columns in the Young diagram of length $j$. These are particularly relevant when looking for a microscopic interpretation of gravity configurations with $\Delta\sim N^2$. This is because giant gravitons, as stressed before, have $\Delta \sim \CO (N)$ and the operator dual to a single giant graviton is a subdeterminant \cite{asad-micha}. In terms of Young diagrams, this operator corresponds to a single column. Thus, the number of columns in the diagram corresponds to the number of giant gravitons.

Given the large degeneracy of states having conformal dimension $N^2$, it is natural to use the statistical mechanics of the $N$ fermions to identify how most of these states look like. Using the correspondence with Young diagrams, this will provide the shape of the typical diagram with these number of boxes. Since the superstar configuration (\ref{eq:ssmetric}) supports $N_C$ giant gravitons, it is natural to consider an ensemble of Young diagrams in which the number of columns is held fixed \cite{vijay-lcurve}\footnote{There is more than one ensemble achieving this, see \cite{mueck} for a discussion on this point.}. Implementing this with a Lagrange multiplier, the canonical partition function equals
\begin{equation}
  Z = \sum_{c_1,c_2,\cdots,c_N =1}^\infty e^{-\beta\sum_j jc_j - \lambda\left(\sum_j c_j - N_C\right)} = \zeta^{-N_C}\,\prod_j \frac{1}{1-\zeta\,q^j}~, \quad \zeta = e^{-\lambda}~.
\end{equation}
The ensemble chemical potential $\beta$ is fixed by requiring
\begin{equation}
  \langle E \rangle = \Delta = q\partial_q \log Z(\zeta,\,q) =
  \sum_{j=1}^N \frac{j\,\zeta\,q^j}{1-\zeta\,q^j}~,
 \label{esuper}
\end{equation}
whereas the Lagrange multiplier $(\lambda)$, or equivalently $\zeta$, is fixed by
\begin{equation}
  N_C = \sum_{j=1}^N \langle c_j\rangle = \sum_{j=1}^N
  \frac{\zeta\,q^j}{1-\zeta\,q^j}~,
 \label{ncsuper}
\end{equation}
where we already computed the expected number of columns of length $j$, i.e. $\langle c_j\rangle$.

\paragraph{Limit curve :} Let us introduce two coordinates $x$ and $y$ along the rows and columns of the Young diagram. In our conventions, the origin $(0,0)$ is the bottom left corner of the diagram, $x$ increases going up while $y$ increases to the right. In the fermion language, $x$ labels the particle number and $y$ its excitation above the vacuum. Determining the typical state consists in computing the curve $y(x)$ describing the shape of the Young diagram. This can be done by identifying this mathematical object with the expectation value
\begin{equation}
  y(x) = \sum_{i=N-x}^N\,\langle c_i \rangle~.
 \label{limitdef}
\end{equation}
In the limit $\hbar\to 0$, $N\to \infty$, keeping the Fermi level $\hbar\,N$ fixed, we can treat $x$ and $y$ as continuum variables, replace the summation by an integral, and derive the limit curve \cite{vijay-lcurve}
\begin{equation}
  y(x) = \frac{\log (1-e^{-\beta N})}{\beta} - \frac{\log (1-e^{-\beta (N-x)})}{\beta} \,.
 \label{continuum}
\end{equation}
For a discussion concerning the size of the fluctuations, see \cite{vijay-lcurve,mueck}.
We will be particularly interested in the $\beta\to 0$ limit of this limit curve. Before studying this further, let us review the classical gravitational description for these half-BPS configurations.

\subsubsection{Gravity description}

All the relevant $\U(1)\times \SO(4)\times \SO(4)$ half-BPS supergravity backgrounds for our discussion were constructed in \cite{llm}. These involve a metric
\begin{equation}
  ds^2 = - h^{-2}
  (dt + V_i dx^i)^2 + h^2 (d\eta^2 + dx^idx^i) + \eta\, e^{G} d\Omega_3^2
  + \eta\, e^{ - G} d \tilde \Omega_3^2\,,
 \label{solmetric}
\end{equation}
and a self-dual five-form field strength $F_{(5)}= F \wedge d\Omega_3 + \tilde{F}\wedge d\tilde{\Omega}_3$, where $d\Omega_3$ and $d\tilde{\Omega}_3$ are the volume forms of the two three-spheres where the two $\SO(4)$s are geometrically realised. The full configuration is
uniquely determined in terms of a {\it single} scalar function $z=z(\eta,x_1,x_2)$ satisfying the linear differential equation (see \cite{llm} for a complete discussion)
\begin{equation}
 \partial_i \partial_i z + \eta \partial_\eta \left( \frac{ \partial_\eta z}{\eta} \right) =0~.
\label{eq:zequation}
\end{equation}
Notice $\Phi(\eta;\,x^1,\,x^2) = z\,\eta^{-2}$ satisfies the Laplace equation for an electrostatic potential in six dimensions being spherically symmetric in four of them. The coordinates $x_1,x_2$ parametrize an $\RR^2$, while $\eta$ is the radial coordinate in the transverse $\RR^4$ in this auxiliary six-dimensional manifold. Thus, the general solution
\begin{equation}
  z(\eta;\,x_1,\,x_2) = \frac{\eta^2}{\pi}\int dx_1'\,dx_2'\,
  \frac{z(0;\,x_1',\,x_2')}{[(x-x')^2 + \eta^2]^2}~,
 \label{eq:zsol}
\end{equation}
depends on the boundary condition  $z(0;\,x_1,\,x_2)$. {\it Regularity} only allows $z(0;\,x_1,\,x_2)=\pm 1/2$. Introducing 
$u(0;\,x_1,\,x_2) = 1/2 - z(0;\,x_1,\,x_2)$, the energy and flux quantisation condition are
\begin{eqnarray}
  \Delta &=& \int_{\RR^2}
  \frac{d^2x}{2\pi\hbar}\,\frac{1}{2}\frac{x_1^2+x_2^2}{\hbar}\,
  u(0;\,x_1,\,x_2)-\frac{1}{2}\left(\int_{\RR^2}
  \frac{d^2x}{2\pi\hbar}\, u(0;\,x_1,\,x_2)\right)^2~,   \label{uDelta} \\
  N &=&   \int_{\RR^2}
  \frac{d^2x}{2\pi\hbar}\, u(0;\,x_1,\,x_2) \label{uN} \,,
\end{eqnarray}
where $\hbar = 2\pi\ell_p^4$ due to the non-standard units carried by $\{x_1,\,x_2\}$.
These expressions resemble {\it expectation values} computed in the phase space of one of the fermions appearing in our gauge theory discussion, suggesting the function $u(0;\,x_1,\,x_2)$ should be identified with the semiclassical limit of the quantum single-particle phase space distributions of the dual fermions.

\subsubsection{Gauge-gravity correspondence \& Coarse graining}

The classical moduli space of configurations described in \cite{llm} was geometrically quantised in \cite{Grant:2005qc,Maoz:2005nk}\footnote{The same methods were applied to the D1-D5 system in \cite{Rychkov:2005ji}.}. The Hilbert space they constructed when restricting to the subspace of BPS configurations was isomorphic to the one describing N fermions in a one dimensional harmonic oscillator appearing in the gauge theory. This automatically guarantees that the gauge theory and gravity counting of states match. In the following, we will focus on the information loss by coarse graining of quantum information.

The matching of the gauge theory and gravity descriptions requires a dictionary. Focusing on $\U(1)$ invariant configurations in the $x_1,\,x_2$ plane, it was proposed in \cite{vijay-lcurve} that in the semiclassical limit $\hbar \to 0$ with $\hbar\,N$ fixed, the integral formulae (\ref{uDelta}), (\ref{uN}) extend to differential relations 
\begin{equation}
  \frac{u(0;r^2)}{2\hbar}\,dr^2 = dx~, \quad \quad
  \frac{r^2\,u(0;r^2)}{4\hbar^2} dr^2 = (y(x)+x)\,dx~.
 \label{eq:egain}
\end{equation}
The first equation relates the number of particles in phase space within a band between $r$ and $r + dr$ to the number of particles as determined by the rows of the associated Young diagram. The second equation matches the energy of the particles in phase space within a ring of width $dr$ to the energy in terms of the Young diagram coordinates. This is equivalent to identifying the $y=0$ plane in the bulk with the semiclassical limit of the phase space of a single fermion \cite{Mandal:2005wv,Takayama:2005yq,Grant:2005qc}.

Combining both equations in (\ref{eq:egain}), we find $y(x) + x = r^2/(2\hbar)$ and taking derivatives,
\begin{equation}
  u(0;r^2) = \frac{1}{1 + y'} \quad
  \Leftrightarrow \quad
  z(0;r^2) = \frac{1}{2}\,\frac{y'-1}{y'+1}~.
 \label{eq:cmatch}
\end{equation}
This establishes a dictionary between the boundary condition $u(0;x_1,x_2)$ in supergravity and
the slope ($y'(x)=dy/dx$) of the Young diagram of the corresponding field theory state. 

We will now rederive Eq. (\ref{eq:cmatch}) from a different perspective, emphasising the intrinsic loss of information involved in the semiclassical limit we are taking. Given the exact quantum mechanical phase space distribution, one can seek a new distribution function that is sufficient to describe the effective response of coarse grained semiclassical observables in states that have a limit as $\hbar \to 0$. The latter will be called the {\it coarse grained} or {\it grayscale} distribution. Since we are interested in single Young diagram states, it can only depend on the energy $E = (p^2 + q^2)/2$. To derive this, let $\Delta E$ be a coarse graining scale such that $\Delta E / \hbar \to \infty$ in the $\hbar \to 0$ limit. The {\it grayscale} distribution $g(E)$ must equal the quotient of the number of fermions with energies between $E$ and $E + \Delta E\,$\footnote{We are using the Husimi distribution $\textrm{Hu}(q,p)$ given its nice properties in the classical limit. For further discussion, see \cite{vijay-lcurve}. For a review on phase space distributions, see \cite{wigner}.}
\begin{equation}
R(E,\Delta E) = 2\pi \hbar \, \int_E^{E + \Delta E} dq \, dp \, \textrm{Hu}(q,p),
\label{Req}
\end{equation}
by the area of phase space between these scales
\begin{equation}
{\rm Area} = \int_{E}^{E+\Delta E} dq \, dp = 2\pi \Delta E\,.
\label{Areaeq}
\end{equation}
Hence, the  {\it grayscale} distribution equals
\begin{equation}
g(E) = 2\pi\hbar \left[ \frac{R(E,\Delta E)}{2 \pi \Delta E }\right].
\label{gdef1}
\end{equation}
Since the number of fermions in this area is related to the continuous coordinate $x$ and the area to the energy, we conclude
\begin{equation}
g(E) = \hbar \frac{\Delta x}{\Delta E} = \frac{\hbar}{\partial E/\partial x} = \frac{1}{1 + y^\prime}\,.
\label{gdef2}
\end{equation}
where the gauge-gravity proposal $E = \hbar (x + y(x))$ was used. 

\begin{figure}[pb]
\centerline{\psfig{file=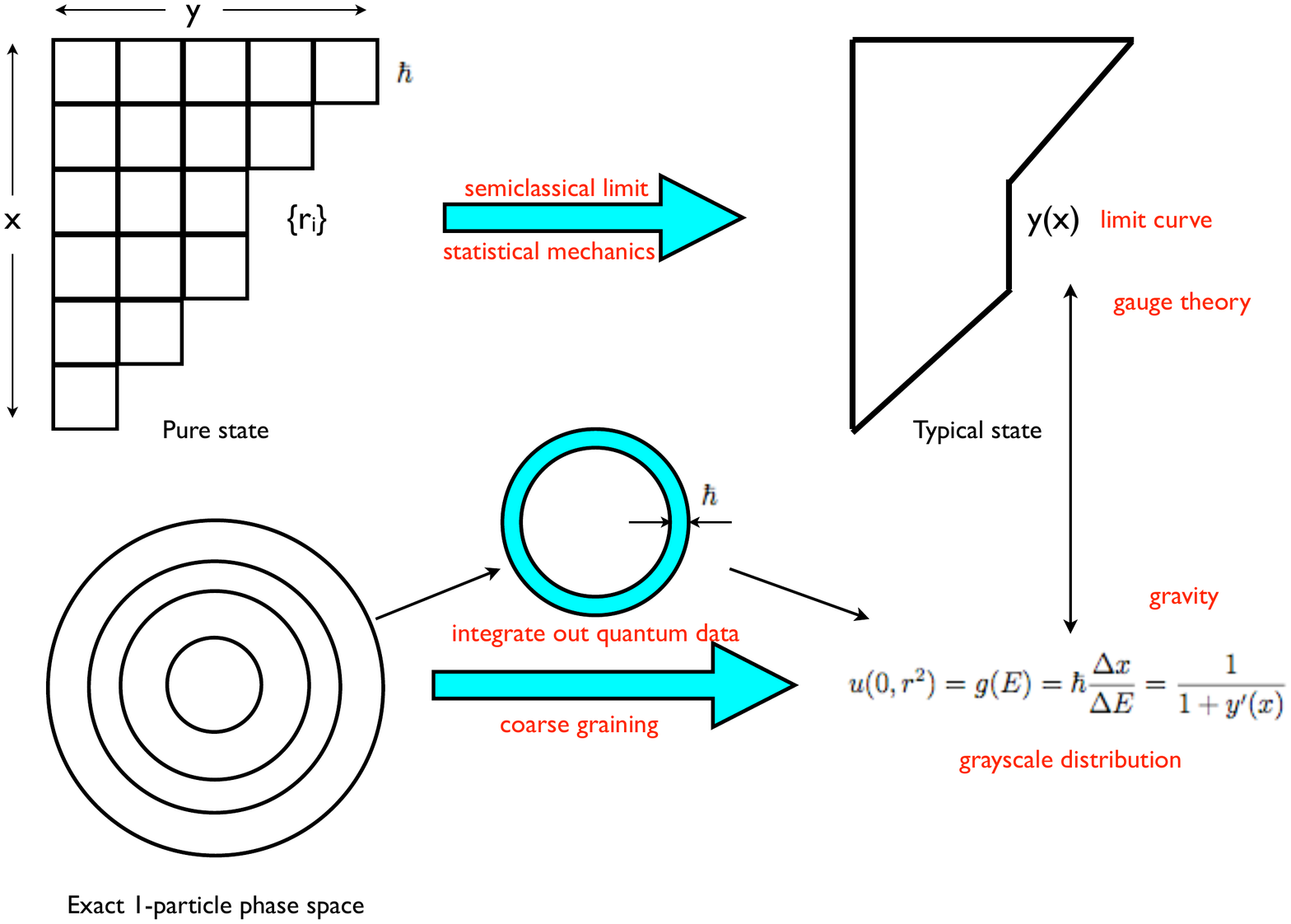,width=15cm}}
\vspace*{8pt}
\caption{From pure states, to typical states and their coarse grained phase space density description providing a gravity boundary condition. \label{fig2a}}
\end{figure}

Remarkably, this is precisely the quantity (\ref{eq:cmatch}) that we proposed on general grounds to determine the classical supergravity half-BPS solution associated to a Young diagram. Thus, explicitly, the proposal is
\begin{equation}
u(0;r^2) = g(r^2/2) = \frac{1}{1 + y^\prime}
\label{proposal2}
\end{equation}
for all half-BPS states that are described by single Young
diagrams in the semiclassical limit.

\paragraph{Matching the superstar geometry :} As argued in \cite{vijay-lcurve}, the entropy of the superstar ensemble is maximised at vanishing chemical potential $\beta$. This is the regime in which we plan to compare the geometry (\ref{eq:ssmetric}) with the one obtained out of the proposal (\ref{proposal2}). First, the $\beta\to 0$ limiting behaviour of the finite chemical potential limit curve (\ref{continuum}) describing the typical state of the superstar ensemble becomes a straight line
\begin{equation}
y = \frac{N_C}{N} x \equiv \omega\, x\,.
\label{sline}
\end{equation}
The grayscale distribution (\ref{gdef2}) will then be a constant fixing the superstar scalar function $z_S(0;r^2)$ to be
\begin{equation}
  z_S(0;r^2) = \left\{ \begin{array}{ll}
               \frac{1}{2}\, \frac{N_C/N - 1}{N_C/N + 1} & {\rm if}\ r^2/2\hbar \le N+N_C, \cr
               \frac{1}{2}                               & {\rm if}\ r^2/2\hbar > N+N_C. 
               \end{array} \right.
 \label{eq:sdroplet}
\end{equation}
Since, within the droplet region, {\it i.e.} $r^2/2\hbar \le N+N_C$, this number is different from $\pm 1/2$, the spacetime is singular. Noting that the coarse grained phase space density derived from our proposal equals
\begin{equation}
u_S(0;r^2) = \frac12 - z_S(0;r^2) = \frac{1}{N_C/N+1},
\end{equation}
in the region of the phase space plane between $r^2/2\hbar = 0$ and $r^2/2\hbar = N + N_C$, and vanishes otherwise, it is straightforward to check, using (\ref{uDelta}) and (\ref{uN}), that (\ref{eq:sdroplet}) reproduces (\ref{eq:superstardata}).

Eq. (\ref{eq:sdroplet}) is a prediction from our proposal and our gauge theory analysis concerning the description of the typical states in the semiclassical limit. We want to reproduce this prediction by explicit analysis of the configuration (\ref{eq:ssmetric}).
If we compare the physical size of the two three-spheres appearing in the superstar metric (\ref{eq:ssmetric}) with their parametrisation in \cite{llm}, we obtain the conditions:
\begin{equation}
  \eta\,e^G =  \sqrt{\gamma}\,r^2~, \quad \quad
  \eta\,e^{-G} = \frac{L^2\,\sin^2\theta_1}{\sqrt{\gamma}}~.
\end{equation}
Using the fact that $z=(1/2)\tanh G$ \cite{llm}, then
\begin{equation}
  z = \frac{1}{2}\frac{r^2\gamma - L^2\sin^2\theta_1}
  {r^2\gamma + L^2\sin^2\theta_1}~.
\end{equation}
Since it is the value $z(\eta=0)$ that is related to the semiclassical distribution function, we must analyse the behaviour of $G$ at $\eta=0$. We observe there are two different regimes where this applies:
\begin{romanlist}[(ii)]
  \item When $\sin\theta_1=0$, $z(\eta=0)=1/2$. Vanishing $\sin\theta_1$ implies the vanishing of the giant graviton distribution. Consequently, it should correspond to absence of fermion excitations in the gauge theory picture. This is precisely reflected in the boundary condition  $z(\eta=0)=1/2$.  
  \item When $r=0$, the giant distribution is non-vanishing. One gets
\begin{equation}
  z(r=0) = \frac{1}{2}\frac{\omega -1}{\omega +1}~.
 \label{eq:zgiant}
\end{equation}
\end{romanlist}
The gravity distribution (\ref{eq:zgiant}) identically matches the one derived from purely field theoretic and statistical mechanical considerations (\ref{eq:sdroplet}). This establishes the singular superstar configuration corresponds to a coarse grained object matching the notion of typical state emerging from the canonical ensemble analysis.

\subsubsection{Measurability, coarse graining and entropy}

The integrability of the system of N fermions and the non-renormalisation properties of this sector of the theory allow us to do better. First, we can be more precise about the kind of information loss occurring when taking the semiclassical limit and the subset of quantum states allowing a reliable gravitational description in that regime. Second, we can derive the semiclassical partition function from a first principle calculation on the gauge theory side that will highlight the definition of coarse graining as a renormalization group transformation in phase space.

To make integrability more explicit, one can either specify a basis of states in terms of the energies $\{ f_1,\cdots f_N \}$ or in terms of the gauge invariant moments 
\begin{equation}
M_k = \sum_{i=1}^N f_i^k = {\rm Tr}(H_N^k/\hbar^k)  ~~~;~~~ k=0,\cdots N \, ,
\label{momentdef}
\end{equation}
where $H_N$ is the $N$ fermion Hamiltonian with the zero point energy removed.  These $M_k$ are conserved charges of the system of fermions in a harmonic potential \cite{surya} and allow a reconstruction of the spectrum $\CF=\{ f_1,\cdots f_N \}$ \cite{info-loss}.

We are interested in reading this information from the bulk. Since the phase space distribution was identified with the scalar function $u(0;r^2)$, we search for the spectrum $\{M_1,\cdots M_N\}$ in its multipole expansion.
After some algebra, one finds \cite{info-loss}
\begin{equation}
u(\rho,\,\theta)
=  2\cos^2 \theta \sum_{l=0}^{\infty}
\frac{\hbar^{l+1}\sum_{f\in \mathcal{F}} A^l(f)}{\rho^{2l+2}}
(-1)^l(l+1)\, \, {_2F_1}(-l,l+2,1;\sin^2 \theta),
\label{eq:expansion}
\end{equation}
where ${_2F_1}$ is the hypergeometric function, $A^l(f)$ is a polynomial of order $l$ in $f$
\begin{equation}
A^n(f) \equiv \sum_{s=0}^f \frac{(-1)^{f-s}2^s f!}{(f-s)!s!}
(s+1)_n \,,
\end{equation}
and $(\alpha)_n = \frac{(\alpha+n-1)!}{(\alpha-1)!}$ is the Pochhammer symbol.  Thus, the data about the underlying state $\mathcal{F}$ enters the l$^{{\rm th}}$ moment in sums of the form
\begin{equation}
\sum_{f\in \mathcal{F}} A^l(f) = \sum_{k=0}^l c_k \, M_k
\end{equation}
where $c_k$ is the coefficient of $f^k$ in the polynomial expansion of $A^l(f)$.   Thus a measurement of the first $N$ multipole moments of the metric functions can be inverted to give the set of charges $M_k$ of the underlying state, from which the complete wave function can be reconstructed. 

The question is whether the above formal considerations survive the semiclassical limit. The latter consists of $\hbar \to 0$ with $\hbar N$ fixed. Moments $M_k$ scale like $M_l = m_l N^{l+1}$. Hence, the multipole expansion reduces to \cite{info-loss}
\begin{equation}
u(\rho,\theta) = 2\cos^2 \theta \sum_{k=0}^{\infty} \frac{2^k \langle M_k\rangle }{\rho^{2k+2}} \, (-1)^k(k+1)\, \,
{_2F_1}(-k,k+2,1;\sin^2 \theta)\,. 
\label{eq:wignerlimit}
\end{equation}
Thus, there is a one--to-- one correspondence between the bulk $u(\rho,\theta)$ multipole moments and the spectrum of the basis states encoded in the set $M_k$.

Since semiclassical observers have finite resolutions, we do not expect them to be able to measure all the required moments to identify the basis state. Indeed, to measure the l$^{{\rm th}}$ multipole in (\ref{eq:wignerlimit}) one needs the  $(2l)^{{\rm th}}$ derivative of the metric functions or any suitable invariant constructed from them. If the measuring device has finite size $\lambda$,      the k$^{{\rm th}}$ derivative of a quantity within a region of size $\lambda$ will probe scales
of order $\lambda/k$.  However, semiclassical devices can only measure quantities over distances larger than the Planck length.  Thus, 
\begin{equation}
\frac{\lambda}{k} > l_p = g_s^{1/4} l_s
\label{bound1}
\end{equation}
Setting the size $\lambda$ to be a fixed multiple of the AdS scale $\lambda = \gamma L$,
this says that
\begin{equation}
k < \gamma N^{1/4}
\end{equation}
for a derivative to be semiclassically measurable. Since we require $\CO(N)$ multipoles to determine the quantum state and $N^{1/4}/N \to 0$ as $N \to
\infty$,  we conclude that semiclassical observers have access to a negligible fraction of the information needed to identify the
quantum state. Reversely, it was shown in \cite{info-loss} that the distribution of low moments is universal, in the sense that the standard deviation to mean ratio of the moments $M_k$ vanishes in the semiclassical limit. Thus, classical configurations have essentially identical low order multipoles, and their differences can not be observed. 

These considerations are fairly generic, as argued in \cite{Balasubramanian:2006iw}. Assuming black holes are quantum mechanically described by a finite number of states, and consequently involve a discrete spectrum, it is clear that different quantum states can in principle be distinguished. The latter would imply there is no information loss. The catch is, once again, the necessary measurements required to tell these different quantum states apart involve Planck scale precision or waiting of the order of $\delta t\sim e^{S}$, due to the Heisenberg uncertainty principle.


So far we focused on the semiclassical measurability of states in the bulk, but nothing was explicitly mentioned about which subset of quantum states allows such a description. This is important because this process is intimately related to the emergence of entropy from the bulk perspective. To gain some insight into this issue, a second quantised formalism was developed in \cite{qgeometry-don} to define an operator $\hat u (\alpha)$ whose expectation value in a generic state $|\Psi\rangle$ equals the one of the one-particle Husimi distribution\footnote{The exact quantum state is an N-particle state. Thus, there is generically information loss when going from this to the one particle description. In the large N limit, this is typically expected to be a subleading effect not emerging in the classical gravitational description. See \cite{qgeometry-don} for a discussion on this matter.}
\begin{equation}
\langle \Psi | \hat{u}(\alpha) | \Psi \rangle  = \pi
\textrm{Hu}^{\rho_1}(\alpha)\,. 
\label{husimi}
\end{equation}
The operator in question is
\begin{equation}
\hat u(\alpha) \equiv b^\dagger(\alpha) b(\alpha)\,.
\end{equation}
Here $b^\dagger(\alpha)$ stands for a fermion creation operator, whereas $|\alpha \rangle$ equals a coherent state localised in some point of phase space  
$\alpha = \frac{x_1+i\,x_2}{\sqrt{2\hbar}}$,
\begin{equation}
|\alpha \rangle = e^{-|\alpha|^2/2} \sum_{n=0}^\infty \frac{\alpha^n}{\sqrt{n!}}|n\rangle \equiv \sum_{n=0}^\infty c_n(\alpha)
|n\rangle\,. 
\label{eq:coherent}
\end{equation}

Coherent states can be thought of states inhabiting a lattice of unit cell area $2 \pi \hbar$ \cite{coherent1,coherent2,coherent3}. Since a semiclassical observer measures
the phase plane at an area scale $\delta A = 2 \pi \hbar M \gg 2
\pi \hbar$, she is only sensitive to a smooth, coarse grained Wigner distribution $0 \leq \hbar W_c = u_c \leq 1$ erasing many details of the underlying precise microstates. The region $\delta A $ consists of $M = \delta A / 2 \pi \hbar$ lattice sites, a
fraction $u_c = \hbar W_c$ of which are occupied by coherent
states. Then the entropy of the local region $\delta A$ is
\begin{equation}
\delta S = \log{\binom{M}{\hbar \, W_c \, M}} \sim  -\frac{\delta
A}{2\pi\hbar} \log{u_c^{u_c} (1-u_c)^{1-u_c}} \,,
\label{localent}
\end{equation}
when $\hbar W_c$ is reasonably far from 0 and 1. For the total entropy this gives \cite{qgeometry-don}
\begin{equation}
S = \int \delta S = -\int dA \, \frac{u_c \log{u_c} + (1-u_c)
\log{(1-u_c)}}{2\pi\hbar} \, . \label{entropyequation}
\end{equation}
Thinking about $u_c = \hbar W_c$ as the probability of occupation of a site by a coherent state, this is simply Shannon's formula for information in a probability distribution \cite{Shannon:1948zz} \footnote{This result was independently obtained by Masaki Shigemori in an unpublished work by considering a gas of fermionic particles in phase space.}. This procedure shows in a rather explicit way how entropy is generated by integrating out quantum data at smaller scales. The large amount of supersymmetry allows us to interpret this as gravitational entropy by providing a bridge connecting the gauge and gravity theory descriptions in the large N limit.

This picture allows us to compute the semiclassical partition function 
\begin{equation}
Z =  \int\mathcal{D} u(x_1,x_2) \, \mu(u)  \,   e^{-\beta (H(u) - \nu N(u) )} \,,
\end{equation}
where the measure $\mu(u)$ reflects not only the
Jacobian in transforming between the supergravity fields and $u$,
but also the number of underlying microscopic configurations that
give rise to the same macroscopic spacetime. Our previous considerations suggest
\begin{equation}
\mu(u) = e^{-\int \frac{dx_1 \, dx_2}{2\pi \hbar} \, \left(u \ln{u}  + (1-u)\ln(1-u)\right) }= e^{S(u)}\,, 
\label{measure}
\end{equation}
where $S$ is understood as the entropy of the  spacetime.   In the
semiclassical limit, a spacetime is nonsingular if $u = 0,1$
everywhere.   In that case, $S(\mu) = 0$ and the measure $\mu$ is
1.   In other words, semiclassical half-BPS spacetimes have an
entropy if and only if they are singular.

Evaluating the partition function by the method of saddle point gives
\begin{equation}
 \ln Z =
 \int \frac{d^2x}{2\pi \hbar} \ln (1 +
e^{-\beta \frac{x_1^2+x_2^2}{2\hbar}+\beta \nu}) = \int_0^{\infty} ds \, \frac{s}{e^{s-\beta\nu}+1} \equiv \frac{1}{\beta}
F_2(e^{\beta\nu})\,, \label{partfn}
\end{equation}
where $s= \beta (x_1^2 + x_2^2)/2$ and $F_2$ is a Fermi-Dirac function. 

\begin{figure}[pb]
\centerline{\psfig{file=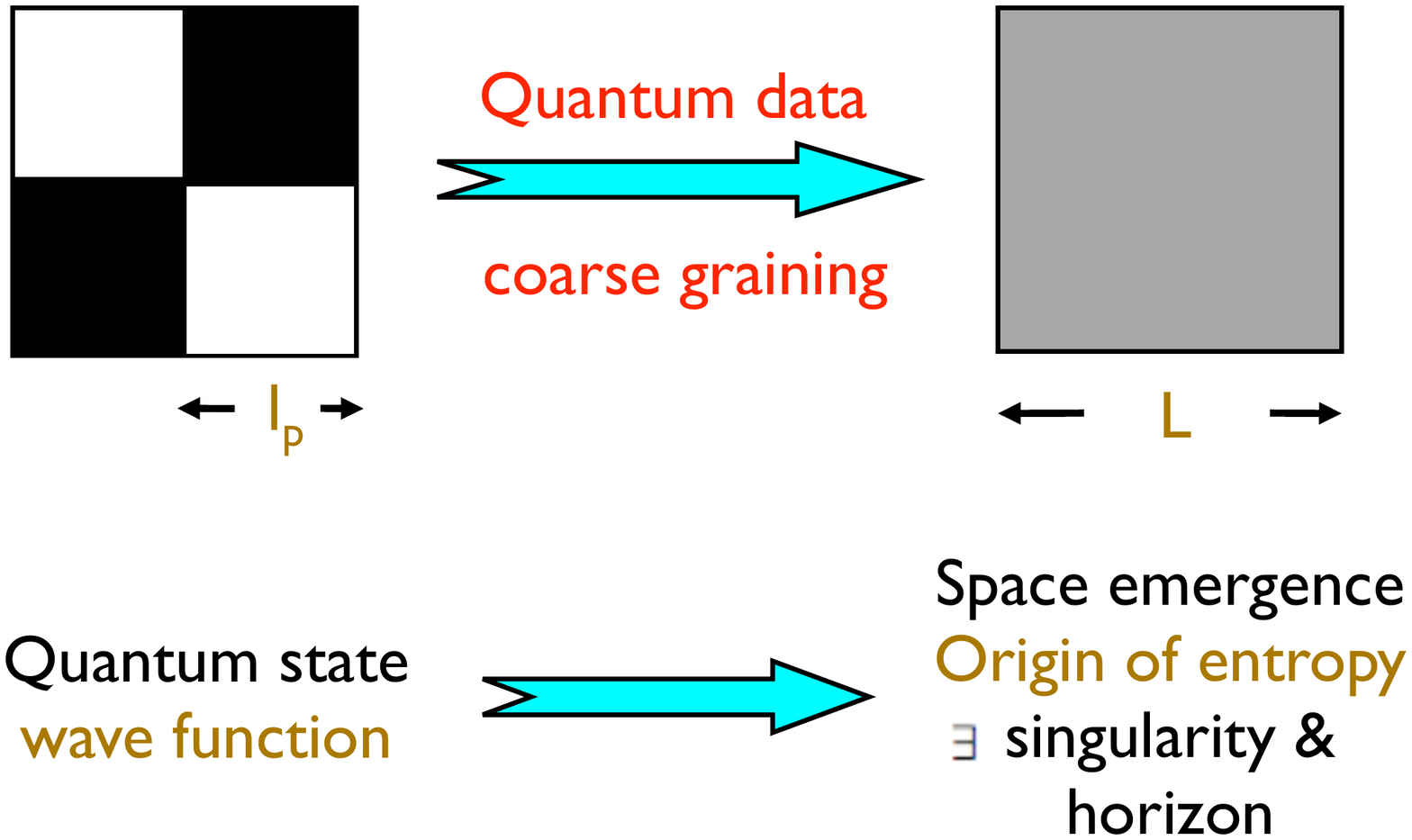,width=15cm}}
\vspace*{8pt}
\caption{RG-transformation in phase space giving rise to entropy. \label{fig2b}}
\end{figure}

This result can be derived from first principles by coarse graining the scale of the fundamental cells in the exact gauge theory partition function. This is defined as a renormalization group (RG) transformation in this space. Consider a lattice whose cells are $M\times M$ (in Planck units). From the microscopic point of view, the energy of each distribution of populated Planck scale cells is different, but in the limit $M\to\infty$, almost all
distributions cluster close to a certain typical distribution in the $M \times M$ cell, and thus observers at these scales will assign the same energy to all of them. Another way to look into this transformation is more analogous to the one taken in the semiclassical limit in gravity
\begin{equation}
  \ell_p\to 0\,, \quad\quad L\to 0\,,\quad \quad \frac{L}{\ell_p}\to \infty
\label{eq:semiclass}
\end{equation}
$L$ is the emergent continuous scale in the classical limit. It can be viewed as $L=M\,\ell_p$.
Before taking the limit, the coarse grained function $u$ will take values $0, \frac{1}{M^2}, \frac{2}{M^2}, \ldots, 1$ in the $M \times M$ cells. This can also be inferred by requiring the phase space to describe $N$ particles. Comparing the two lattices of sizes $M\times M$ and $1\times 1$, in Planck units, one finds that
\begin{equation}
N =  \sum_{\{x_1,\,x_2\}} u(x_1,\,x_2) = M^2\,\sum_{\{x_1^M,\,x_2^M\}} u^M(x_1^M,\,x_2^M)\,,
\end{equation}
allowing to derive that
\begin{equation}
u^M(x_1^M,\,x_2^M) = \frac{1}{M^2}\, \!\!\!\! \sum_{\stackrel{\{x_1,\,x_2\}}{\in\{x_1^M,\,x_2^M\}}} \!\!\!\! u(x_1,\,x_2)\,,
\end{equation}
where variables with superscript $M$ are defined in the $M\times
M$ lattice and in the second equality we are summing over all
Planck-scale lattice sites inside a single $M\times M$ cell
labelled by $(x_1^M,\,x_2^M)$. This sum computes the fraction of
populated sites in the coarse grained cell. Finally, the sum over all possible 
$u^M$ configurations at cell location $(x_1^M,x_2^M)$ becomes
\begin{equation}
\sum_{u^M} e^{-f\, u^M} = 1 + \binom{M^2}{1} e^{-f \frac{1}{M^2}} + \binom{M^2}{2} e^{-f \frac{2}{M^2}} + \ldots + e^{-f\frac{M^2}{M^2}} = \left( 1 + e^{\frac{-f}{M^2}} \right)^{M^2}.
\end{equation}
where $f(x_1,x_2) = \beta\left( (x_1^2 + x_2^2)/2  - \nu\right)/(2\pi \hbar)$.
The factors in front of each exponential count how many ways a given value of $u^M$ in the coarse grained lattice can be attained in terms of the Planck scale lattice. These combinatorial factors equal the ones already argued for in the semiclassical considerations in (\ref{localent}). This is the precise origin of entropy when implementing the coarse graining transformation in the limit (\ref{eq:semiclass}).The complete partition function becomes
\begin{equation}
Z^{M\times M}  =  \prod_{x_1,x_2 \in M \mathbb{Z}} \left( 1 + e^{-\frac{\beta}{M^2} (\frac{x_1^2+x_2^2}{2} - \nu )} \right)^{M^2}
=  \frac{M^2}{\beta} F_2(e^{\frac{\beta\nu}{M^2}})\,.
 \label{eq:MMpartfn}
\end{equation}
This derivation reproduces the semiclassical computation if we identify $\beta$ in \eqref{partfn} with the rescaled potential $\beta/M^2$ due to the renormalization group transformation. One could interpret the computation \eqref{eq:MMpartfn} as a derivation for the entropy formula \eqref{localent}.

What this derivation highlights is that as soon as one observer has {\it no} access to Planck scale physics, the measured coarse grained phase space density will typically be fractional. The advantage of the system discussed above is that we can also explicitly see this in gravity in terms of singular configurations. This is a satisfactory explanation for the source of entropy in the semiclassical description.

\subsection{Large black holes, fuzzball and ensembles}
\label{sec:lbhole}

Testing the ideas behind the fuzzball conjecture has been reasonably successful for small supersymmetric black holes, the example discussed above being one particular example. See \cite{fuzzball,samir-rev,bena-warner-rev,kostas-marika,vijay-jan} for discussions and references involving other set-ups. It is natural to wonder how general and testable these ideas are for generic supersymmetric and non-supersymmetric black holes.

Even in the context of small supersymmetric black holes, it was pointed out in \cite{Sen:2009bm} that the existence of classical fuzzball configurations depends on the U-dual frame being considered. Even more, it was noticed there, that for all known examples, whenever such solutions do not exist, the small black hole develops a horizon through higher order corrections, whereas when it can be argued that the horizon scale remains of vanishing size, fuzzball configurations already exist at tree level. Sen conjectured this to be a generic fact \cite{Sen:2009bm}. This observation starts emphasising the importance of both $\alpha^\prime$ and $g_s$ corrections in these considerations, since both are not U-duality invariant.

What about large black holes ? There exist candidate supersymmetric fuzzball configurations starting with the multi-center configurations in \cite{bates-denef,bena-warner,gimon-levi}, including the scaling solutions \cite{Denef:2007vg,Bena:2006kb}, where the coordinate distance between such centers goes to zero, and also more recent work \cite{Bena:2010gg} involving configurations with arbitrary functions. Their (classical) moduli spaces are much more complex than the ones for small black holes. Hence our understanding is far from complete (see \cite{vijay-jan} for a summary on the state of this important matter). But if fuzzball ideas were to be borrowed to these large black holes, one would ideally expect, at least in an euclidean path integral approach, that the partition function equals \cite{kostas-marika}
\begin{equation}
  Z = \int_{f_a} D[g_a]\,D[\Phi_a]\,e^{-I_{f_a}} = e^{-I_{f_a} + S} = e^{-I_{BH}}\,,
\label{eq:parfuzzball}
\end{equation}
where the sum is {\it only} carried over the fuzzball configurations $f_a$, with metric $g_a$, matter fields $\Phi_a$ and $I_{BH}$ is the action for the black hole having the same mass and charges as the set $\{f_a\}$. Notice that the absence of a horizon for all $f_a$ implies that the on-shell Euclidean action should satisfy
\begin{equation}
  I_{f_a} = \beta\left(E-\mu_i\,Q_i\right)\,,
\end{equation}
where the set $\{E,\,Q_i\}$ stands for the mass and charges carried by the black hole. Notice the above argument also used there is a total degeneracy $\CD=e^{S}$ obtained after quantisation of the space of classical fuzzball solutions.

Using this formalism provides intuition, but its assumptions may not be generically fulfilled, especially in the semiclassical regime where they are technically applied. It is still useful to present some of the puzzles on the subject. The identity \eqref{eq:parfuzzball}  assumes both the existence of solutions to the equations of motion of the effective action governing gravity and that their number explains the entropy of the original macroscopic black hole.

Let us put these assumptions in perspective with our previous discussions \cite{Simon:2009mf}. First, our knowledge on this effective action typically reduces to its tree level classical part. Thus, geometric quantisation deals with quantisation of a classical moduli space of configurations of a classical theory which is known to receive both $\alpha^\prime$ and $g_s$ corrections. Since the most conservative expectation is that information about individual microstates is stored in Planck scale physics, the use of the classical action is typically not justified to begin with. Thus, we expect {\it not} to be able to reproduce the entropy of the black hole from these considerations. Indeed, preliminary work in this direction  \cite{deBoer:2008zn,deBoer:2009un} confirms this expectation.

Second, there are matters of principle arising. Large lorentzian black holes typically have curvature singularities in their deep interior. Thus,  they are {\it not} solutions to the classical equations of motion. Their {\it euclidean} continuations, however, are smooth when suitable boundary conditions are imposed, at the expense of removing the interior of the geometry.
These euclidean configurations are {\it saddle points} of the semiclassical partition function and provide the dominant contributions to the macroscopic entropy $d_{\text{macro}}$ by construction. 
But saddle points are believed {\it not} to provide a semiclassical description of states with well-defined mass and spin in a quantum gravity Hilbert space \cite{farey-tale}. Thus, in an euclidean formalism, which is intrinsically canonical in nature, ensemble wise, the euclidean black hole definitely knows about the total number of states, but the information about the microstates seems to be lost (or it is not manifest in our current understanding).  

The above remarks are consistent with the different available formulations of the AdS/CFT correspondence. The euclidean path integral allows to compute the partition function, and to extract the total number of states by saddle point approximation, but we also know that the use of lorentzian geometry is essential to capture the difference between microstates through the expectation values of the different gauge invariant operators encoded in the boundary fall-off conditions of the different bulk fields \cite{lor-ads/cft}.

Furthermore, this discussion also highlights our ignorance on how to define path integrals more accurately. Indeed, in the examples were candidate fuzzball configurations are known, their euclidean continuations involve {\it complex} metrics, whereas their on-shell actions remain real\footnote{This observation may not be that surprising since we know of examples in quantum mechanics in which the saddle point approximation involves a complex configuration. It seems still meaningful to appreciate its conceptual consequences beyond its purely technical nature.}. Clearly, it is important to understand the space of euclidean configurations one needs to sum over. A different way of stressing this point is to notice that "thermal" circles in euclidean black holes are contractible, whereas the ones for known fuzzball configurations are not, suggesting topology can play a role in these considerations. 

It is interesting to revisit our half-BPS sector of ${\cal N}=4$ SYM discussion. This describes a small black hole in the bulk and it was argued in \cite{Simon:2009mf} that quantum corrections would
not generate a macroscopic horizon. The large symmetry in this sector of the theory allowed us to match the semiclassical partition function with a first principle (quantum mechanical) derivation.
The outcome of this matching was the derivation of the non-trivial measure $\mu (u)$ in (\ref{measure}). The correct partition function was obtained by summing over smooth and singular configurations in the semiclassical limit. If we would have ignored the non-trivial measure, the partition function would have reduced to
\begin{equation}
\ln \tilde{Z} = \int \frac{d^2x}{2\pi } \ln \left( \frac{1-e^{(-\beta \frac{x_1^2+x_2^2}{2} + \beta \nu)}}{\beta
\frac{x_1^2+x_2^2}{2} - \beta \nu} \right) = \frac{1}{\beta} \int_0^{\infty} ds \ln \left( \frac{1-e^{-s-\beta \nu}}{s-\beta \nu}
\right).
\end{equation}
There are two important points to be made. First, this integral diverges at the upper limit. Second, it would only reproduce (\ref{partfn}) if one restricts the partition sum to be over
smooth geometries ($u = 0,1$) with $u$ taking constant values within elementary cells at the Planck scale.  Even though this would geometrically mimic the coherent state analysis in quantum mechanics, its validity is certainly doubtful in a manifest semiclassical treatment applicable all the way to the Planck scale.


It is interesting to point out the different role that different ensembles also play in Sen's independent approach to explain the entropy of supersymmetric extremal black holes with charges $Q$ having AdS${}_2$ throats from an entirely macroscopic perspective  \cite{Sen:2008vm,Sen:2009vz}. Sen's proposal is
\begin{equation}
  d_{\text{macro}}(Q) = \sum_s \sum_{
      Q_i,\, Q_{\text{hair}}}^{\sum_{i=1}^s Q_i + Q_{\text{hair}} = Q} \Big\{\prod_{i=1}^s d_{\text{hor}}(Q_i)\Big\}\,d_{\text{hair}}(Q_{\text{hair}};\,\{Q_i\})\,.
 \label{eq:senp}
\end{equation}
The s-th term represents the contribution from an s-centered black hole configuration; $d_{\text{hor}}(Q_i)$ stands for the degeneracy associated with the horizon of the i-th black hole center carrying charge $Q_i$; and $d_{\text{hair}}(Q_{\text{hair}};\,\{Q_i\})$ stands for the hair degeneracy, i.e. smooth black hole deformations supported outside the horizon and sharing the same asymptotics. 

Sen's prescription uses a mixture of formulations. Indeed, whereas the contribution from the degrees of freedom localised at the horizon is captured by an euclidean path integral, both the contribution from horizonless configurations, through geometric quantisation, and hair modes employ entirely lorentzian methods. At any rate, a better understanding on how to formulate gravitational path integrals more rigorously is clearly desirable from many points of view.

Even though the above arguments strongly suggest the fuzzball programme should not work at tree level for non-supersymmetric configurations, this does not forbid, a priori, the existence of {\it non-typical} non-extremal fuzzball configurations solving the classical equations of motion. The first known examples of these were found in \cite{simon-vishnu}. Remarkably, there exists an interesting body of work for some of these non-extremal configurations giving evidence that some features of these fuzzball ideas are still realised in less symmetric situations \cite{Chowdhury:2007jx}. More recently, there has also been some progress in finding non-extremal fuzzball like configurations \cite{Bena:2009qv,Bena:2009fi} and explicit multi-center extremal non-BPS solutions (see \cite{Dall'Agata:2010dy} and references there in).

\section{Extremal black holes}\label{sec:three}

In the seminal work of Strominger \& Vafa \cite{Strominger:1996sh} the entropy of a certain supersymmetric black hole was accounted for by identifying its degrees of freedom with those of a 2d CFT. Later, it was realised \cite{Strominger:1997eq} that matching the universality of the 2d CFT Cardy's formula \eqref{eq:cardy} with the universality of the Bekenstein-Hawking entropy formula \eqref{eq:beh-haw} follows from the seminal work of Brown \& Henneaux \cite{Brown-Henneaux} analysing the asymptotic symmetry group in the AdS${}_3$ region emerging near the horizon of the original black hole.

The above synthesises the two most common approaches used in the string theory community to explain the macroscopic entropy of a given black hole. The first one is microscopic in nature. One maps black hole charges to D-brane charges (possibly) wrapping some internal cycles in some compact manifold. These provide an open string (gauge theory) description in which one counts the number of microstates compatible with the given conserved charges. The matching with the gravitational macroscopic result relies on the existence of non-renormalisation theorems guaranteeing the number of states does not change as the gauge coupling increases, which is what generates the gravity description. The second method is {\it semiclassical} in nature. It constructs a Hilbert space out of the study of the asymptotic symmetry group of a given spacetime. The emergence of a 2d conformal field theory realising two Virasoro algebras allows one to use Cardy's result \eqref{eq:cardy} to account for the entropy of the black hole in terms of the number of operators carrying its charges.

The main goal of this section is to review part of the recent work devoted to extend both of these approaches to extremal, non necessarily BPS, black holes
\begin{romanlist}[(ii)]
\item first, the constituent model developed in \cite{GLS07,GLS08} for extremal non-BPS d=4 static black holes in the STU model, 
\item second, the Kerr/CFT correspondence \cite{kerrcft} and its generalisation to extremal black holes/CFT \cite{Hartman:2008pb}.
\end{romanlist}

\subsection{Non-BPS extremal microscopics}
\label{sec:nonbpsmicro}

\subsubsection{The theory and its charges}

In this section, we will work in the STU-model \cite{Cremmer:1984hj,Duff:1995sm,Behrndt:1996hu}, {\it i.e.} $N=2$ supergravity in four dimensions with
$n_V=3$ vector supermultiplets coupling through the prepotential $F = \frac{X^1\,X^2\,X^3}{X^0}$.
The bosonic terms in the action are \cite{Ceresole:1995jg}:
\begin{equation}
\CS = \frac{1}{8\pi G_N} \int d^4x \CL = \frac{1}{8\pi G_N} \int d^4x \Big[-
\frac{R}{2} + G_{a\bar b} \partial_\mu z^a \partial^\nu \bar z^b + \textrm{Im}\,
\big(\CN_{\Lambda\Sigma} \CF^{-\Lambda\mu\nu}
\CF_{\mu\nu}^{-\Sigma}\big)\Big]~,
\end{equation}
where $\CF^{\pm\Lambda}_{\mu\nu} = \CF^{\Lambda}_{\mu\nu} \pm
\frac{i}{2} \varepsilon_{\mu\nu\rho\sigma}\CF^{\Lambda\,\mu\nu}$. We will work in the gauge $X^0=1$ and rewrite the remaining projective coordinates as $X^i = z^i = x^i - i y^i$ ($i=1,2,3$).

Let us understand its degrees of freedom. The STU-model is a subsector of type IIA string theory compactified on $T^6 = T^2 \times T^2 \times T^2$. The scalars $z^i = x^i - i y^i$ are the complexified K\"{a}hler moduli of the three $T^2$'s, {\em i.e.} $x^i$ stands for the flux treading the 2-torus and $y^i$ for its volume. Besides the four dimensional metric, the theory contains four gauge fields. Their electric $Q_\Sigma$ and magnetic $P^\Lambda$ charges are
\begin{equation}
Q_\Sigma =  \frac{1}{4\pi}
\int_{S^2_\infty} \CG_\Sigma~, \qquad P^\Lambda = \frac{1}{4\pi}
\int_{S^2_\infty} \CF^\Lambda~,
\label{eqn:physchar}
\end{equation}
where the symplectic dual field strength is $\CG_{\pm\Lambda}^{\mu\nu} = -i \delta \CL/\delta
\CF^{\pm\Lambda}_{\mu\nu} = \overline\CN_{\Lambda\Sigma} \CF^{+\Sigma\,\mu\nu}$. The electric charges correspond to $D0$-branes $(Q_0)$ and $D2$-branes
wrapping the i-th $T^2$ $(Q_i)$, while the magnetic charges correspond to $D6$-branes $(P^0)$ wrapping the entire 6-torus and $D4$-branes wrapping the dual $T^4$, {\em i.e.} $P_1$ corresponds to D4-branes wrapping the second and third 2-tori.

Physical charges (\ref{eqn:physchar}) are organised in symplectic
pairs:
\begin{equation}
\Gamma \equiv (P^\Lambda,Q_\Sigma)~.
\end{equation}
They have units of length and are related to dimensionless
quantised charges by some dressing factors. We use the conventions in \cite{GLS07} where the
asymptotic volume moduli is normalised to $y^i|_\infty = 1$ but the asymptotic B-fields
$x^i_\infty = B^i = \frac{1}{V_i}\int_{V_i} B $ are kept as free
variables. Then the dressing factors are just numerical factors
\begin{equation}
P^\Lambda = C^\Lambda\,p^\Lambda~,\qquad Q_\Sigma = C_\Sigma \,q_\Sigma~,
\end{equation}
which are essentially the masses of the underlying branes:
\begin{eqnarray}
&&
C^0 = 2^{3/2}G_N M_{D6} = \sqrt{G_N v_6}~,\qquad
C^i = 2^{3/2} G_N M_{D4} =\sqrt{G_Nv_6}\cdot \frac{1}{v_i}~, \\
&&
C_0 =2^{3/2} G_N M_{D0} = \sqrt{\frac{G_N}{v_6}}~,
\qquad C_i = 2^{3/2} G_N M_{D2} = \sqrt{\frac{G_N}{v_6}}
\cdot v_i ~. \nonumber
\end{eqnarray}
Here $v_i$ are the volumes of the $T^2$'s measured in string units $v_i=V_i/(2\pi l_s)^2$.
The overall compactification volume is $v_6=v_1 v_2 v_3$ and the $4d$
Newton's constant $G_N = l_s^2g_s^2/8v_6$.

Different classes of STU black hole solutions are classified by the quartic invariant $I_4$ inherited from the $\CN=8$ theory
\begin{equation}
\label{quartic}
I_4(\Gamma) = 4Q_0P^1P^2P^3 - 4P^0Q^1Q^2Q^3- \left(P^\Sigma Q_\Sigma\right)^2  + 4\sum_{i<j} P^iQ_i P^jQ_j~.
\end{equation}
Whenever $I_4(\Gamma) > 0$, solutions are BPS, while $I_4(\Gamma) < 0$ single center solutions cannot be BPS \cite{ferraramald,Ferrara:2007pc}. All extremal black holes, be they BPS or not, exhibit an attractor mechanism \cite{attractor}. This is reviewed in \cite{Ferrara:2008hwa}. See also \cite{GLS07} for a discussion on attractors for non-BPS extremal solutions and its relation to the existence of flat directions. The entropy of all extremal black holes equals
\begin{equation}
S = \frac{\pi}{G_N}\sqrt{ | I_4(\Gamma)|}~.
\label{eqn:entrop}
\end{equation}
Our main interest below will be in non-BPS solutions in $\CN=2$ $(I_4<0)$.

\subsubsection{The most general static non-BPS extremal solution}

The most general single center spherically symmetric black hole solutions in the STU-model are characterised by their charge vector $\Gamma = (P^I, Q_I)$, the
asymptotic value of the complex moduli $z^i=x^i - i y^i$, their mass $M$ and their angular momentum $J$. This is a 16 parameter family of solutions. Restricting our attention to extremal static configurations reduces the number of parameters to 14, since $J=0$ and $M_{\text{ext}}=M_{\text{ext}}(P^I, Q_I, z^i)$. Furthermore, the STU-model has an $\SL(2,\RR)^3$ duality symmetry acting nontrivially on these parameters. This allows us to consider a seed solution with just $14-9=5$ parameters with the understanding that the most general charge vector and asymptotic moduli can be restored if needed, by acting with these dualities \cite{Cvetic:1996zq}. 

The above was the procedure followed in \cite{GLS07}. It was shown there that there exists a canonical duality frame, the ${\overline{D0}}-D4$ frame, where the solution simplifies\footnote{The same solution was obtained in a different U-duality frame in \cite{dellagata}.}. In this frame the five parameters of the seed solution are four non-vanishing charges $Q_0$, $P^i$ and the diagonal pseudoscalar $z^i = B-i$ (with the same $B$ for $i=1,2,3$). This extended the work presented in \cite{hotta} without B-field. We take $Q_0<0$ and $P^i>0$, so that $I_4(\Gamma)<0$. Thus, supersymmetry will be broken. With these choices, the four dimensional metric of the seed solution is \cite{GLS07} :
\begin{equation}
  ds^2 = - e^{2U}\,dt^2 + e^{-2U} \left(dr^2 + r^2\,(d\theta^2 + \sin^2\theta\,d\phi^2)\right)\,,
  \label{eq:metric}
\end{equation}
with conformal factor
\begin{equation}
  e^{-4U} = -4H_0H^1H^2H^3 - B^2\,,
\end{equation}
depending on four harmonic functions
\begin{equation}
\sqrt{2}H_0 =
  -(1+B^2) + \frac{\sqrt{2}Q_0}{r}  \quad ,~\sqrt{2}H^i = 1+\frac{\sqrt{2}P^i}{r}~.
\end{equation}
Notice the conformal factor is positive definite because $Q_0<0$.
The scalar moduli $z^i$ are written in terms of the harmonic functions as
\begin{equation}
 z^i = \frac{B-i\,e^{-2U}}{s_{ijk}H^jH^k}\,.
 \label{eq:moduli}
 \end{equation}
 The asymptotic behaviour is $z^i\to B-i$ as $r\to\infty$ in accord with the duality frame we
 have chosen.

Decomposing the field strength and the symplectic dual field strength into electric
and magnetic components as :
\begin{equation}
  d\vec{\CA}^I = E^I\,dt\wedge dr + d\vec{a}^I\,,
   \quad \quad d\vec{\CA}_J = E_J\,dt\wedge dr + d\vec{a}_J\,,
   \label{eq:formansat}
\end{equation}
we find the electric fields supporting the seed solution \eqref{eq:metric} are \cite{GLS08}
\begin{eqnarray}
  E^0 &=& -\frac{e^{2U}}{r^2}\,\frac{1}{y_1y_2y_3}\,\left(Q_0-\frac{1}{2}s_{ijk}x^ix^jP^k\right)\,, \label{eq:es0} \\
  E^i &=& -\frac{e^{2U}}{r^2}\,\frac{1}{y_1y_2y_3}\,\left(x^i\,Q_0 - (x_i^2+y_i^2)\,s_{ijk} x_j\,P^k-\frac{1}{6}P^i\,s_{jkl}x^jx^kx^l\right)\,, \label{eq:esi}
\end{eqnarray}
where there is no summation over the free index $i$ in the expression for $E^i$.
The magnetic fields are
\begin{equation}
  \vec{a}^I = -P^I\,\cos\theta\,d\phi\,, \quad \vec{a}_J = -Q_J\,\cos\theta\,d\phi\,.
  \label{eq:magfields}
\end{equation}
The dual electric fields are
\begin{eqnarray}
  E_0 &=&-\frac{e^{2U}}{r^2}\,\frac{1}{y^1y^2y^3}\,\left(-x^1x^2x^3\,Q_0 + \frac{1}{2}x_iP^i\,s_{ijk}(x_j^2+y_j^2)(x_k^2+y_k^2)\right)\,, \label {eq:ess0} \\
  E_i &=& -\frac{e^{2U}}{r^2}\,\frac{s_{ijk}}{2y^1y^2y^3}\,\left(x_jx_k\,Q_0-(x_j^2+y_j^2)\,(x_k^2+y_k^2)\,P^i -2x_ix_j\,(x_k^2+y_k^2)\,P^j\right)\,.\nonumber
\end{eqnarray}
Again, there is no summation over the free index $i$ in the expression for $E_i$. These
give the forces on magnetic probes.

\paragraph{The Non-BPS Mass Formula : } Expanding the warp factor at infinity we find the mass:
\begin{equation} 
2G_N M_{\rm Non-BPS} = \frac{1}{\sqrt{2}}\,\big(|Q_0| + \sum_i P^i (1 + B^2)\big)~.
 \end{equation}
This has a simple interpretation. It is the sum of the masses of four individually half-BPS constituents, $\overline{D0}$ and D4-branes with the B-field taken into account for each constituent independently. Interestingly, this would suggest the non-BPS black
hole is a {\it marginal} bound state at threshold viewed as a collection of $\overline{D0}$-brane and $D4$-brane constituents placed on top of each other with no binding energy.

It is instructive to compare the above mass to the BPS mass formula
\begin{equation}
 2G_N \,M_{\rm BPS} = \frac{1}{\sqrt{2}}\Big| Q_0 +
 \sum_i P^i\,(1 + iB)^2\Big| ~.
 \label{BPSmass}
\end{equation}
There exists a non-vanishing gap between the squares of the two masses 
\begin{equation}
 8G_N^2 \, (M^2 - M^2_{BPS}) = 4\big|Q_0\big| \sum_i P^i  > 0~.
 \label{eq:massgap} 
\end{equation}
Thus the additional energy associated with a non-BPS state is always
strictly positive.

\subsubsection{Testing the model : probe \& supergravity calculations}

The lack of binding energy means constituents can be arbitrarily separated. Thus it should be
possible to bring in additional quanta from infinity, without being
subject to any force. This has two consequences :
\begin{romanlist}[(ii)]
\item Probe $\overline{D0}$-branes and $D4$-branes wrapping any two torii must feel no force in the non-BPS black hole background.
\item There must exist multicenter configurations with total charge distributed among constituents whose location is unconstrained.
\end{romanlist}

\paragraph{Probe analysis : } The potential felt by a static $Dp$-brane at a constant position due to a fixed supergravity background is given by its effective Lagrangian density, up to a sign :
\begin{equation}
  V_{Dp} = T_p \left[ e^{-(\phi-\phi_\infty)}\,\sqrt{-\det (G+B)} - \sqrt{2}\eta\,A_{p+1}\right] \equiv T_p\left( V_{\text{DBI}} + V_{\text{WZ}}\right)\,,
  \label{eq:DBI}
\end{equation}
where $\eta$ parameterises whether we are describing a $Dp$ or an
$\overline{Dp}$ brane. We have in mind infinitesimal constituents being added and so
it is justified to use the probe approximation where distortion of the background due to the probe is neglected. 

Let us briefly discuss the DBI contribution. The dilaton in \eqref{eq:DBI} is the 10D one, with its asymptotic value absorbed in the tension of the brane. Since the 4D dilaton is a component of a hypermultiplet, it has no radial dependence. Hence, all the 10D dilaton non-triviality is uniquely acquired through the volume of $T^6$, {\em i.e.} $e^{-2(\phi_4-\phi_{4\infty})}=e^{-2(\phi-\phi_\infty)}V_6=1$. It is convenient to evaluate the combination :
\begin{equation}
  e^{-(\phi-\phi_\infty)}\,\sqrt{-g_{tt}} = \frac{1}{\sqrt{y^1y^2y^3}} e^U =
  \frac{2\sqrt{2}H^1 H^2 H^3}{(-I_4 - B^2)}
  \,.
  \label{eq:effdil}
\end{equation}
Here we replaced the charge dependence in (\ref{quartic}) by the corresponding harmonic functions so that $I_4$ above is
\begin{equation}
I_4 = 4H_0H^1H^2H^3 - 4H^0H_1H_2H_3 - \left(\sum_I H_I H^I \right)^2
+ 4 \sum_{i<j} H^i H_i H^j H_j~,
\label{eq:i4def}
\end{equation}
which for the seed solution reduces to :
\begin{equation}
I_4 \to\;
4H_0H^1H^2H^3\,.
\end{equation}
If we remember the $B$-field appearing in \eqref{eq:DBI} is the spatially varying $B$-field,
whose components on each $T^2$ we hitherto denoted $x$, then for a single $T^2$, the contribution to the potential is
\begin{equation}
\sqrt{\det (G+B)}  \to \sqrt{x^2 + y^2} = |z|   \,.
\end{equation}

We are left with the WZ contribution. Its contribution to the force (in units of the brane tension $T_p$) is simply:
\begin{equation}
- \frac{\partial V_{WZ}}{\partial r} = \sqrt{2}\eta \frac{\partial A_{p+1}}{\partial r}
= -\sqrt{2}\eta E\,.
\label{eq:WZforce}
\end{equation}
The electric field one should use in this expression depends on the identity of the probe:
it is $E^0$, $E^i$ for $D0$, $D2$ branes and $E_0$, $E_i$ for $D6$, $D4$ branes.
The electric fields generated by the $\overline{D0}$-D4 background
were given in (\ref{eq:es0}-\ref{eq:ess0}).

It was shown in \cite{GLS08} that $\overline{D0}$ and any $D4$-brane probe with {\it no} fluxes on their worldvolumes were satisfying the no-force condition, whereas any other probe involving
$D2$ and/or $D6$-branes, with or without fluxes turned on, did feel a force. Thus, these are the natural constituents making up the black hole. The analogous claims for BPS black holes do not hold due to the existence of a non-trivial binding energy in the supersymmetric branch  \cite{GLS08}.

\paragraph{Multi-center extremal non-BPS supergravity configurations :} In \cite{Gaiotto:2007ag}, the moduli space of static multi-centered extremal non-BPS configurations was studied for the particular case in which all D4-brane charges were equal. Their main conclusions were
\begin{romanlist}[(ii)]
\item There are no constraints on the location of each center.
\item There is a constraint on the charge vector $\Gamma_i$ carried by each i-th center.
\end{romanlist}
It is reassuring to check that the only charge vectors solving these constraints are precisely the ones matching our constituents (or linear combinations thereof). 

Since our original statement applies to different D4-brane charges and different values for the moduli, due to U-duality invariance, we expect there should exist more general multi-centered extremal non-BPS solutions generalising the results in \cite{Gaiotto:2007ag}. Notice these statements only apply to static extremal non-BPS configurations. The different centers are thus mutually local and carry no angular momentum. These properties do not hold in more general multi-centered extremal non-BPS configurations constructed in \cite{Bena:2009ev,Bena:2009en}. These do carry angular momentum and their centers are constrained to satisfy a cubic "bubble" equation.

\subsubsection{The D0-D6 frame}

The microscopic picture developed in the canonical duality frame must hold at any point of the U-duality orbit. This follows from the U-duality invariance of the total mass and the lack of binding energy. We will explore such model in different frames to gain further insights into its stability and its comparison with the branch of BPS black holes.

In this subsection, I discuss the D0-D6 frame. The five parameters of the seed solution are mapped to the $D0$-brane charge $Q_0$, $D6$-brane charge $P^0$, and three independent $B$-fields $B_{1}, B_{2}, B_{3}$ along the three $T^2$'s. The explicit map (constructed in section 5 of \cite{GLS07}) depends prominently on three parameters $\Lambda_i$ (part of the original $\SL(2,\RR)^3$) related to the parameters of the $D0-D6$ frame by 
\begin{eqnarray}
\Lambda_1\Lambda_2\Lambda_3 &=&  \frac{P^0}{Q_0}\,,\label{eq:prodlambda} \\
\frac{1}{2}[ \Lambda_1(1 + B_{1}^2) - \Lambda_1^{-1} ]&=& \frac{1}{2}[\Lambda_2(1 + B_{2}^2) -
\Lambda_2^{-1} ]= \frac{1}{2} [ \Lambda_3(1 + B_{3}^2) - \Lambda_3^{-1}] \,. 
\label{eq:sugcons} 
\end{eqnarray}
The constraint \eqref{eq:sugcons} arises from requiring the $B$-field in the
$\overline{D0}-D4$ seed solution to be the same on the three $T^2$'s. 

The mass of the D0-D6 black hole is :
\begin{eqnarray} \hspace{-30pt}
2^{3/2}\,G_N\,M &=&  \frac{P^0}{4}\Big(1 +
(\Lambda_1^{-1} + B_1)^2\Big)^{1/2} \Big(1 +
(\Lambda_2^{-1} + B_2)^2\Big)^{1/2}\Big(1 + (\Lambda_3^{-1} + B_3)^2\Big)^{1/2} \nonumber \\
&+& \frac{P^0}{4}\Big(1 + (\Lambda_1^{-1} + B_1)^2\Big)^{1/2} \Big(1 +
(\Lambda_2^{-1} - B_2)^2\Big)^{1/2}\Big(1 + (\Lambda_3^{-1} - B_3)^2\Big)^{1/2} \\
&+& \frac{P^0}{4}\Big(1 + (\Lambda_1^{-1} - B_1)^2\Big)^{1/2} \Big(1 +
(\Lambda_2^{-1} + B_2)^2\Big)^{1/2}\Big(1 + (\Lambda_3^{-1} - B_3)^2\Big)^{1/2} \nonumber \\
&+& \frac{P^0}{4}\Big(1 + (\Lambda_1^{-1} - B_1)^2\Big)^{1/2} \Big(1 +
(\Lambda_2^{-1} - B_2)^2\Big)^{1/2}\Big(1 + (\Lambda_3^{-1} + B_3)^2\Big)^{1/2}
\nonumber.
\end{eqnarray}
This mass formula is consistent with the sum of masses of four individual half-BPS constituents with charge vectors
\begin{eqnarray}
\label{constvector}
  \Gamma_{I} &=&
\frac{1}{4}  \Big(P^0;-P^0/\Lambda_1,-P^0/\Lambda_2,-P^0/\Lambda_3;Q_0;P^0/(\Lambda_2\Lambda_3),
P^0/(\Lambda_1\Lambda_3),P^0/(\Lambda_1\Lambda_2)\Big)\label{cvecone} \nonumber \\
  \Gamma_{II} &=&
\frac{1}{4}  \Big(P^0;-P^0/\Lambda_1,P^0/\Lambda_2,P^0/\Lambda_3;Q_0;P^0/(\Lambda_2\Lambda_3),
-P^0/(\Lambda_1\Lambda_3),-P^0/(\Lambda_1\Lambda_2)\Big) \label{cvectwo} \nonumber \\
  \Gamma_{III} &=&
\frac{1}{4}  \Big(P^0;P^0/\Lambda_1,-P^0/\Lambda_2,P^0/\Lambda_3;Q_0;-P^0/(\Lambda_2\Lambda_3),
P^0/(\Lambda_1\Lambda_3),-P^0/(\Lambda_1\Lambda_2)\Big)\label{cvecthree} \nonumber \\
  \Gamma_{IV} &=&
\frac{1}{4}  \Big(P^0;P^0/\Lambda_1,P^0/\Lambda_2,-P^0/\Lambda_3;Q_0;-P^0/(\Lambda_2\Lambda_3),
-P^0/(\Lambda_1\Lambda_3),P^0/(\Lambda_1\Lambda_2)\Big)\,, \label{cvecfour} \nonumber
 \end{eqnarray}
each charge vector being the image under the U-duality transformation of the half-BPS constituents in the canonical frame.This is a manifestation of the U-duality invariance of the constituent model. Notice the total charge vector 
$$
\Gamma =  \Gamma_{I}  + \Gamma_{II}+  \Gamma_{III} + \Gamma_{IV} =(P^0;\vec{0}; Q_0; \vec{0})
$$ 
is that of the $D0-D6$ black hole, as it should.

The microscopic picture emerging from these charge vectors extends the known model \cite{Taylor:1997ay,Larsen:1999pu} in the absence of external B-fields based on coincident $D6$-branes. Their idea was to reproduce the $D0$ and $D6$-brane charges by wrapping four $D6$-branes on  $T^2\times T^2\times T^2$ with flux assignments :
\begin{align}
  (F_{12},\,F_{34},\,F_{56})^I &= (f_1,\,f_2,\,f_3)~,  \label{eq:fluxone}\\
  (F_{12},\,F_{34},\,F_{56})^{II} &= (f_1,\,-f_2,\,-f_3)~,
   \label{eq:fluxtwo}\\ (F_{12},\,F_{34},\,F_{56})^{III} &=
  (-f_1,\,f_2,\,-f_3)~,  \label{eq:fluxthree}\\
  (F_{12},\,F_{34},\,F_{56})^{IV} &= (-f_1,-\,f_2,\,f_3)\,.
 \label{eq:fluxfour}
\end{align}
The superindex $\{I,\,II,\,III,\,IV\}$ enumerates the four $D6$'s while the subindex 
in the individual fluxes $f_i$ refers to the torus in which they are thread. It is interesting to point out that the minimal size for the gauge group to cancel the induced $D2$ and $D4$-brane charges is four, precisely matching the number of independent constituents in our model. The induced $D0$-brane charge from the worldvolume flux $F$ is the third Chern class 
\begin{equation}
n_0 = -\frac{n_6}4\,\frac{1}{6(2\pi)^3}\int {\rm tr} F\wedge F \wedge F = -\frac{n_6V_6
f_1f_2f_3}{(2\pi)^3}~,
\label{eq:inddzero}
\end{equation}
so we have
\begin{equation}
\frac{P^0}{Q_0} = \frac{M_6}{M_0}\frac{n_6}{n_0} = -\frac{V_6}{
(2\pi)^6\alpha^{\prime 3}} \frac{(2\pi)^3}{V_6 f_1f_2f_3}
 =  - \frac{1}{(2\pi\alpha^\prime)^3f_1f_2f_3}~.
 \end{equation}
In order to induce the correct total $D0$-brane charge, the fluxes must be chosen so that
\begin{equation}
(2\pi\alpha^\prime)^3 f_1f_2f_3  = - \frac{Q_0}{P^0}~.
\label{eq:prodflux}
\end{equation}

The non-abelian open string description reproduces our proposed microscopic charge vectors $\Gamma$ if we make the identification
\begin{equation}
\label{matchf}
\Lambda_i =  - \frac{1}{2\pi\alpha' f_i}~.
\end{equation}
Furthermore, the mass of the non-abelian system with flux
\begin{equation}
M = T_6 \int {\rm Tr} \sqrt {{\rm det} \left[ G+ (2\pi\alpha^\prime F- B)\right]}~,
\end{equation}
agrees with the one computed in gravity (in (5.56) of \cite{GLS07}). The dynamical mechanism behind the matching between these weakly coupled calculations in gauge theory and gravity is not understood.

The non-abelian description does {\it not} determine the individual fluxes $\Lambda_i$, but only their product. Thus, it does not explain the supergravity constraint (\ref{eq:sugcons}). Furthermore, in the absence of fluxes, the system is known to be metastable \cite{Taylor:1997ay}. Given the lack of supersymmetry, it is natural to wonder about the stability of the system. I discuss both issues in a different duality frame in the next subsection.

\subsubsection{Perturbative stability and supersymmetry breaking}

Acting with T-duality three times, once along each of the $T^2$s, our four constituents become D3-branes intersecting at angles. This U-duality frame is particularly useful to analyse the perturbative stability of the system. This is because the lightest states in the spectrum of open strings stretching between any pair of intersecting D3-brane constituents is well known to contain 
four complex scalars in spacetime with masses (for more details see {\it e.g.} \cite{rabadan}):
\begin{align}
\alpha^\prime   m_1^2 &= \frac{1}{2\pi}\left(-\theta_1+\theta_2 + \theta_3\right)\,, \nonumber \\
\alpha^\prime   m_2^2 &= \frac{1}{2\pi}\left(\theta_1-\theta_2 + \theta_3\right)\,, \nonumber \\
\alpha^\prime   m_3^2 &= \frac{1}{2\pi}\left(\theta_1+\theta_2 - \theta_3\right)\,, \nonumber \\
\alpha^\prime   m_4^2 &= 1 -\frac{1}{2\pi}\left(\theta_1+\theta_2 + \theta_3\right)\,,
 \label{eq:masses}
\end{align}
where $0 \le \theta_i \le \pi$ stands for the relative angle in the ith $T^2$.

This spectrum generically contains tachyons, interpreted as a classical instability. Given the set of fluxes dictated by supergravity, two of the relative angles are always equal and the third one vanishes \cite{GLS08}, for any pair of D3-branes. The only non-trivial relative angle of a given D3-brane pair is \cite{GLS08}
\begin{equation}
 \cot \vartheta^{AB}_i  = \cot(\phi^A_i - \phi^B_i) = - \frac{1}{2}\, [\Lambda_i(1 + B_i^2) -
  \Lambda_i^{-1}] ~,\label{fluxes}
\end{equation}
where $A, B= I, II, III, IV$ label the D3-brane pair and we used $ \cot \phi_i = 2\pi\alpha^\prime{\cal F}_i = - (\Lambda^{-1}_i + B_i)$. Thus, for the set of fluxes dictated by supergravity, the spectrum is free of tachyons. Hence, our extremal non-BPS black holes are perturbatively stable and we could interpret the constraints \eqref{eq:sugcons} as a stability condition in the $D0-D6$ frame. Notice there are more general solutions to the absence of tachyons in the open string analysis, but they are not realised in supergravity. It is not understood why this is the case.

\paragraph{Supersymmetry breaking :} Some of the results presented above suggest the existence of some kind of non-renormalisation when varying the coupling of the system. This is unexpected because of the lack of supersymmetry, but it is interesting to remember the precise way in which the latter is broken in these systems. Consider a pair of constituent $D3$-branes of type $A, B$ situated at the relative angles $\{\vartheta^{AB}_i\}$. A candidate supersymmetry $|s_1~,s_2~,s_3\rangle$ preserved by one of these is also preserved by the other when \cite{micha,vijay-rob}
\begin{equation}
s_1 \vartheta_1^{AB}  + s_2 \vartheta_2^{AB} + s_3 \vartheta_3^{AB} \;=\; 0\;
\text{mod}\; 2\pi~.
\label{eq:susycond}
\end{equation}
Here we used the eigenvectors $S_i | s_1~,s_2~,s_3\rangle =\frac{1}{2}s_i \,| s_1~,s_2~,s_3\rangle~$, with $S_i$ being the generators of rotations $S_1 = i/2\, \Gamma^1\Gamma^2~, S_2 = i/2\, \Gamma^3\Gamma^4~,
S_3 = i/2\, \Gamma^5\Gamma^6$.

%
%

The analysis of these conditions done in \cite{GLS08} concluded that 
\begin{romanlist}[(iii)]
\item any pair of $D3$-branes preserves eight supersymmetries. 
\item any three constituent subset of $D3$-branes preserves four supersymmetries.
\item it is the addition of the fourth constituent that breaks supersymmetry.
\end{romanlist}

The perturbative open string analysis evidently focusses on pairs of 
constituent $D$-branes. Since each of the pairs preserves some supersymmetry, the absence of tachyons from the open string spectrum was anticipated on general grounds. 
The complete spectrum of the non-BPS black hole includes collective states that depend
for their existence on the presence of three and four branes. Only the last kind, depending 
on all four constituent branes, are sensitive to supersymmetry breaking. Since the 
classical non-BPS black hole entropy vanishes unless all four constituent branes are 
present, one expects numerous modes of this type. The finite entropy and sensible 
thermodynamics do not suggest any instability among these more exotic modes so one 
would expect no tachyons in this sector either. It would be interesting to confirm this expectation.

\subsubsection{Summary and comparison with BPS solutions}

The microscopic model for static extremal non-BPS black holes based on four mutually local half-BPS constituents described above is different from the analogous microscopic considerations emerging for BPS black holes \cite{denef0,Denef:2002ru,vijay-eric-tommy,strominger-denef,Gimon:2007mha}. I already alluded to the existence of an strictly positive mass gap in (\ref{eq:massgap}). Even though this was computed in the canonical frame, masses are {\it invariant} under U-duality transformations. Thus, the existence of the gap holds quite generically in moduli space. This is important because of the potential decay channels that non-BPS solutions can possess \cite{GLS07,GLS08}.

Using the main properties satisfied by multi-centre BPS configurations \cite{bates-denef,bena-warner,gimon-levi}, the main differences among both branches are summarised below :
\begin{itemlist}
\item[1.] Their mass is BPS, which is always {\it strictly} smaller than that of the non-BPS solution with otherwise identical quantum numbers (see \cite{GLS07}).
\item[2.]
The charge vectors of the 1/2-BPS constituents are mutually non-local, {\it i.e.} they have non-zero 
intersection number $\langle\Gamma_i,\,\Gamma_j\rangle \neq 0$. (The four constituents of the non-BPS black holes are mutually local.) 
\item[3.]
The BPS constituents are a finite distance apart, an scale essentially determined by the charge intersection numbers, {\it i.e.} $R_{ij} = |\vec{x}_i-\vec{x}_j| \propto \langle\Gamma_i,\,\Gamma_j\rangle$ \cite{denef0}. (The constituents of the non-BPS black holes can move freely in the 
supergravity approximation.)
\item[4.] 
These BPS states only exist in part of the moduli space. There is a co-dimension one wall of threshold stability in moduli space beyond which they disappear from the spectrum.
See \cite{wittend0d6} for a thorough discussion on this point in the D0-D6 U-duality frame, \cite{Mihailescu:2000dn} for an analysis of supersymmetry and \cite{Lee:2008ha,Castro:2009ac} for the explicit construction of the different two-centre BPS solutions in the STU model. (The non-BPS black holes exist everywhere in moduli space.) 
\item[5.]
The mutual non-locality of the charges generally requires angular momentum in the multi-centre BPS solutions. (this is not necessary for static non-BPS black holes.)
\end{itemlist}
Taking these considerations into account, one concludes BPS solutions cannot be continuously connected to 
any non-BPS stationary solution through the wall of marginal stability. Instead, there can exist decay from the 
non-BPS branch to the BPS branch on the part of moduli space where BPS solutions exist. 
The transition will release energy, entropy and generally also angular momentum. This indicates a first order 
transition between the two branches. 

When all B-fields vanish, the mass formulae in both branches is related by analytical
continuation from $Q_0<0$ to $Q_0>0$. This was used in \cite{emparan-horowitz} to provide a microscopic counting for the entropy of extremal neutral black holes. The latter was extended to extremal rotating Kaluza-Klein (D0-D6) black holes in the fast rotating regime in \cite{Emparan:2007en}. Both calculations are based on a model of intersecting D3-branes and assume that the entropy of low energy excitations is a local property of the intersection and is independent of whether the branes wrap the torus along minimal or non-minimal rational cycles. However, the fact that when turning on general B-fields these mass expressions are no longer so easily related suggests that the physics of the two branches is qualitatively different, in a manner reminiscent of a system with distinct phases. 

Let me stress the comments above apply to static extremal non-BPS black holes. Recently, some progress was achieved in the construction of general extremal non-BPS configurations, including non-vanishing angular momentum \cite{Bena:2009ev}. For these, there exist multi-centered configurations whose locations are non-trivially constrained by the moduli given the true bound nature of the system \cite{Bena:2009ev,Bena:2009en}. It would be very interesting to extend the microscopic description reviewed here to these systems.

\subsection{Extremal black holes \& Conformal field theory}
\label{sec:ext-cft}

After the microscopic considerations of the previous subsection, I now consider the semiclassical approach initiated by Brown \& Henneaux in \cite{Brown-Henneaux}. One of the virtues of this seminal work is to provide a semiclassical construction of a Hilbert space in a classical gravity theory given some boundary conditions. The heuristic idea is as follows. Given a reference metric $g$ (global AdS${}_3$ in \cite{Brown-Henneaux}), one determines the subset of non-trivial diffeomorphisms $\zeta$ preserving some set of boundary conditions $h$ (at infinity in \cite{Brown-Henneaux})
\begin{equation}
  {\cal L}_\zeta \left(g + h\right) \sim h\,,
\end{equation}
where ${\cal L}_\zeta g$ stands for the Lie derivative of the metric $g$ along the vector field $\zeta$.
These are understood as normalisable excitations of the background metric $g$. By non-trivial here, one means the associated conserved charge $Q_\zeta[g]$ does not vanish. One is interested in computing the algebra closed by these conserved charges under Dirac brackets, since the states in this semiclassical approximation will fit into representations of the latter. Thus one needs surface integrals defining them in terms of the given diffeomorphism $\zeta$ and the background metric $g$. Here, I follow the covariant formalism developed in  \cite{barnichbrandt,barnichcompere}, based on \cite{sev-covariant} and  further developed in \cite{barnichstokes,comperethesis}. The charges generating $\zeta$ are 
\begin{equation}
Q_{\zeta}= \frac{1}{8 \pi G}\int_{\partial\Sigma} k_{\zeta}[h,g]\,,
\end{equation} 
where $G$ is Newton's constant, $\partial\Sigma$ the boundary of a spatial slice and
\begin{eqnarray}
k_\zeta[h,g] &=&\frac{1}{2} \Big[ \zeta_\nu\nabla_\mu h
  - \zeta_\nu \nabla_\sigma h_\mu{}^\sigma +
  \zeta_\sigma\nabla_\nu h_\mu{}^\sigma + \frac{1}{2} h \nabla_\nu\zeta_\mu
- h_\nu{}^\sigma \nabla_\sigma\zeta_\mu\nonumber\\
&&\qquad\qquad\qquad + \frac{1}{2} h_{\nu\sigma}
(\nabla_\mu\zeta^\sigma + \nabla^\sigma\zeta_\mu)\Big] \, {*(dx^\mu\wedge
dx^\nu)}\,,\label{kdef}
\end{eqnarray}
All raised indices are computed using $g_{\mu\nu}$. The Dirac bracket algebra of the asymptotic symmetry group is then computed by varying the charges
\begin{equation} 
\{Q_{\zeta_m},Q_{\zeta_n}\}_{D.B.} =Q_{[\zeta_m,\zeta_n] }+
\frac{1}{8\pi G}\int_{\partial\Sigma}k_{\zeta_m}[{\cal L}_{\zeta_n} g,g]\,.
\label{eq:bccharge}
\end{equation}
Notice the resulting algebra can include a central term \cite{barnichbrandt} if the last term does not vanish.

Recently, this philosophy was applied to extremal Kerr
\begin{equation}
  ds^2 = -\frac{\Delta}{\rho^2}\left(d\hat t-a\sin^2\theta\,d\hat \phi\right)^2 + \frac{\sin^2\theta}{\rho^2}\left((r^2+a^2)\,d\hat\phi-a\,d\hat t\right)^2 + \frac{\rho^2}{\Delta}\,dr^2 + \rho^2\,d\theta^2\,,
\end{equation}
where $\Delta=(r-a)^2$ and $\rho^2=r^2+a^2\cos^2\theta$. The Bekenstein-Hawking entropy equals
\begin{equation}
  S = 2\pi\,J\,,\qquad \qquad \text{with} \quad \quad J=\frac{M^2}{G}\equiv G\,M^2_{\rm ADM}\,.
\label{eq:ekerrs}
\end{equation}
Taking a near horizon limit \cite{bh} 
\begin{equation}
  t = \frac{\lambda\,\hat t}{2M}\,, \qquad y = \frac{\lambda\,M}{r-M}\,, \qquad \phi = \hat \phi - \frac{\hat t}{2M}\,, \qquad \qquad \lambda\to 0
\label{eq:nhkerr}
\end{equation}
keeping $(t,\,y,\,\phi,\,\theta)$ fixed, leads to the near-horizon extreme Kerr (NHEK) geometry
\begin{equation}
  ds^2 = 2GJ\Omega^2\left(\frac{-dt^2+dy^2}{y^2} + d\theta^2 + \Lambda^2\left(d\phi + \frac{dt}{y}\right)^2\right)\,,
\end{equation}
where $\Omega^2\equiv(1+\cos^2\theta)/2$ and $\Lambda\equiv 2\sin\theta/(1+\cos^2\theta)$. This has an enhanced $\SL(2,\RR)\times \U(1)$ isometry group acting on the fixed polar angle $\theta$ 3d slices, whose geometry is that of a quotient of warped AdS${}_3$ and describes an S${}^1$ bundle over an AdS${}_2$ base.

To study the semiclassical excitations around NHEK, one studies its asymptotic symmetry group. To do that we impose the boundary conditions \cite{kerrcft}
\begin{equation}\label{strictbc}
\left(
  \begin{array}{ccccc}
 h_{\tau\tau}= \mathcal{O}({r^2}) & h_{\tau\varphi}= \mathcal{O}({1}) & h_{\tau\theta}= \mathcal{O}(r^{-1}) &h_{\tau r}= \mathcal{O}(r^{-2})  \\
 h_{\varphi \tau}=h_{\tau\varphi} & h_{\varphi\varphi}= \CO(1) &h_{\varphi\theta}= \mathcal{O}(r^{-1})  &h_{\varphi r}= \mathcal{O}(r^{-1})  \\
   h_{\theta \tau}=h_{\tau\theta} & h_{\theta\varphi}=h_{\varphi\theta} & h_{\theta\theta}= \mathcal{O}(r^{-1}) &h_{\theta r}= \mathcal{O}(r^{-2}) \\
   h_{r\tau}=h_{\tau r} & h_{r\varphi}=h_{\varphi r} & h_{r\theta}=h_{\theta r} & h_{rr}= \mathcal{O}(r^{-3}) \\
  \end{array}
\right)\ , 
\end{equation}
where the radial coordinate $r$ in the global AdS${}_2$ coordinates was used
\begin{eqnarray}
  y &=& \left(\cos\tau\,\sqrt{1+r^2} + r\right)^{-1}\,, \nonumber \\
  t &=& y\sin\tau\,\sqrt{1+r^2}\,, \\
  \phi &=& \varphi + \log\left(\frac{\cos\tau + r\sin\tau}{1+\sin\tau\,\sqrt{1+r^2}}\right)\,. \nonumber
\end{eqnarray}
The most general diffeomorphism preserving these boundary conditions is \cite{kerrcft}
\begin{equation*}
\zeta = \left(-r\epsilon^\prime(\varphi) + \CO(1)\right)\partial_r + \left(C+\CO(r^{-3})\right)\partial_\tau + \left(\epsilon(\varphi) + \CO(r^{-2})\right)\partial_\varphi + \CO(r^{-1})\partial_\theta\,,
\end{equation*}
where $\epsilon(\varphi)$ is an arbitrary smooth function of the periodic boundary coordinate $\varphi$ and $C$ is an arbitrary constant. Expanding $\epsilon(\varphi)$ into Fourier modes and defining dimensionless quantum versions of the $Q$s by $\hbar L_n \equiv Q_{\zeta_n}+\frac{3J}{2}\delta_{n}$ plus the usual rule of Dirac brackets to commutators as $\{.,.\}_{D.B.}\to-i/\hbar\,[.,.]$,  the quantum charge algebra is then \cite{kerrcft,Compere:2009dp}
\begin{equation} 
[L_m, L_n]= (m-n) L_{m+n} + \frac{J}{\hbar}m(m^2 - 1)\delta_{m+n,0}\,. 
\label{eq:kerrvirasoro}
\end{equation} 
This is a Virasoro algebra with central charge
\begin{equation}\label{cc} 
  c_L = \frac{12J}{\hbar} \, . 
\end{equation}

\subsubsection{CFT origin of the gravitational entropy}

If the gravitational entropy of extremal Kerr allows a microscopic interpretation in terms of a chiral CFT as suggested by the chiral Virasoro algebra emerging from the previous semiclassical considerations, one is left to determine the temperature of the mixed state describing the NHEK geometry. To identify it, remember the state of an scalar quantum field in the Kerr background after integrating out its interior is given by a density matrix with eigenvalues
\begin{equation}\label{ggf}
e^{-\hbar \frac{\omega-\Omega_{H} m}{T_H} }\,, \quad \quad \text{with} \quad \quad \Omega_H=\frac{a}{2Mr_+}\,, \quad T_H=\frac{r_+-M}{4\pi\,M\,r_+}
\end{equation}
Here $\Omega_H$ is the angular velocity of the horizon and $T_H$ is its Hawking temperature. We can relate these eigenvalues to the ones associated to the Killing vector fields $\partial_\phi$ and $\partial_t$ naturally appearing in the near-horizon region through the identity
\begin{equation} 
e^{-i\omega \hat t+i m\hat \phi}=e^{-\frac{i}{\lambda}(2M\omega -{m}) t+i m\phi}=
e^{-in_Rt+i n_L \phi}\, 
\end{equation} 
where 
\begin{equation}
   n_L\equiv m,~~~~~~~~n_R\equiv \frac{1}{\lambda}(2M\omega -{m})\,.
\end{equation}
In terms of these variables the Boltzmann factor (\ref{ggf}) is 
\begin{equation} 
e^{-\hbar\frac{\omega-\Omega_H m}{T_H} }=e^{-\frac{n_L}{T_L}
-\frac{n_R}{T_R}}\,, 
\end{equation} 
where the dimensionless left and right
temperatures are 
\begin{equation} 
T_L= \frac{r_+ -M}{2\pi\,(r_+-a)},~~~~~T_R=\frac{r_+ -M}{2\pi \lambda r_+}\,. 
\end{equation} 
In the extremal limit $M^2\to GJ$, these reduce to \cite{kerrcft}
\begin{equation} 
T_L=\frac{1}{2\pi},~~~~~T_R=0\,. 
 \label{eq:tlkerr}
\end{equation} 
The left-movers are then thermally populated with the Boltzmann distribution at temperature $1/2\pi$:
\begin{equation} 
  e^{-2\pi n_L}\,,
\end{equation} 
while only the purely reflecting modes survive the limit since $\omega = m/(2M)$. Thus,
even though extreme Kerr has zero Hawking temperature, the quantum fields
outside the horizon are not in a pure state.

Assuming the existence of a unitary chiral CFT with central charge \eqref{cc}, one is tempted to appeal to Cardy's formula to account for the CFT entropy\footnote{Cardy's formula requires the temperature to be large. See \cite{Jejjala:2009if} for a justification on the validity of Cardy's regime for extremal Kerr.}
\begin{equation} 
  S_{\text{CFT}} = \frac{\pi^2}{3}c_L T_L\,. 
\end{equation}
Using (\ref{eq:tlkerr}) and (\ref{cc}), one reproduces the entropy of extremal Kerr (\ref{eq:ekerrs})  \cite{kerrcft}
\begin{equation} 
  S_\text{micro}=\frac{2 \pi J}{\hbar}=S_{BH}\,.
\end{equation}
Notice this approach uses the symmetries emerging in the semiclassical analysis and the universality of Cardy's formula, but does not provide any explicit microscopic description of the system. This is a common feature of this kind of considerations.

This evidence was used in \cite{kerrcft} to conjecture a new duality between quantum gravity in the near horizon of extremal Kerr and (a chiral half of) a two dimensional conformal field theory, the so called Kerr/CFT correspondence. By considering near-extremal Kerr, it was shown in \cite{superradiance} that the superradiant scattering of an scalar field by a near-extreme Kerr black hole was fully reproduced by a two dimensional conformal field theory in which the black hole corresponds to a thermal state and the scalar field to a specific operator in the dual CFT, extending the standard AdS/CFT framework.

In the following, I briefly discuss how this same structure emerges for any extremal black hole giving rise to the extremal black hole/CFT conjecture \cite{Hartman:2008pb,otherwork}. For preliminary work on the subject regarding the entropy of near-extremal black holes and the AdS${}_2$/CFT${}_1$ correspondence, see \cite{NavarroSalas:1999up}. For a more complete set of references on the subject, see the recent review \cite{Bredberg:2011hp}.

\subsubsection{General extremal black holes \& conformal field theory}
\label{sec:gebh}

Consider any asymptotically globally AdS (or Minkowski) extremal black hole solution to a general theory of $D=4,5$ Einstein gravity coupled to some arbitrary set of Maxwell fields $F^I$ and neutral scalars $\phi^A$. Extremality requires the existence of more than one charge besides mass. That is, either angular momentum, as in Kerr, or electric/magnetic charges. It was shown in \cite{harvey} that assuming the black hole has a regular horizon and an $\RR\times \U(1)^{D-3}$ isometry group, Einstein's equations guarantee the corresponding near-horizon geometry is
\begin{eqnarray}
&& ds^2 = \Gamma(\rho) \left[ -r^2 dt^2 +\frac{dr^2}{r^2}  \right] + d\rho^2 +\gamma_{ij}(\rho) (dx^i +k^i rdt)(dx^j +k^j rdt) \nonumber \\
&&F^I= d[  e^I rdt + b^I_i(\rho)(dx^i +k_I^i rdt) ]  \nonumber \\
&& \phi^A = \phi^A(\rho) \label{eq:gexthorizon}
\end{eqnarray}
where $i=1, \dots D-3$, $r=0$ is the horizon, $\Gamma(\rho)>0$, $\{\partial/ \partial t,\, \partial / \partial x^i\}$ are Killing vector fields, and $k^i,e$ are constants. The precise form of the $\rho$-dependent functions depends on the subset of field equations that have not yet been integrated. This metric has several S${}^1$ bundles over AdS${}_2$, the latter spanned by $\{t,\,r\}$. Thus, there exists an enhancement of symmetry to $\SO(2,1)\times \U(1)^{D-3}$. The presence of these bundles will be crucial to extend the previous  semiclassical considerations leading to the existence of the chiral Virasoro algebra \eqref{eq:kerrvirasoro} for extremal Kerr. If the initial black hole only carries electric charges, this can be viewed as "rotation" using a convenient KK reduction from higher dimensions. Thus, the conclusions below would end up being the same \cite{Hartman:2008pb}.

To see how the ideas developed for extremal Kerr extend to more general situations, consider
the most general near horizon geometry in 5d :
\begin{eqnarray}
ds_5^2 &=& A(\theta) \left( -r^2 dt^2 + \frac{dr^2}{r^2} \right)
+ F(\theta) d\theta^2 + B_1(\theta)\, \tilde e_1^2 +
B_2 (\theta) (\tilde e_2 + C(\theta)\, \tilde e_1)^2\,,\nonumber\\
\tilde e_1 &=& d\phi_1 + k_1 r\, dt\,,\qquad
\tilde e_2 = d\phi_2 + k_2 r\, dt\,,\label{metform1}
\end{eqnarray}
where $A$, $B_i$, $C$ and $F$ are functions of the latitude coordinate
$\theta$ (the analogue of $\rho$ in \eqref{eq:gexthorizon}). The metric can be viewed as an $S^3$ bundle over AdS$_2$ and its Bekenstein--Hawking entropy is
\begin{equation}
S_{BH}= \frac{1}{4} \int d\theta \sqrt{B_1 B_2 F} \int d\phi_1
d\phi_2\,.
\end{equation}

It was shown in \cite{Chow:2008dp} that this near horizon geometry has a pair of commuting diffeomorphisms that generate two commuting Virasoro algebras
\begin{eqnarray}
\zeta^{\scriptscriptstyle{1}} = - \textrm{e}^{-{{\rm i}} n \phi_1}\, \frac{\partial}{\partial\phi_1} - {{\rm i}} n\, r\,
   \textrm{e}^{-{{\rm i}} n \phi_1}\, \frac{\partial}{\partial r}\,,\nonumber\\
\zeta^{\scriptscriptstyle{2}} = - \textrm{e}^{-{{\rm i}} n \phi_2}\, \frac{\partial}{\partial\phi_2} - {{\rm i}} n\, r\,
   \textrm{e}^{-{{\rm i}} n \phi_2}\, \frac{\partial}{\partial r}\,.
\end{eqnarray}
Using the covariant formalism reviewed before, one can compute the central charges $c_i$ in these Virasoro algebras \cite{Chow:2008dp}
\begin{equation}
c_i = \frac{3}{2\pi} k_i \int d\theta \sqrt{B_1 B_2 F} \int d\phi_1
d\phi_2 = \frac{6 k_i S_{BH}}{\pi}\,, \quad i=1,\,2
\end{equation}
They reproduce the entropy of the original black hole, through Cardy's formula,
\begin{equation}
S_{BH} = \frac{\pi^2}{3} c_1 T_1 = \frac{\pi^2}{3} c_2 T_2\,,
\end{equation}
if the constants $k_1$ and $k_2$ are related to the CFT temperatures by
\begin{equation}
k_i=\frac{1}{2\pi T_i}\,.
\label{eq:tkcft}
\end{equation}
It is reassuring to check that these are precisely the values that these constants take when we consider the near horizon of a given extremal black hole. Indeed, in that case,
\begin{equation}
T_i = \lim_{r_+ \to r_0} \frac{T_{\textrm{H}}}{\Omega_i^0 - \Omega_i}
= - \frac{T^{\prime \, 0}_\textrm{H}}{\Omega^{\prime \, 0}_i}\,,
\label{Tidef}
\end{equation}
where $\Omega_i$ describe the angular velocities of the black hole at the horizon, $T_H$ its Hawking temperature and quantities with a $0$ label refer to their extremal values. These results can be generalised to higher dimensions \cite{Chow:2008dp}.

These constants $T_i$ appear naturally when expanding quantum fields in eigenmodes of the asymptotic charges. For an scalar field, after tracing over the interior of the black hole, the vacuum is a diagonal density matrix with eigenvalues
\begin{equation}
  e^{-\left(\omega - \Omega_1m_1-\Omega_2m_2\right)/T_H}\,.
\end{equation}
Expanding $T_H = T_H^\prime\,x$ and $\Omega_i = \Omega_i^0 + \Omega_i^\prime\,x$, where $x$ measures the distance to the extremal point, one concludes the density matrix after the extremal limit is given in terms of
\begin{equation}
  e^{-\frac{m_1}{T_1} -\frac{m_2}{T_2}}\,,
\end{equation}
with $T_i$ defined as in (\ref{Tidef}). Thus these quantities can be interpreted as the Frolov--Thorne temperatures \cite{frotho} associated with two CFTs, one for each azimuthal angle $\phi_i$.  Furthermore, the above requires the relation $\omega=\Omega_1^0m_1+\Omega_2^0m_2$ among the different quantum numbers. These are the modes that are fully reflected from the black hole (no energy absorbed by the black hole) \cite{bh}. 

The existence of more than one CFT description may appear to be surprising. But we are well aware of this same fact for the black holes described in \cite{Strominger:1996sh}. There, the entropy only depends on two of the three charges carried by the black holes, and depending on the U-duality frame being used, there are different available CFTs. In the current context, the existence of a lattice of CFTs was argued for in \cite{Loran:2009cr}. It is not clear whether the $\SL(2,\,\RR)$ transformations acting on the moduli characterising the 5d near horizon geometry can be interpreted as a U-duality transformation. Embedding these systems in string theory, as in \cite{Jejjala:2009if}, could clarify this point.

The superradiant scattering of an scalar field by these backgrounds also matches the analogous calculation in a chiral 2d CFT, using the appropriate dual operator \cite{Cvetic:2009jn}. This provides similar evidence to the one reported for extremal Kerr. It is interesting to point out the work in \cite{Becker:2010dm}, where besides providing further evidence for this correspondence, bulk correlators are computed for asymptotically flat black holes using the same recipe developed in the AdS/CFT correspondence, perhaps pointing towards the existence of new holographic relations for this different asymptotics.

Before closing this discussion, it is interesting to mention the potential connection between the results reviewed and further existent work in the literature. Prior to the Kerr/CFT conjecture, it was already observed that a quantum theory of gravity in 2d with negative cosmological constant coupled to an electric field could allow a non-trivial central charge under a suitable set of boundary conditions \cite{Hartman:2008dq}. These were responsible for twisting the energy momentum tensor $T_{\pm\pm}$ generating 2d conformal transformations by the $\U(1)$ gauge current $j_\pm$ generating $\U(1)$ gauge transformations as
\begin{equation}
  \tilde{T}_{\pm\pm} = T_{\pm\pm} \pm \alpha\,\partial_\pm j_\pm\,.
\end{equation}
The constant $\alpha$ depends on the details of the theory (the AdS${}_2$ radius $\ell$ and the electric field $E$ in this case). Notice this twisting is responsible for generating a non-trivial central charge, since the original $T_{\pm\pm}$ has $c=0$, as is customary in two spacetime dimensions. This mechanism was explicitly realised in a holographic formulation of AdS${}_2$ black holes cross-checked from Kaluza-Klein compactification of the standard AdS${}_3$/CFT${}_2$ dictionary \cite{Castro:2008ms}. This may essentially be the same phenomena that is happening in the generic extremal near horizon geometry described in this section. If the starting extremal black hole has a compact horizon, its near horizon geometry \eqref{eq:gexthorizon} will allow, in principle, an effective description in terms of 2d gravity coupled to matter fields. One crucial property of the solutions to this theory is that all effective electric fields in 2d diverge at infinity. This is the reason why standard 2d conformal transformations must be accompanied by a $\U(1)$ gauge transformation to preserve the physical boundary conditions analysed in \cite{Hartman:2008dq}. In other words, the singular behaviour of the electric field at infinity is the origin of the twisting. Interestingly, this twisting is reminiscent of the large gauge transformation that takes place in any near horizon limit for any extremal black hole. As one can see from \eqref{eq:nhkerr}, this limit involves two transformations
\begin{itemlist}
\item[1.] An IR limit, due to the red shift inherited by exploring the near horizon region of the starting extremal black hole, i.e. $r= r_h + \lambda\,y$ with $\lambda\to0$
\item[2.] A large gauge transformation $t=\frac{\tau}{\lambda}$ and $\phi=\varphi + \Omega^0\frac{\tau}{\lambda}$.
\end{itemlist}
The large gauge transformation acts non-trivially on the generators of isometries $\partial_t$ and $\partial_\phi$ on any quantum field propagating in this fixed background. At the level of the full theory, these symmetries are generated by the energy momentum tensor $T_{\pm\pm}$ and the $\U(1)$ gauge current $j_\pm$. Thus, the large gauge transformation implements the CFT twisting described in \cite{Hartman:2008dq} at the level of the geometry. In the particular case where there is an available AdS/CFT description in the UV, one is tempted to view this near horizon limit as effectively implementing a non-trivial RG-flow in the field theory, by integrating out the spacetime outside of the black hole horizon. The non-trivial singular large gauge transformation required to keep the solution on-shell identifies the appropriate hamiltonian in the IR. For related discussions having condensed matter applications in mind, see the recent work \cite{RG-CMT}.

\subsubsection{Comments on the existence of non-trivial dynamics} 
\label{sec:dynamics}

There are several arguments challenging whether the extremal BH/CFT correspondence has any dynamical content on it and just contains the degeneracy of the "vacuum" state
\begin{romanlist}[(iii)]
\item The existence of AdS${}_2$ fragmentation \cite{fragmentation} in the two-dimensional Einstein-Maxwell-Dilaton theory with a negative cosmological constant states that, at least classically, any matter excitation satisfying the null energy condition with support in the AdS${}_2$ base will back-react strongly, modifying the spacetime boundary structure. This suggests there should be no states charged under the $\SL(2,\RR)$ isometry group whenever this energy condition is satisfied and the physics we are interested in are described by the same effective 2d Einstein-Maxwell-Dilaton theory.
\item There are {\it no} normalisable linear perturbations of the NHEK geometry \cite{Horowitz-NHEK,Reall-NHEK} satisfying the boundary conditions described in \cite{kerrcft}. In \cite{Horowitz-NHEK}, it was further argued that the same conclusion holds for non-linear perturbations, which would match the above expectation in the semiclassical approximation.
\item Sen's work on extremal black holes reproducing microscopic results from a purely macroscopic point of view \cite{sen-0,sen-1}, also concluded that when applying the straight AdS/CFT correspondence to the particular AdS${}_2$/CFT${}_1$ correspondence, the dual conformal quantum mechanics only includes the degeneracy of the vacuum, as also argued below \cite{Sen:2008yk,Gupta:2008ki}. 
\end{romanlist}
For asymptotically AdS black holes, this expectation can be understood as follows.
Extremal black holes represent complicated mixed states in the dual
UV CFT. Their excitations will have a gap if this CFT is non-singular and defined on
the cylinder $\mathbb R\times$S${}^{d-1}$. At sufficiently low energies above, less than the size of the gap, there will be no dynamics left, and no non-trivial theory remains. In the next subsection, we will see how this mechanism operates in AdS${}_3$/CFT${}_2$.

This argument already suggests a couple of ways to circumvent its conclusion
\begin{romanlist}[(ii)]
\item if the field theory lives on a {\it non-compact space}, its spectrum will be continuous. Thus the effective two dimensional Newton constant in AdS${}_2$ will vanish, allowing us to bypass the fragmentation argument. This feature has appeared prominently in some recent applications of the AdS/CFT to condensed matter systems \footnote{The amount of literature here is immense. We refer the reader to a subset of reviews and references therein \cite{AdS-CMT}.}. 
\item reduce the gap of the dual CFT by taking a {\it large central
charge or large $N$ limit}, since the gap typically scales with an
inverse power of $N$ or $c$. For finite size extremal black holes
this would lead to a divergent entropy for a fixed temperature $T$. To obtain a finite entropy, it is tempting to consider large $N$ limits together with a {\it vanishing horizon} limit. 
\end{romanlist}

Interestingly, it had previously been observed that under certain circumstances, whenever
the horizon area of {\it extremal} black holes can be tuned to zero, their near-horizon geometries develop local AdS${}_3$ throats \cite{bh,Balasubramanian:2007bs,Fareghbal:2008ar,Fareghbal:2008eh}. This is remarkable for at least two reasons
\begin{itemlist}
\item[1.] Given the AdS${}_3$/CFT${}_2$ correspondence, these particular points in the moduli space of extremal black holes may provide independent derivations for the existence of an IR CFT
description of the black hole degrees of freedom.
\item[2.] Since these configurations are continuously connected to large extremal black holes where our previous considerations apply, one may identify the operator deforming the 2d CFT dual to AdS${}_3$ and hope to be able to identify the finite deformation induced by it on the initial 2d CFT. This way, one could in principle try to derive whether the extremal black hole/CFT conjecture holds.
\end{itemlist} 
A general caveat about these classical configurations is its singular nature, and what their fate is when corrections are included in the bulk. There is an important distinction to be made in the cases discussed so far in the literature: when the system is near-extremal but far from BPS, the near horizon geometry involves a non-supersymmetric {\it pinching} $\ZZ_N$ orbifold of AdS${}_3$\footnote{The action of this orbifold at the AdS${}_3$ boundary is like the one of a conical defect. It would be interesting to see whether the techniques developed in \cite{Martinec:2001cf} to compute the worldsheet string perturbative spectrum can be extended to this case, and whether there is any interesting structure emerging in the large N limit.}, whereas in the near BPS situation the transverse space decompactifies, but has an smooth 3d AdS throat.

Even though this direction will not be reviewed in these notes, it is worth mentioning two different approaches that have been followed
\begin{itemlist}
\item[1.] Given a near extremal black hole in AdS providing a well defined UV CFT dual description, one interprets the near horizon limit as a {\it large N IR limit} of the original CFT  focusing in some low energy excitations of a definite sector of its Hilbert space selected by the {\it large gauge transformations} accompanying the near horizon limit. This is the approach followed in \cite{Balasubramanian:2007bs,Fareghbal:2008ar,Fareghbal:2008eh}. In this context, the conjectured CFT appearing in the extremal black hole/CFT correspondence emerges as an effective description for these excitations.
\item[2.] When no UV dual description is available, embedding the given black hole into string theory may provide with the existence of some points in the U-duality orbit where there exists a CFT dual. Tuning the charges of the black hole in that U-dual frame may allow the emergence of a local AdS${}_3$ throat in the near horizon limit. One then identifies its central charge and temperature, matching the bulk entropy. Moving away from the point in charge space where the AdS${}_3$ exists, one computes the linear deformation in the geometry allowing to identify the dual marginal operator deforming its dual 2d CFT dual. Ideally, the finite integration of this marginal deformation would connect the 2d CFT dual to AdS${}_3$ to the one emerging in the extremal BH/CFT correspondence. This was the approach followed in \cite{guica,compere}. 
\end{itemlist}

Despite these observations, one may still be interested in investigating whether the chiral CFT structure emerging in the strict extremal limit hides the existence of a non-chiral CFT as non-extremality is turned on. This is the direction pursued in \cite{finn-ale1,finn-ale2} for small non-extremality and in \cite{alex-ale,Becker:2010dm} for finite non-extremality. The conclusion in all these works is affirmative, but further work is required to settle this very important question regarding non-extremal black holes. Both, research in extremal black hole microscopics and its applications to strongly coupled condensed matter systems suggest that the emergence of these CFTs appears at scales below a certain cut-off. Beyond it, extra degrees of freedom are necessary, and whether their interactions allow a CFT description remains an open question.

\subsection{The AdS${}_3$ perspective}

The physics of AdS${}_3$ provides an excellent arena where to test the ideas previously described. First, AdS${}_3$ allows a perturbative description as a 2d CFT string worldsheet \cite{ads3}. Second, due to the lack of bulk degrees of freedom in three dimensions, black holes in 3d gravity with a negative cosmological constant correspond to quotients of global AdS${}_3$ \cite{BTZ-henneaux}. The latter also allow a perturbative worldsheet description \cite{btz-worldsheet}. Third, due to the well-established AdS${}_3$/CFT${}_2$ correspondence, the system has a UV description in terms of a 1+1 non-chiral CFT, prior to any low energy (near horizon) being considered.  Finally, we already know that the two Virasoro algebras in this CFT are realised, in the semiclassical approximation, as a set of non-trivial diffeomorphisms preserving some set of boundary conditions defining the AdS${}_3$ asymptotics \cite{Brown-Henneaux}.

In the following, our goal will be to understand the previous IR limits in terms of AdS${}_3$/CFT${}_2$. Let us first review both the bulk and CFT description of BTZ black holes \cite{BTZ-original}. These are asymptotically AdS$_3$ spacetimes with metric %
\begin{equation}\label{BTZ-metric}%
ds^2 = -\frac{(r^2-r_+^2)(r^2-r_-^2)}{r^2 \ell^2} dt^2
 +  \frac{\ell^2 r^2}{(r^2-r_+^2)(r^2-r_-^2)} dr^2
 + r^2 (d\phi - \frac{r_+r_-}{\ell r^2} dt)^2 \,.
\end{equation}%
The periodicity $\phi\sim\phi + 2\pi$ makes them a quotient of AdS${}_3$ \cite{BTZ-henneaux}. Their ADM mass and angular momentum are
\begin{equation}\label{mass-J}%
M\ell=\frac{r_+^2+r_-^2}{8G_3\ell}\ ,\quad J=\frac{r_+r_-}{4G_3\ell}\,.%
\end{equation}%
These depend on the radius of AdS${}_3$ $\ell$, 3d Newton's constant $G_3$ and both the inner and outer horizons, $r_-$ and $r_+$, respectively.

Their dual interpretation is in terms of thermal states in a 1+1 non-chiral CFT with left and right temperatures
\begin{equation}\label{TL-TR}
T_L=\frac{r_++r_-}{2\pi\ell} \,,\qquad T_R=\frac{r_+-r_-}{2\pi\ell}\,,
\end{equation}
related to the Hawking temperature of the BTZ black hole $T_H$ by $\frac{2}{T_H}=\frac{1}{T_L}+\frac{1}{T_R}$.

The connection between the gravity and CFT descriptions is most easily reviewed following the
asymptotic symmetry group analysis of Brown and Henneaux \cite{Brown-Henneaux}. To do so, it is more convenient to work with lightlike coordinates $\hat u=t/\ell-\phi$ and $\hat v=t/\ell+\phi$, in which the global AdS$_3$ metric is $ds^2=\ell^2(\frac{dr^2}{r^2}-2 r^2 d{\hat u} d{\hat v})$.
The boundary conditions at large $r$ are \cite{Brown-Henneaux}
\begin{equation}\label{Brown-Henneaux-b.c.}%
 \delta g_{\hat u\hat u}\sim \delta g_{\hat v\hat v}\sim
\delta g_{\hat u\hat v}\sim {\cal O}(1), \qquad \delta g_{rr}\sim
{\cal O}\left(r^{-4}\right),\quad \delta g_{r\hat u}\sim \delta g_{r\hat
v}\sim {\cal O}\left(r^{-3}\right)\ .%
\end{equation}%
That order one fluctuations in $\delta g_{\hat u\hat u}$, $\delta g_{\hat v\hat v}$ correspond to normalisable modes in the 2d CFT can be inferred by writing the BTZ black hole metric in these coordinates and examining its asymptotics. In this way, order ${\cal O}(1)$ fluctuations in $\delta g_{\hat u\hat u}$, $\delta g_{\hat v\hat v}$ are seen to independently change the mass and angular momentum in the dual $2d$ CFT. 

The set of non-trivial charges constructed out of the diffeomorphisms preserving these boundary conditions close two commuting Virasoro algebras
\begin{eqnarray}
  \left[L_m,\,L_n\right] &=& \left(m-n\right)\,L_{m+n} + \frac{c}{12}m(m^2-1)\delta_{m+n,0}\,, \nonumber \\
  \left[{\bar L}_m,\,{\bar L}_n\right] &=& \left(m-n\right)\,{\bar L}_{m+n} + \frac{{\bar c}}{12}m(m^2-1)\delta_{m+n,0}\,, 
\label{eq:virasoro}
\end{eqnarray}
with central charges 
\begin{equation}
  c = {\bar c} = \frac{3\ell}{2G_3}\,.
\end{equation}
The generators $L_n$ and ${\bar L}_n$ are given by
\begin{equation}
  L_n - \frac{c}{24}\,\delta_{n,0} = e^{in\hat{v}}\,\frac{\partial}{\partial \hat{v}}\,, \quad\quad
  {\bar L}_n - \frac{{\bar c}}{24}\,\delta_{n,0} = e^{in\hat{u}}\,\frac{\partial}{\partial \hat{u}}\,,
\label{eq:virasorovf}
\end{equation}
whereas their zero modes are related to the bulk charges by
\begin{equation}\label{L0-barL0}%
L_0-\frac{c}{24}=\frac{M\ell+J}{2}\,,\quad \quad
\bar L_0-\frac{{\bar c}}{24}=\frac{M\ell-J}{2}\,.
\end{equation}
The states in the Hilbert space thus constructed arrange themselves into representations of these Virasoro algebras. It is then a universal result that the number of highest weight operators/states carrying conformal dimension $\Delta = L_0 + \bar{L}_0$ and spin $L_0-\bar{L}_0$ is given by Cardy's formula \eqref{eq:cardy} \cite{Cardy:1986ie}. Using the dictionary described above
\begin{equation}\label{entropy-CFT-gravity}%
S_{\text{Cardy}}= 2\pi\sqrt{\frac{c}{6}\left(L_0-\frac{c}{24}\right)} +
 2\pi\sqrt{\frac{{\bar c}}{6}\left(\bar{L}_0-\frac{{\bar c}}{24}\right)}= \frac{2\pi r_+}{4G_3} = S_{\text{B-H}}\,,
\end{equation}
one always reproduces the Bekenstein-Hawking formula \eqref{eq:beh-haw} using the BTZ metric \eqref{BTZ-metric}.

\subsubsection{Near horizon of extremal BTZ black holes}

Consider the subset of extremal BTZ black holes for which the inner horizon coincides with the outer one. Denote the horizon by $r_h=r_+=r_-$, then
\begin{equation}
  ds^2 = -\frac{(r^2-r_h^2)^2}{r^2\,\ell^2}\,dt^2 + \frac{\ell^2\,r^2}{(r^2-r_h^2)^2}\,dr^2 + r^2\,\left(d\phi - \frac{r_h^2}{r^2}\,\frac{dt}{\ell}\right)^2\,.
\label{eq:btz}
\end{equation}
The UV dual description of this limit, $M\ell = J$, involves setting the right-movers to their ground state
\begin{equation}
{\bar L}_0 = \frac{c}{24} ~;~~~~ T_R = 0\,,
\end{equation}
while the left moving temperature $T_L = \frac{r_h}{\pi\ell}$ and $L_0$ remain arbitrary. 

Following the philosophy discussed for generic extremal black holes in four and five dimensions, we want to recover the same physics from the low energy limit involved when taking the near horizon limit of these black holes. To study the latter, introduce new coordinates
\begin{equation}\label{uvrho}%
\hat u=t/\ell-\phi\ ,\qquad \hat v=t/\ell+\phi,\qquad
r^2-r_h^2=\ell^2
e^{2\rho}\,, %
\end{equation}%
in which the metric (\ref{eq:btz}) takes the form
\begin{equation}
ds^2=r^2_h \,  d\hat u^2+\ell^2 \,  d\rho^2-\ell^2 e^{2\rho} \, d\hat u \,  d\hat v\ .%
\label{Ext-BTZ}
\end{equation}
The near horizon limit consists of taking $\rho_0\to-\infty$
\begin{equation}
\rho=\rho_0+ r,\quad u=\hat u \, \frac{r_h}{\ell},\quad v=
e^{2\rho_0} \frac{\ell}{r_h} \, \hat v , \quad \{ u,v \} \sim \{u -
2\pi \frac{r_h}{\ell}, v + 2\pi \frac{\ell}{r_h} e^{2 \rho_0} \}
\label{NHlimit-BTZ}
\end{equation}
while keeping $r,\ u,\ v$ and $r_h$ fixed. The resulting near horizon metric
\begin{equation}
 ds^2 = \ell^2(  du^2 + dr^2 - e^{2r} \, du \, dv )
\label{NHlimit-BTZmet1}%
 \end{equation}
is locally identical to (\ref{Ext-BTZ}) but the boundary periodicities in the limit are
\begin{equation}
 \{u, v \} \sim \{u - 2\pi \frac{r_h}{\ell}, v \} \, .
 \end{equation}
Thus, the boundary of (\ref{NHlimit-BTZmet1}) ($r \to \infty$) is a
``null cylinder'' -- it has a metric conformal to $du \, dv$, the
standard lightcone metric on a cylinder, but has a single compact {\it
null} direction ($u$). 

The metric (\ref{NHlimit-BTZmet1}) is the spacelike self-dual orbifold \cite{selfdual,asad} \footnote{The importance of this geometry for the physics of extremal black holes was already emphasised some time ago in \cite{andyads2,asad}.}, an S${}^1$ fibration over an AdS${}_2$ base with isometry group $\SL(2,\RR)\times \U(1)$, which is more easily seen when written as%
\begin{equation}
  ds^2 = \frac{\ell^2}{4}\frac{d\rho^2}{\rho^2} - \frac{\rho^2}{r_h^2}\frac{d\tau^2}{\ell^2} + r_h^2\left(d\varphi + \frac{\rho}{r_h^2}\,\frac{d\tau}{\ell}\right)^2 = \frac{\ell_3^2}{4}\frac{d\rho^2}{\rho^2}+ 2\frac{\rho}{\ell} d\tau d\varphi+ r_h^2 d\varphi^2\,.
\label{eq:c1}
\end{equation}

To understand the physical meaning of a null boundary cylinder, introduce a UV cut-off in the dual CFT by considering bulk surfaces of fixed large $r$
\begin{equation}
  ds^2 = du^2 - e^{2r} \,du \, dv\,.
 \label{NHmetsimp} 
\end{equation} 
It was shown in \cite{asad} that (\ref{NHmetsimp}) is conformal to a boosted cylinder. As $r \to \infty$ the boost becomes infinite, precisely matching the procedure defined by Seiberg \cite{Seiberg} for realising the Discrete Light Cone Quantization (DLCQ) of a field theory\footnote{The precise definition of DLCQ  in quantum field theory is rather subtle. As emphasised in \cite{Hellerman:1997yu}, amplitudes computed in these theories diverge order by order in perturbation theory due to strong interactions among longitudinal zero modes. This quantization scheme was argued to be well defined non-perturbatively.}. This suggests the spacelike self-dual orbifold is dual to the DLCQ of the original 1+1 non-chiral CFT. We will see later the latter only keeps one chiral sector of the original UV CFT, in agreement with our previous Kerr/CFT considerations. The only parameter of the spacelike self-dual orbifold metric, $r_h$, is then related to the value of the light-cone momentum $p^+$ defining the DLCQ sector
\begin{equation}\label{LC-momentum}%
p^+=\frac{c}{6} \left(\frac{r_h}{\ell}\right)^2\,.
\end{equation}
This conclusion can also be reached by computing its boundary stress-tensor \cite{vijay-simon}. The spacelike self-dual orbifold (\ref{eq:c1}) has a finite temperature \cite{Balasubramanian:2009bg}\footnote{There are different ways of arguing the existence of this temperature. From the global version of the spacelike self-dual orbifold \cite{asad,vijay-simon} containing two disjoint causally connected boundaries, the finite temperature originates from entanglement entropy after integrating out part of the space leading to the single boundary metric (\ref{eq:c1}), pretty much in the same way Rindler space has a finite temperature when viewed as a local patch of the full Minkowski spacetime.} 
\begin{equation}\label{T-self-dual}
T_{\rm{self-dual}}=\frac{r_h}{\pi \ell}=\sqrt{\frac{6p^+}{\pi^2 c}}\,,
\end{equation}
agreeing with $T_L$ in the UV description. This same temperature could have been derived using the general discussion in subsection~\ref{sec:gebh}. In particular, using the relation \eqref{eq:tkcft} between the CFT temperatures $T_i$ and the constants $k_i$ appearing in the near horizon extremal geometries using the same normalisation as in \eqref{eq:gexthorizon}.

The $r_h\to 0$ limit of the spacelike self-dual orbifold sends the temperature (\ref{T-self-dual}) to zero and yields the metric
\begin{equation}\label{null-orbifold}%
ds^2=r^2 dx^+dx^-+\ell^2\frac{dr^2}{r^2}\,,\qquad x^-\sim
x^-+2\pi\,, 
\end{equation}%
where we have conveniently renamed $\rho=r^2$, $\varphi=x^-$
and $\tau=2\ell x^+$. The causal character of the compact direction $x^-$ has changed, from an everywhere spacelike direction (except at the boundary) to an everywhere null direction. Thus, (\ref{null-orbifold}) should be identified with the null self-dual orbifold \footnote{Readers interested in the supersymmetric properties of this orbifold, see \cite{selfdual,FigueroaO'Farrill:2004yd}. In particular, \cite{FigueroaO'Farrill:2004yd} discusses the embedding of this orbifold in higher dimensional supergravities stressing the importance of the fermion chirality to assess the supersymmetry of this quotient.}. By construction, this spacetime contains closed lightlike curves but it has the same boundary as (\ref{eq:c1}). Thus, it can be viewed as a different state belonging to the same DLCQ CFT. Since the $r_h\to 0$ limit corresponds to $p^+\to 0$, the null self-dual orbifold should correspond to the $p^+=0$ sector of the DLCQ CFT.

Since taking $r_h\to 0$ in (\ref{eq:btz}) corresponds to the massless BTZ black hole 
\begin{equation}
ds^2 = r^2d\tilde x^+d\tilde x^- +\ell^2 \frac{dr^2}{r^2}\quad \quad
\tilde x^\pm = \phi \pm t/\ell\,, \qquad \phi\sim \phi+2\pi\,,
 \label{eq:massless}
\end{equation}
it is natural to view the null self-dual orbifold as its near horizon geometry. Indeed, consider%
\begin{equation}
r=\epsilon \rho\,,\qquad \tilde x^-=x^-\,,\qquad \tilde
x^+=\frac{x^+}{\epsilon^2}\,,\qquad \epsilon\to 0\,. %
\label{eq:limit1}
\end{equation}%
The lightlike direction $\tilde x^+$ effectively decompactifies, while $x^-$ remains compact $x^-\sim x^-+2\pi$. Thus, the near horizon limit \eqref{eq:limit1} of a massless BTZ black hole is the null self-dual AdS${}_3$ orbifold \eqref{null-orbifold}.

\paragraph{Exciting the null self-dual orbifold :} If our interpretation is correct, the spacelike self-dual orbifold \eqref{eq:c1} should be viewed as an excitation over the null self-dual orbifold (\ref{null-orbifold}). In particular, injecting some chiral momentum into the system keeping its causal null cylinder boundary should correspond to the spacelike self-dual orbifold. This is achieved by adding some wave to the conformally flat metric
\begin{equation}
ds^2 = \frac{\ell^2}{z^2}\left[dx^+dx^- + kz^2(dx^-)^2+
dz^2\right]\,.
\end{equation}
Since there are no propagating degrees of freedom in d=3, the latter is locally AdS${}_3$, and it is indeed isometric to the spacelike self-dual orbifold \eqref{eq:c1}, with $r^2_h$ being replaced with $k\ell^2$. 

All these observations are consistent with the well-known fact that
extremal BTZ is a chiral excitation above the massless BTZ black
hole \cite{Cvetic:1998jf,Brecher:2000pa}. This is the three dimensional counterpart of the easiest constructions of non-relativistic gravity duals to DLCQ CFTs in higher dimensions \cite{Goldberger:2008vg,Maldacena:2008wh}, the main difference here being the non-dynamical character of 3d gravity. Notice the only non-singular non-relativistic gravity dual corresponds to the sector of large $p^+$, as is customary in gauge/gravity theory dualities and Matrix theory\cite{BFSS,Balasubramanian:1997kd}.

\paragraph{The pinching $\ZZ_N$ orbifold: } There exists a second inequivalent near horizon limit one could take from (\ref{eq:massless})
\begin{equation}
r=\epsilon\rho\,,   \quad \quad \hat{x}^\pm = \epsilon\,x^\pm\,.
\label{eq:pinching}
\end{equation}
The resulting geometry is locally AdS${}_3$
\begin{equation}
  ds^2 = \rho^2d\hat{x}^+\,d\hat{x}^- + \ell^2\,\frac{d\rho^2}{\rho^2}\quad \quad \hat{x}^\pm\sim \hat{x}^\pm + 2\pi\epsilon\,.
\end{equation}
I will refer to it as a pinching AdS${}_3$ orbifold \cite{evh-btz} given its action on the boundary, which becomes a \emph{pinching cylinder}, $R\times S^1/\mathbb{Z}_N$, with $N=1/\epsilon$. In the bulk though, the quotient is that of a massless BTZ with periodicity scaling to zero \footnote{The same structure appears in the near horizon of extremal black holes with vanishing horizon. See \cite{evh-others} for different examples of its appearance.}. One way of getting some intuition for what this may mean is to consider the identity \cite{evh-btz}
\begin{equation}\label{BTZ-orbifolding}%
\textrm{BTZ}(M\lambda^2,J\lambda^2;2\pi)\equiv\textrm{BTZ}(M,
J;2\pi\lambda)\,, %
\end{equation}%
This states that a BTZ black hole with pinching periodicity $2\pi\lambda$ is classically equivalent to a BTZ black hole with standard periodicity but mass and angular momentum scaled by $\lambda^2$. At any rate, all these geometries are singular, and a proper understanding of the physics for these values of the parameters requires to go beyond the classical gravitational approximation. Since the pinching orbifold is {\it not} equivalent to the null self-dual orbifold, this establishes that different near horizon limits can indeed capture different physics, as we will argue below.

\begin{figure}[pb]
\centerline{\psfig{file=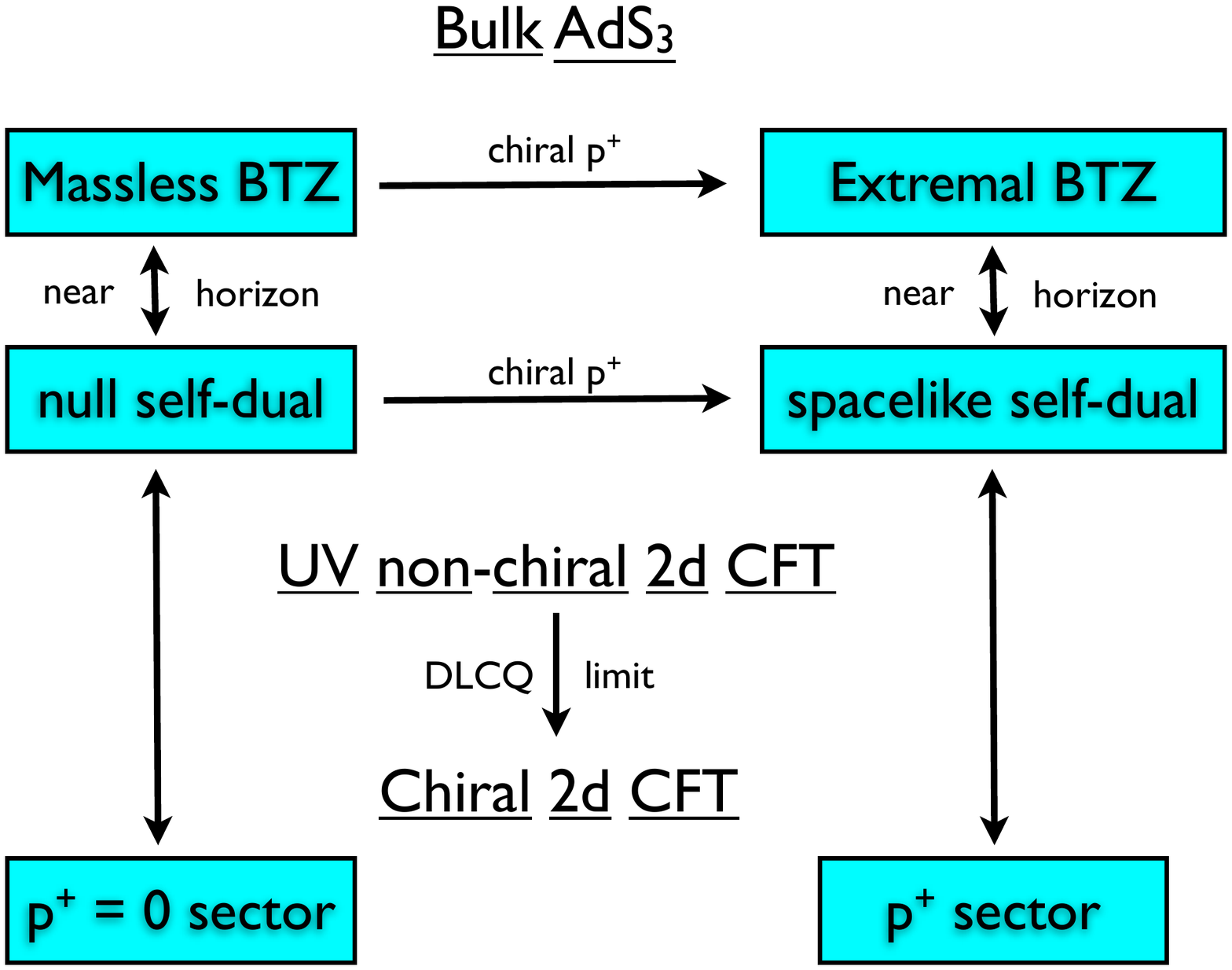,width=15cm}}
\vspace*{8pt}
\caption{Relation between different local AdS${}_3$ geometries, their near horizon limits and their dual interpretations. \label{fig4}}
\end{figure}

\subsubsection{Low energy IR limits of 2d non-chiral CFTs}
\label{2d-CFT-section}%

Let us interpret the bulk near horizon limits in the dual CFT theory. Consider a non-singular non-chiral $2d$ CFT on a cylinder of radius $R$
\begin{equation}
ds^2 = -dt^2 + d\phi^2 = -du' \, dv'   ~~~;~~~ u' = t - \phi,  \  v' = t + \phi\,.
\label{cylinder}
\end{equation}
Since $\phi\sim \phi + 2\pi R$, the light-like coordinates satisfy $\{u' , v'\} \sim
\{u' - 2\pi R, v' + 2\pi R \}$. Let $P^{u'}$ and $P^{v'}$ denote momentum operators in the $v'$
and $u'$ directions respectively.   Their eigenvalues
 \begin{equation}
 P^{v'} =
\left(h+n-\frac{c}{24}\right)\frac{1}{R},\qquad P^{u'} =
\left(h-\frac{c}{24}\right)\frac{1}{R}\,, \qquad  n\in \mathbb Z %
\end{equation}
are given in terms of the quantised momentum $n$ along the S${}^1$,
the $2d$ central charge $c$ and an arbitrary value of $h$ with $h
\geq 0$ and $h + n \geq 0$ due to unitarity constraints. These are related to the eigenvalues
of the standard operators $L_0,\bar{L_0}$ used in radial
quantisation on the plane by $\bar L_0=h+n$ and ${L}_0 = h$. 

Let us first show the DLCQ of this 2d non-chiral CFT is a 2d chiral CFT. 
Following Seiberg \cite{Seiberg}, we will study the consequences of the kinematics of an infinite boost on the discrete spectrum of the theory. We do this because the boundary structure of the near horizon bulk geometry was interpreted above as an infinitely boosted cylinder. Consider a boost with rapidity $\gamma$ and take the double scaling limit
\begin{equation}\label{boost}%
u'  \to e^\gamma u', \qquad v' \to e^{-\gamma} v' \,, \qquad \gamma\to \infty\,, \qquad R_-\equiv R\,e^\gamma\,\text{fixed}
\end{equation}%
The metric is invariant but the cylinder periodicities become
\begin{equation}%
\left(\begin{array}{cc} \phi \\ t \end{array}\right)\sim
\left(\begin{array}{cc} \phi \\
t \end{array}\right)+\left(\begin{array}{cc} 2\pi R \\ 0
\end{array}\right) \ -{\mathrm{infinite\ boost}}
\rightarrow \ \left(\begin{array}{cc} u' \\ v'
\end{array}\right)\sim \left(\begin{array}{cc} u' \\ v'
\end{array}\right)+ \left(\begin{array}{cc} -2\pi  R_-  \\
2\pi R_- e^{-2\gamma}
\end{array}\right)
\label{tpmy-periodicity}
\end{equation}%
We can now identify $\{u',v'\}$ with the light-like boundary coordinates
of AdS${}_3$ in \eqref{NHlimit-BTZ} via $u' = u (\ell / r_+) R_-$ and
$v' = v (r_+ / \ell) R_-$. Comparing (\ref{NHlimit-BTZ}) and
(\ref{tpmy-periodicity}), we conclude that the action of the near
horizon limit on $u,v$ precisely reproduces the identifications
induced by the double scaling limit (\ref{boost}). Thus, the dual to 
the near-horizon geometry of the extremal BTZ black hole should be the DLCQ 
of the 1+1 dimensional CFT dual to AdS${}_3$.

Let us study the states that survive the double scaling limit (\ref{boost}).
First, because of the kinematics of the DLCQ boosts,
\begin{equation}\label{Ppm-limit}%
 P^{v'} =
\left(h+n-\frac{c}{24}\right)\frac{e^{-\gamma}}{ R},\qquad P^{u'} =
\left(h-\frac{c}{24}\right)\frac{e^{\gamma}}{ R}\,. %
\end{equation}
Keeping $P^{u'}$ finite in the  $\gamma \to \infty$ limit requires
$h=c/24$. This leads to
\begin{equation}
P^{v'} =n\cdot \frac{e^{-\gamma}}{ R}=\frac{n}{R_-}\ .%
\end{equation}
The DLCQ limit generates an infinite energy gap in the right-moving sector. Thus, keeping only finite energy excitations, it freezes to its ground state. The energy gap in the left-moving sector is kept finite. All physical finite energy states only carry momentum along the compact null direction $u'$. Therefore, the Hilbert space ${\cal H}$ of the DLCQ  of the original 2d non-chiral CFT is
\begin{equation}
 {\cal H} = \{  |{\rm anything}\rangle_L\otimes |c/24\rangle_R\}\,.%
\end{equation}
The chirality of the DLCQ theory spectrum can also be seen by studying which subset of the original AdS${}_3$ Virasoro generators \eqref{eq:virasoro} remains under the double scaling limit (\ref{boost}) \cite{Balasubramanian:2009bg}.

The null self-dual orbifold is now easily interpreted. Since it has the same boundary structure as the spacelike self-dual orbifold, it also corresponds to a state in the DLCQ theory. But, it describes its vanishing momentum $p^+=0$ sector. 

The physical interpretation of the pinching $\ZZ_N$ orbifold must be different. Since the latter corresponds to sending the radius $R$ of the limiting boundary cylinder to zero, $R\sim \epsilon\to 0$, it certainly generates an infinite gap in the untwisted sector for both chiral sectors of the initial 2d CFT. The only surviving untwisted finite excitations are those corresponding to $h=n=0$. Thus, given a CFT with a fixed central charge $c$, this near horizon limit freezes out both left and right moving sectors, leaving us with the Hilbert space:
\begin{equation}\label{double-decoupled-Hilbert}
 {\cal H} = \{  |c/24 \rangle_L\otimes |c/24\rangle_R\}\,.
\end{equation}
But this simple argument does, a priori, not capture the full perturbative string spectrum. It would be interesting to extend the results in \cite{Martinec:2001cf} for this case, clarifying whether there exists any massless twisted modes and whether their dynamics simplifies in the $N\to \infty$ limit.

\subsubsection{Asymptotic symmetries and the chiral Virasoro algebra}

Our arguments above suggest that half of the available UV Virasoro generators become irrelevant for the IR physics captured by the near horizon extremal geometry. One way of checking this would be to study how these generators \eqref{eq:virasorovf} transform under the infinite Lorentz boost defining the DLCQ limit of the original 2d non-chiral CFT. Instead, one can study the asymptotic symmetry group preserving the near horizon geometry.

The problem is then to identify the subset of non-trivial diffeomorphisms preserving the boundary conditions defining an asymptotically spacelike self-dual orbifold. Since these spaces are locally AdS${}_3$, the analysis must be very similar to the one in \cite{Brown-Henneaux}. The main physical insight comes from the observation that a general deformation of $\delta g_{\hat u\hat u}$,
$\delta g_{\hat v \hat v}$ would be non-extremal and would thus
excite both chiral sectors of the UV dual CFT.  By contrast, we want
to restrict to extremal excitations. Imposing the extremality condition $L_0=c/24$ 
requires a more stringent boundary condition on the variations in $g_{\hat v \hat
v}$. In \cite{Balasubramanian:2009bg}, it was suggested to replace the boundary
condition on $g_{\hat v \hat v}$ by
\begin{equation}\label{new-gvv}
\delta g_{\hat{v}\hat{v}}\sim {\cal O}(r^{-2}) \,. 
\end{equation}
The connection to the Brown-Henneaux diffeomorphisms is now made
explicit: the diffeomorphisms generated by $\zeta=\zeta^\alpha
\partial_\alpha$ are exactly of the form
\begin{subequations}\label{BH-diffeos}%
\begin{align}
\zeta^u & =    2 f(u) +\frac{1}{2r^2}g''(v)
+ {\cal O}(r^{-4})  \\
\zeta^{v} & =  2 g(v) + \frac{1}{2r^2}f''(u)+ {\cal O}(r^{-4}),\\
\zeta^r & =  -r\left(f'(u)+ g'(v)\right) + {\cal O}(r^{-1}) %
\end{align}
\end{subequations}
They satisfy the constraint 
\begin{equation}
 g'''(v)=0 \quad \Longrightarrow \quad g(v)=A+B\,v + C \, v^2
\label{B-condition}\ .
\end{equation}
implementing the boundary condition (\ref{new-gvv}). Thus, one set of allowed diffeomorphisms is
specified by a periodic function $f(u)=f(u+2\pi)$. The analysis of
generators of these diffeomorphisms follows directly from those of
Brown and Henneaux , leading to a \emph{chiral Virasoro algebra}
at central charge $c= 3\ell/2G_3$. The remaining three parameter family of diffeomorphisms in
(\ref{B-condition}) describes the $\SL(2,\RR)$ isometries of the spacelike
self-dual orbifold and act trivially on the Hilbert space \cite{Balasubramanian:2009bg}.

Notice this construction mimics the phenomena reported for extremal Kerr and for general extremal black holes in d=4,5 dimensions described in section~\ref{sec:gebh}. Indeed, one starts with an extremal BTZ black hole, whose near horizon geometry consists of an S${}^1$ fibration over AdS${}_2$. The latter has isometry group $\SL(2,\,\RR)\times \U(1)$. This gets enlarged to a full chiral Virasoro, but the infinite asymptotic symmetry algebra extends its $\U(1)$ global part, whereas the $\SL(2,\,\RR)$ acts trivially. 

The virtue of the AdS${}_3$ set-up is the existence of a well-defined UV dual CFT description allowing us to interpret the near horizon limit as an IR limit that turns out to be equivalent to a DLCQ limit. This gives some validity to the general arguments given in previous sections, but it does not clarify whether the Kerr/CFT correspondence is correct. Indeed, one of the original motivations in \cite{Balasubramanian:2009bg} was to argue that the mechanism behind the Kerr/CFT conjecture was three dimensional in nature. Some further supporting evidence was given in \cite{finn-ale2}, where it was explicitly checked that the twisted CFT advocated in \cite{Hartman:2008dq} was consistent with an AdS${}_3$ reduction to AdS${}_2$\footnote{For a different emphasis on how to use the AdS${}_3$/CFT${}_2$ correspondence to learn how to formulate the AdS${}_2$/CFT${}_1$ correspondence, see \cite{Gupta:2008ki}.}.

\subsubsection{Large N limits, double scaling limits and existence of dynamics}

The double scaling limit discussed in subsection~\ref{sec:dynamics} is easy to implement in the AdS${}_3$ context \cite{evh-btz}. In gravity, one is forced to consider a vanishing horizon limit, 
$r_\pm\to \epsilon r_\pm$ with $\epsilon\to 0$, while scaling Newton's constant $G_3\to \epsilon G_3$, to keep the full entropy finite. One can achieve this on BTZ metrics by combining a near horizon limit with this double scaling limit
\begin{equation}\label{NH-near-massless}%
r_\pm=\epsilon \rho_\pm\,,\quad r=\epsilon\rho_++\epsilon\rho\,,\quad t=\epsilon^{-1}\tau\,, \quad   \phi=\epsilon^{-1}\psi\,,\quad G_3=\epsilon \tilde G_3 \quad
\epsilon\to 0\,.
\end{equation}%
This reproduces the double scaling limit one could have considered in terms of 2d CFT data. Indeed, the latter corresponds to $R\to 0,\,c\to\infty,\, cR=\textrm{fixed}$, where $R$ stands for the radius of the original cylinder. Notice the scaling of $R$ is related to the presence of a pinching orbifold in the gravity construction. Both transformations achieve $c\to c/\epsilon$,  $L_0-\frac{c}{24}\to \epsilon(L_0-\frac{c}{24})$ and $\bar{L}_0-\frac{c}{24}\to \epsilon(\bar{L}_0-\frac{c}{24})$ as
required.

One can gain some intuition about the different 2d CFT's appearing in this discussion by thinking about the CFT dual to the D1-D5 system. This 2d CFT can be described by a 2d sigma model with $N=(4,4)$ supersymmetry with a target which can be thought of as a suitable
symmetric product ${\rm Sym}^{N_1N_5}({\cal M}_4)$. Rescaling the central charge is like rescaling $N_1N_5$. Therefore, one expects a relation of the form
\be \label{j11}
{\rm CFT}_{\rm new} \approx {\rm Sym}^{K}({\rm CFT}_{\rm old}).
\ee
This clearly only makes sense when $K=1/\epsilon$ is an integer. The new CFT
has a long string sector which is directly inherited from the old theory. Its Virasoro
algebra is related to that of the original CFT using standard orbifold technology. Explicitly,
given a set of generators $\{L_n\}$, consider the subalgebra with generators\footnote{Notice the transformation for the generator $l_0$ is due to the fact that we were working on the plane. Indeed, if we would have worked on the cylinder, the transformation is the expected one :
$$
  l^{\text{cyl}}_n \equiv \frac{1}{K} L^{\text{cyl}}_{nK}\,, \quad n\neq 0\,, \quad \quad
  l^{\text{cyl}}_0 \equiv \frac{1}{K} L^{\text{cyl}}_{0}\,.
$$
We now see that the transformation quoted on the plane makes sure the above cylinder transformation brings us back to the plane.}
\begin{equation}
  l_n \equiv \frac{1}{K}\,L_{nK}\,, \quad n\neq 0\,, \quad \quad \quad l_0 \equiv \frac{1}{K}(L_0-\frac{c}{24}) + \frac{c}{24} K\,.
\end{equation}
It is then straightforward to see that $l_n$ also form a Virasoro algebra with central charge $c^\prime = cK$ and
that the spectrum of $l_0$ has a spacing of $1/K$ compared to that of $L_0$.

Since the long string sector in an orbifold theory tends to dominate the entropy, this provides a natural explanation for the constancy of the entropy in the bulk. Further work is required to clarify the fate of these constructions given the $K\to \infty$ nature of the limit and the role twisted states must play in the resolution of the singular classical geometries corresponding to the nearly massless BTZ and its near horizon geometries involved in these limits.

\section{Holography, Thermodynamics \& CFTs}

The {\it holographic principle} suggests the relation between gravity and thermodynamics should go beyond black hole physics. In this section, this extension will be briefly considered focusing on the following aspects
\begin{romanlist}[(iii)]
\item One may argue the holographic principle captures some of the non-local aspects of General Relativity responsible, for example, for black hole entropy, a concept intrinsically tied to a global property of spacetime, such as the existence of an event horizon. But General Relativity is formulated in terms of Einstein's equations through the Equivalence principle. How is the local perspective provided by the latter related to thermodynamics ?
\item We previously saw how coarse graining of microscopic information gave rise to the emergence of classical spacetime and curvature singularities\footnote{For an idea relating the entanglement of degrees of freedom in quantum gravity with the connectivity of emergent classical spacetime, see \cite{VanRaamsdonk:2010pw}.}. The holographic principle does not necessarily require the existence of classical spacetime in both sides of the boundary where it is applied.
Can one infer some fundamental property of gravity by examining the known laws of physics (matter) taking place close to this boundary ?
\item Semiclassical methods applied to extremal black holes claimed the emergence of 
some effective conformal field theory description justifying the universality of the Bekenstein-Hawking formula \eqref{eq:beh-haw} through the universality of the Cardy formula \eqref{eq:cardy}. The holographic principle suggests this relation may be more general. Can we provide any evidence in favour of this expectation ?
\end{romanlist}

\begin{figure}[pb]
\centerline{\psfig{file=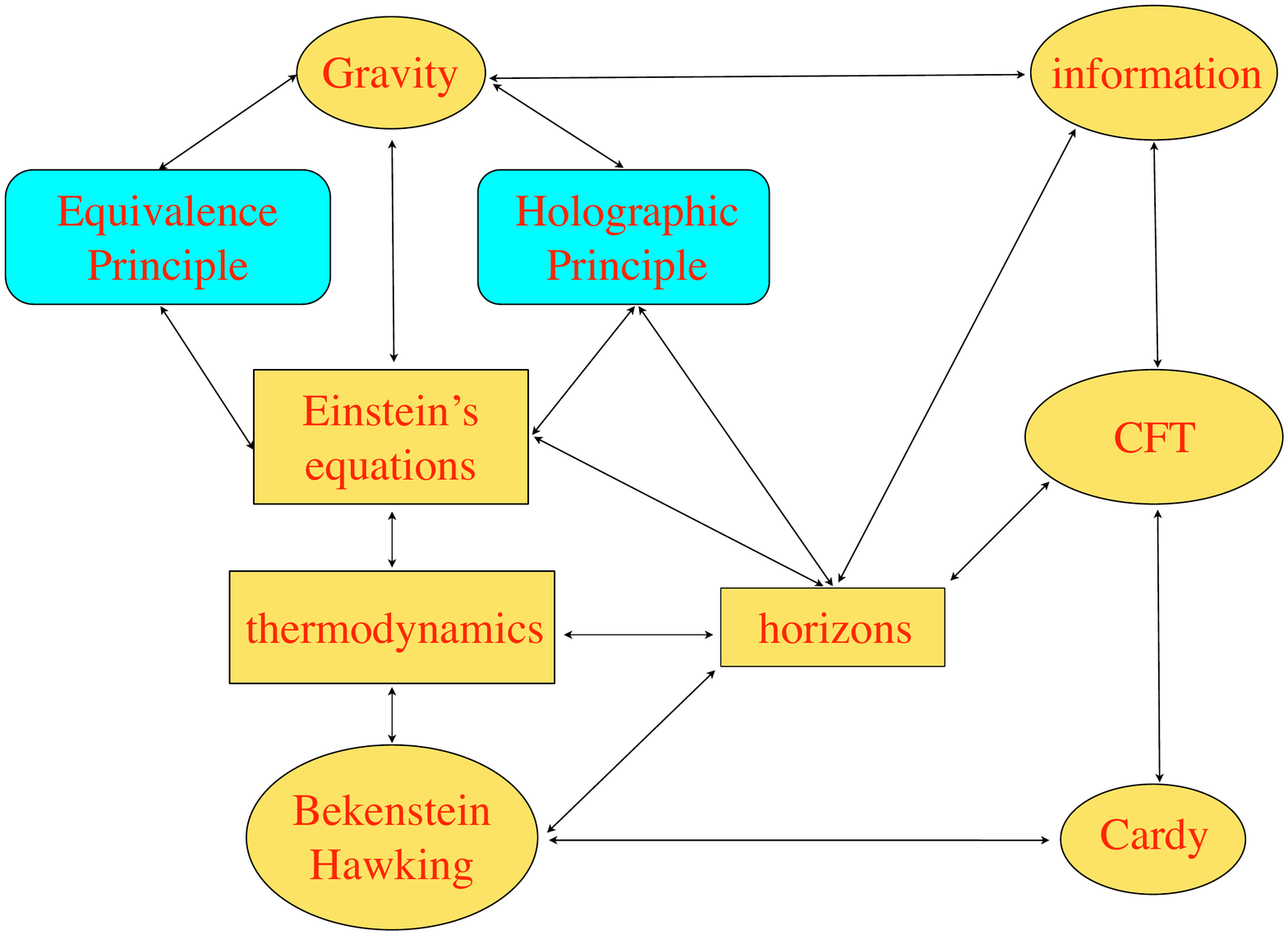,width=15cm}}
\vspace*{8pt}
\caption{Different perspectives on gravitational first principle approaches. \label{fig6}}
\end{figure}

\subsection{Local observers, thermodynamics and Einstein's equations}

Despite the {\it locality} of Einstein's equations, the holographic principle captures some non-local features of gravity which are otherwise deeply hidden in the standard formulation of General Relativity based on the {\it Equivalence Principle}. The latter guarantees that any neighbourhood of a point looks like flat spacetime. If we believe classical gravity is thermodynamical in nature in general, how can we reconcile both perspectives ?

Jacobson analysed this local perspective proving that Einstein's equations appear as an equation of state \cite{jacobson} under some set of assumptions
\begin{itemlist}
\item[1.] {\it Heat} is identified with energy flux across a causal horizon. The latter affects the gravity field, but it is unobservable from outside the horizon. This is a sensible identification because heat, in thermodynamics, stands for transfer of energy among microscopic constituents of the system which are unobservable for the macroscopic observer. In gravity, one identifies the "system" with the degrees of freedom outside a causal horizon. The latter provides a causal barrier replacing the more common diathermic wall in thermodynamics.
\item[2.] The holographic principle holds. Consequently, there exists a universal entropy density $\alpha$ per unit horizon area, i.e. $\delta S = \alpha\, \delta A$.
\item[3.] The Equivalence Principle holds. Taking the perspective of a locally non-inertial observer, due to Unruh's effect \cite{Unruh:1976db}, we know the local vacuum fluctuations of a quantum field are thermal, with temperature $T=\hbar\,\kappa/2\pi$, $\kappa$ being the acceleration of the local observer.
\item[4.] Existence of {\it local equilibrium} to ensure the use of equilibrium thermodynamics. This translates into a vanishing condition for the expansion and shear of the local causal horizon.
\end{itemlist}

Technically, the equivalence principle allows one to work in some locally flat region around a point $p$. Our "system" is identified with the past horizon of a small spacelike 2-surface ${\cal P}$ through $p$. There exists an approximate Killing vector field $\chi^a$ generating boosts orthogonal to ${\cal P}$. This is precisely the local hamiltonian, which allows to identify the temperature of the vacuum fluctuations predicted by Unruh's effect to be as above, with $\kappa$ being the acceleration along the Killing orbit. 

Heat is clearly computed as the flux integral along the past horizon
\begin{equation}
  \delta Q = \int T_{ab}\chi^a\,d\Sigma^b\,.
\end{equation}
Both $\chi^a$ and $d\Sigma^b$ can be written in terms of the tangent vectors to the horizon $k^a$ and the affine parameter $\lambda$. The Clausius relation $\delta S = \delta Q/T$ following from equilibrium thermodynamics relates this to the entropy and the holographic principle relates the latter to the area of the local causal horizon, $\delta S = \alpha\,\delta A$. 
Finally, the change in area $\delta A$ is controlled by the propagation of congruences of null geodesics emanating from the horizon
\begin{equation}
  \delta A = \int \theta\, d\lambda\,dA\,.
\end{equation}
This is controlled by Raychaudhuri's equation
\begin{equation}
  \frac{d\theta}{d\lambda} = -\frac{1}{2}\theta^2 - \sigma^2- R_{ab}k^ak^b\,,
\end{equation}
which simplifies in our discussion due to the local equilibrium conditions imposed, i.e. both the square of the shear $\sigma^2$ and the expansion $\theta^2$ vanish at $p$. Thus their contribution to the rate of variation of the expansion is higher order. Hence, at lowest order $\theta=-\lambda\,R_{ab}k^ak^b$.

It was shown in \cite{jacobson} that equating both integrals one derives an identity for the integrands
\begin{equation}
  T_{ab}k^a\,k^b = \frac{\hbar\alpha}{2\pi}\,R_{ab}k^ak^b\,,
\end{equation}
for any value of the affine parameter $\lambda$. Its general solution depends on an arbitrary function $f$
\begin{equation}
  R_{ab}+ f\,g_{ab} = \frac{2\pi}{\hbar\,\alpha} T_{ab}\,,
\end{equation}
which gets fixed, using the divergence free nature of $T_{ab}$ and the contracted Bianchi identity, to $f=-R/2 + \Lambda$. This gives rise to the condition
\begin{equation}
  R_{ab} - \frac{R}{2}\,g_{ab} + \Lambda\,g_{ab} = \frac{2\pi}{\hbar\,\alpha}\,T_{ab}\,,
\end{equation}
which identifies the arbitrary constant $\Lambda$ as the cosmological constant. As usual, appealing to the correspondence principle, and matching with Poisson's equation in the newtonian limit allows one to find $G = \frac{1}{4\hbar\alpha}$, which agrees with the Bekenstein-Hawking relation, {\em i.e.} $\delta S = \delta A/\left(4\hbar\,G\right)$.

It is reassuring that by properly interpreting the physics of causal connection associated with local horizons and borrowing concepts from quantum field theory which are believed to be applicable as long as the curvature of spacetime is much larger than $\ell_p$, one learns that thermal equilibrium can only be maintained if the distortion of the causal structure of spacetime caused by the gravitational lensing due to matter energy is consistent with Einstein's equations.

Given the effective field theory nature of General Relativity, it is natural to wonder what the effects of higher order corrections to the Einstein-Hilbert action are to the above scheme. It was argued and proposed in \cite{Eling:2006aw} that such corrections are responsible for bringing the "system" out of equilibrium. The nature of the changes depends on the corrections turned on, but they can generically involve non-vanishing shear viscosity of the horizon and/or bulk viscosity entropy production.

\subsection{Classical gravity as an entropic force} 

There are several arguments suggesting that the notion of spacetime is an emergent concept. 
The AdS/CFT correspondence confirms this expectation by explicitly realising the holographic principle matching 4d quantum physics into 5d bulk physics. In this correspondence, the radial coordinate $r$ plays the role of some coarse graining scale which is typically interpreted as some field theory renormalization group energy scale \cite{Susskind:1998dq}. One of the virtues of the AdS/CFT is the resummation of many complicated strong interactions in terms of classical curved spacetime geometry. The process that goes from one description to the other exchanges an open string description, where there is {\it no} spacetime, with a closed string one, where the degrees of freedom are gravitational\footnote{For recent discussions on the conditions that a CFT must satisfy to have a gravity dual, see \cite{Heemskerk:2009pn,ElShowk:2011ag}.}.

In black hole physics, there is also a privileged radial coordinate measuring the coordinate distance to the event horizon. In all our previous discussions, a horizon provided a causal separation between our "system" and the rest of the universe. But either in our fuzzball  considerations, or in our semiclassical considerations, or in our applications of the holographic principle, either locally or globally, there was {\it no} necessity in assuming there is a reliable classical description in terms of gravitational degrees of freedom beyond the horizon. 

Combining these two ideas, it is tempting to consider the somehow vague possibility of coarse graining all the physical data up to a certain scale/distance $r$. Assuming that for larger values
of $r$ a classical spacetime description exists, one may study whether some important lesson can be learnt about gravity in this set-up by analysing the known laws of physics involving the interaction of the screen (surface where we stopped the coarse-graining) with the rest of matter existing outside of the screen.

This set-up was considered by Erik Verlinde in \cite{erik}\footnote{Similar ideas had been considered by T.~Padmanabhan \cite{Padmanabhan:2009vy}. Technically, his analysis is analogous. Conceptually, the emergence of spacetime and the potential relevance of coarse graining was not emphasised.}. His analysis emphasises the universal character of gravity may be intrinsically linked with information. More precisely,  it is the information associated with matter and its location that generates gravity through the changes in these variables. In Verlinde's approach, {\it gravity is an entropic force}.

I will very briefly review his argument for non-relativistic spherically symmetric and time translationally invariant matter distributions. Verlinde's further assumptions are
\begin{itemlist}
\item[1.] information is stored on surfaces, or screens, whose detailed dynamics are unknown. For many purposes, these screens can be thought of as stretched horizons in black hole physics, but applicable to more general holographic situations. See \cite{Bousso:2002ju} for a precise definition of holographic screen in a generic classical gravity context.
\item[2.] there exists a unique emergent direction, the radial one, along which the coarse graining occurs, as in the AdS/CFT correspondence set-up.
\item[3.] extending Bekentein's argument in black hole physics \cite{Bekenstein:1973ur}, one assumes the variation in the entropy of a given screen when a particle of mass $m$ is a Compton distance away from it, is linear in the distance $\Delta x$
\begin{equation}
  \Delta S = 2\pi\,k_B\,\frac{mc}{\hbar}\,\Delta x\,.
\end{equation}
\item[4.] the energy of the system is evenly distributed among all available bits in which the area of the screen can be discretised, $N=Ac^3/G\hbar$
\begin{equation}
  E = \frac{1}{2}Nk_B\,T
\end{equation}
\item[5.] Gravity is entropic originating from changes in the locations of matter through the relation
\begin{equation}
  F\,\Delta x = T\,\Delta S\,.
\end{equation}
Thus, the screen has a temperature, and by similar arguments to the ones reviewed previously, one takes this to be Unruh's temperature
\begin{equation}
  k_B\,T = \frac{\hbar\,a}{2\pi\,c}\,,
\end{equation}
with the important conceptual difference that this temperature is now required to achieve an acceleration on matter, and it is {\it not} due to the non-inertial nature of matter.
\end{itemlist}

It is now a simple matter of putting all these ingredients together, using further that the area of a given spherical screen is $A=4\pi\,R^2$ and $E=M\,c^2$, where $M$ stands for the total mass already available on the screen, that solving for $F$ reproduces Newton's law
\begin{equation}
  F = \frac{G\,M\,m}{R^2}\,.
\end{equation}

These arguments can be generalised to more general matter distributions and extended to relativistic set-ups. We refer the readers to Verlinde's original work \cite{erik} for further details.

\subsection{Horizons \& emergent CFTs}

Many of the relations discussed so far, summarised in figure \ref{fig6}, may encourage us to ask whether there is some universal relation between the Bekenstein-Hawking formula \eqref{eq:beh-haw} and Cardy's formula \eqref{eq:cardy}. For extremal black holes, we reviewed some of the arguments and evidence for the existence of a conformal field theory in the IR capturing the degeneracy of the vacuum state and potentially keeping some information about the dynamics of the system below some cut-off. The holographic principle would suggest this is a much more general phenomena, including {\it any} horizon in General Relativity. Jacobson's argument \cite{jacobson} suggests a strong link between the equivalence principle, thermodynamics and Einstein's equations based on the conceptual identification of a {\it local} horizon with a causal boundary, the latter replacing the standard notion of a diathermic wall in thermodynamics. Verlinde's arguments \cite{erik} emphasise the notion of screen as an information storage device appearing as a boundary between parts of our physical system allowing a spacetime description or not. But these arguments do not tell us whether there is any approximate microscopic description compatible with the macroscopic laws inferred from their assumptions and analysis. Since the notion of screen is intimately related to that of a stretched horizon, the conjecture that any horizon in General Relativity may allow an approximate CFT at low enough energies capturing the effective dynamics of the relevant degrees of freedom on the screen could provide a more precise link relating the {\it universality} of Bekenstein-Hawking entropy formula \eqref{eq:beh-haw} and the {\it universality} of Cardy's formula \eqref{eq:cardy} in 2d CFT. 

The covariant semiclassical methods reviewed in section \ref{sec:ext-cft} can provide a technical tool to test this idea. Indeed, the existence of a non-trivial central charge when computing the Dirac bracket between conserved charges generating the asymptotic symmetry group crucially depends on the existence of a boundary \eqref{eq:bccharge}. Following all our previous considerations, it is natural to explore this construction when the boundary is identified with a horizon itself. This is the approach originally considered by Carlip \cite{carlip} and Solodukhin \cite{sol1}. 

In Carlip's approach \cite{carlip}, one assumes the existence of a local {\it Killing} horizon and computes, using a covariant formalism, the algebra of conserved charges of the subset of diffeomorphisms preserving such horizon structure. Given some technical assumptions, the main results of his analysis are the existence of a {\it chiral} Virasoro algebra with central charge and energy
\begin{equation}
  \frac{c}{12} = \frac{A}{8\pi\,\kappa\,G}\,, \quad \quad \text{and} \quad \quad  L_0 - \frac{c}{24} = \frac{A\,\kappa}{8\pi\,G}\,,
\label{eq:carlip}
\end{equation}
where $A$ stands for the area of the Killing horizon and $\kappa$ for the surface gravity of the horizon surface. Using Cardy's formula would reproduce the holographic principle prediction \eqref{eq:beh-haw}. Notice this dictionary would also be consistent with the identification of the CFT temperature with Unruh's temperature \cite{Unruh:1976db} $T=\kappa/(2\pi)$, which makes sense given the local nature of the argument. Just as in Jacobson's argument \cite{jacobson} the normalisation of the tangent vector to the horizon $\chi^a\propto k^a$ was not relevant to derive Einstein's equations, here the entropy is invariant under the transformations 
$c\to \kappa\,c$ and $L_0-\frac{c}{24}\to \left(L_0-\frac{c}{24}\right)\,\kappa^{-1}$. These would correspond to picking a different hamiltonian and CFT temperature\footnote{
This approach has recently been reconsidered in an attempt to extend the extremal BH/CFT correspondence to non-extremal black holes in \cite{recent-carlip}.}.

In Solodukhin's approach \cite{sol1}, one assumes  the existence of an {\it apparent} horizon in an 
spherically symmetric configuration
\begin{equation}
  ds^2 = \gamma_{ab}\,dx^adx^b + r^2(x^a,\,x^b)\,d\Omega_2^2\,,
 \label{mansatz}
\end{equation}
described by a curve ${\mathcal H}$ satisfying
\begin{equation}
  \gamma^{ab}\nabla_ar\,\nabla_b r\,|_{{\mathcal H}} = 0\,.
\end{equation}
Since any 2d metric is conformally flat, $-2e^{2\sigma}dx^+dx^-$, one can choose coordinates so that locally the curve ${\mathcal H}$ becomes $x^-=0$. Then, the above condition reduces to
\begin{equation}
  \partial_+ r(x^+,\,x^-)|_{x^-=0} = 0\,
\end{equation}
Expanding the radius close to the horizon $r = r_h + \lambda (x_+)\,x_- + \CO(x_-^2)$, we find
$\partial_+ r =\lambda^\prime\,x_- + \CO(x_-)^2$. It can then be shown that the vector field $\xi^\mu$ with only non-trivial component $\xi^+ = g(x_+)$ preserves the existence of the apparent horizon \cite{sol1}. To argue the existence of a CFT in the vicinity of the horizon, Solodukhin writes the Einstein-Hilbert action for the class of metrics \eqref{mansatz}. This is a 2d effective action involving the 2d metric and one scalar describing the radius $r$. This is rewritten as a Liouville theory (after some redefinition of the scalar field), and shows that the energy momentum tensor for such theory has vanishing trace very close to the horizon. As usual, the energy momentum tensor generates conformal transformations. Using the Poisson algebra from the 2d Liouville theory, one computes the commutator of these generators, finding a non-trivial central charge \cite{sol1}
\begin{equation}
  c= 3q^2\,S_{BH}\,.
\end{equation}
The dependence on $q^2$ is somehow analogous to the dependence on $\kappa$ in \eqref{eq:carlip}. Here, it appears due to some arbitrariness in the field redefinition involved when rewriting the effective 2d gravity theory as a Liouville theory, but it cancels out when computing the entropy because the energy $L_0$ scales like $q^{-2}$ \cite{sol1}. 

Notice Solodhukin's argument is general and applies to apparent, and even dynamical, horizons. Either way, these semiclassical considerations suggest the emergence of an approximate 1+1 chiral CFT at low enough energies, that is, close enough to the local horizon in spacetime. It is interesting to point out that given such an effective CFT description of the degrees of freedom on a horizon, the energy of the system satisfies some notion of equipartition, as was assumed in Verlinde's derivation of Newton's law \cite{erik}. Indeed, from Cardy's formula 
\begin{equation}
  L_0 - \frac{c}{24} = \frac{1}{2} S\,T \quad \Rightarrow \quad E = \Delta - c/24 \propto Nk_B\,T
\end{equation}
where we have introduced the notion of number of bits $N$ as $N=A/G$. 

These considerations require a more precise formulation but they are encouraging since they suggest a potentially much stronger version of a holographic relation between gravity and CFT.

\subsection{Open \& related questions}

There are many important foundational questions that remain open. Given the topics covered in these notes, it is natural to highlight the following subset
\begin{romanlist}
\item Develop new ideas to understand the microscopics of non-extremal black holes and whether there exists any relation with the physics of Rindler space (its near-horizon geometry).
\item Extend the microscopic model for extremal black holes described in section \ref{sec:nonbpsmicro} in the presence of angular momentum, clarifying its relation to \cite{emparan-horowitz,Emparan:2007en}. Develop their open string dual description in terms of quiver theories by extending Denef's work in the BPS branch to this sector \cite{Denef:2002ru}. This could ideally help to connect these microscopic considerations with the emerging conformal field theories derived out of semiclassical considerations, as discussed in subsection \ref{sec:ext-cft}.
\item Develop gauge theory techniques to compute correlations functions in the AdS/CFT correspondence in non-trivial heavy states to achieve a more accurate mathematical formulation of the information paradox \cite{Hawking:1976ra} along the lines outlined in subsection~\ref{sec:typstates}.
\item Develop the formulation of holography in de Sitter and Minkowski spacetimes. Apply some of the black hole ideas reviewed here to more general holographic scenarios such as time dependent ones, cosmology and their classical singularities. It is tempting to speculate that the Big Bang singularity is a consequence of the loss of quantum information regarding the initial state of the universe. Entanglement entropy \cite{e-entropy} and its AdS/CFT formulation \cite{Ryu:2006bv} could provide some technical tools to handle some of these time dependent questions. 
\end{romanlist}

It is worth mentioning two results regarding this last point :
\begin{romanlist}
\item It was shown in \cite{Cai:2005ra,Akbar:2006kj} that Einstein's equations in Friedmann-Robertson-Walker spacetimes can be reinterpreted in terms of a 1st law of thermodynamics, if energy is properly defined, as in  \cite{Misner:1964je}, and the holographic principle is applied to apparent horizons. In some vague sense, time in this set-up plays a similar technical role as the radial coordinate in some of our previous considerations.
\item A potential example of a holographic duality involving a flat spacetime was presented in 
 \cite{Bredberg:2011jq} , and further supported in \cite{Compere:2011dx}, by working out a precise
relation between incompressible non-relativistic fluids in $(d+1)$ dimensions satisfying the Navier-Stokes equation and $(d+2)$-dimensional Ricci flat metrics. More precisely, based on previous work \cite{Fouxon:2008tb,Bhattacharyya:2008kq,Eling:2009pb,Bredberg:2010ky} and working in a hydrodynamic non-relativistic limit,  a bulk metric can be constructed using the hydrodynamic expansion methods developed in \cite{Bhattacharyya:2008jc} and reviewed in \cite{Rangamani:2009xk} for any non-relativistic fluid satisfying its equations of motion in one less dimension.
\end{romanlist}

Many of the conceptual ideas discussed in these notes are compatible with the {\it membrane paradigm} framework  \cite{membrane}, where the causal properties of an event horizon suggested formulating the latter as an effective membrane. For a nice discussion regarding the connection between these and the fuzzball ideas, see \cite{samir-membrane}. 

There is a large body of literature giving evidence, from different perspectives, that classical gravity is thermodynamical in nature. The universal connection between thermodynamics and statistical mechanics led the way to the notion of holography. The duality between open and closed strings led to the discovery of the AdS/CFT correspondence. They all definitely shed new light in the fascinating quest to understand what gravity \& spacetime are. The future will bring, no doubt, further surprises \& revelations in the resolution of some of the more fundamental puzzles in theoretical physics.

\section*{Acknowledgments}

JS would like to thank V.~Balasubramanian, A.~Castro, B.~Czech, J.~de Boer, G.~Compere, J.M.~Figueroa-O'Farrill, E.~Gimon, Y-H.~He, V.~Hubeny, V.~Jejjala, M.~Johnstone, P.~Kraus, K.~Larjo, F.~Larsen, J.~Lucietti, D.~Marolf, M.~Rangamani, S.~Ross, A.~Sen, M.~Sheikh-Jabbari for collaboration and discussions on some of the topics reported in this review. JS would like to thank the international Summer School SAM 09 in Frascati for an stimulating environment and the research institutions including the Weizmann Institute, the KITP and the universities of Michigan, Pennsylvania, Imperial College, Queen Mary, Amsterdam, Durham, Johannesburgh, Oviedo, Oxford, Brussels and McGill where seminars covering the material presented here were given. The work of J.S. was partially supported by the Engineering and Physical Sciences Research Council [grant number EP/G007985/1].




\end{document}